\def\year{2020}\relax
\documentclass[letterpaper]{article} 
\usepackage{aaai20}  
\usepackage{times}  
\usepackage{helvet} 
\usepackage{courier}  
\usepackage[hyphens]{url}  
\usepackage{graphicx} 
\usepackage{graphicx}  
\usepackage{amsfonts}
\usepackage{amssymb}
\usepackage{amsmath, amsthm} 
\usepackage{subcaption}
\usepackage{algorithm, algorithmic}
\usepackage{tabularx}
\usepackage{microtype}
\usepackage{booktabs}

\newtheorem{definition}{Definition}
\newtheorem{theorem}{Theorem}

\newcommand\blfootnote[1]{%
	\begingroup
	\renewcommand\thefootnote{}\footnote{#1}%
	\addtocounter{footnote}{-1}%
	\endgroup
}

\title{Norm-Explicit Quantization: Improving Vector Quantization for Maximum Inner Product Search }
\author{
	\Large 
	{Xinyan Dai*, 
		Xiao Yan*, 
		Kelvin K. W. Ng, 
		Jie Liu, 
		James Cheng}\\ 
		The Chinese University of Hong Kong\\
	\{xydai, xyan, kwng6, jliu, jcheng\}@cse.cuhk.edu.hk
}
\begin{document}

\maketitle

\begin{abstract}
		
Vector quantization (VQ) techniques are widely used in similarity search for data compression, computation acceleration and etc. Originally designed for Euclidean distance, existing VQ techniques (e.g., PQ, AQ) explicitly or implicitly minimize the quantization error. In this paper, we present a new angle to analyze the quantization error, which decomposes the quantization error into norm error and direction error. We show that quantization errors in norm have much higher influence on inner products than quantization errors in direction, and small quantization error does not necessarily lead to good performance in maximum inner product search (MIPS). Based on this observation, we propose norm-explicit quantization (NEQ) --- a general paradigm that improves existing VQ techniques for MIPS. NEQ quantizes the norms of items in a dataset explicitly to reduce errors in norm, which is crucial for MIPS. For the direction vectors, NEQ can simply reuse an existing VQ technique to quantize them without modification. We conducted extensive experiments on a variety of datasets and parameter configurations. The experimental results show that NEQ improves the performance of various VQ techniques for MIPS, including PQ, OPQ, RQ and AQ.

\end{abstract}

\section{Introduction}\label{sec:intro}

\blfootnote{*Co-first authors are ranked alphabetically. Correspondence to Xiao Yan.}

Given a dataset $\mathcal{X}\subset\mathbb{R}^d$ that contains $n$ vectors (also called items) and a query $q\in\mathbb{R}^d$, maximum inner product search (MIPS) finds an item $x^{*}$ that has the largest inner product with the query,
\begin{equation}\label{equ:mips}
x^{*}=\arg \max_{x\in\mathcal{X}}{q^{\top}x}.
\end{equation}
The definition of MIPS can be easily extended to top-$k$ inner product search, which is used more commonly in practice. MIPS has many important applications such as recommendation based on user and item embeddings~\cite{koren:mf}, multi-class classification with linear classifier~\cite{dean:obj}, and object matching in computer vision~\cite{felzens:obj}. Recently, MIPS is also used for Bayesian interference~\cite{mussmann:mipsBayes}, memory network training~\cite{chandar:mipsmemory} and reinforcement learning~\cite{jun:mipsreinforcement}.

\textbf{Vector quantization (VQ).} VQ quantizes items in the dataset with $M$ codebooks~$\mathcal{C}^1, \mathcal{C}^2,\dots,\mathcal{C}^M$. Each codebook $\mathcal{C}^m$ contains $K$ codewords and each codeword is a $d$-dimensional vector, i.e., $\mathcal{C}^m=\{c^m[1],c^m[2],...,c^m[K]\}$, $c^m[k]\in\mathbb{R}^d$ for $1\le m \le M$ and $1\le k \le K$. Denote $i^m_x$ as the index of the codeword in codebook $\mathcal{C}^m$ that item $x$ maps to, then $x$ is approximated by $\tilde{x}=\sum_{m=1}^{M}c^m[i^m_x]$. Therefore, the inner product between query $q$ and item $x$, i.e., $q^{\top}x$, is approximated by $q^{\top}\tilde{x}=\sum_{m=1}^{M}q^{\top}c^m[i^m_x]$. There are a number of VQ algorithms with different quantization strategies and codebook learning procedures, such as product quantization (PQ)~\cite{h:pq}, optimized product quantization (OPQ)~\cite{kaiming:opq}, residual quantization (RQ)~\cite{chen:rq} and additive quantization (AQ)~\cite{babenko:aq}. We describe them in greater details in Section~\ref{sec:background}.

VQ can be used for~\textit{data compression},~\textit{fast inner product computation} and~\textit{candidate generation} in MIPS. For data compression, the $M$ codeword indexes \{$i^1_x,i^2_x,...,i^M_x$\} is stored instead of the original $d$-dimensional vector $x$, which enables storing very large datasets (e.g., with 1 billion items) in the main memory of a single machine~\cite{johnson:faiss}. When the inner products between query $q$ and all codewords are precomputed and stored in look-up tables, the approximate inner product of an item (i.e., $q^{\top}\tilde{x}$) can be computed with a complexity of $O(M)$ instead of $O(d)$. With two codebooks, VQ can use the efficient multi-index algorithm~\cite{babenkol:imi} to generate candidates for MIPS. Note that VQ is orthogonal to existing MIPS algorithms, such as tree-based methods~\cite{koenigstein:retrival,ram:cone}, locality sensitive hashing (LSH) based methods~\cite{neyshabur:simple-lsh,shrivastava:alsh}, proximity graph based method~\cite{morozov:graphmips} and pruning based methods~\cite{li:fexipro,teflioudi:lemp}. These algorithms focus on generating good candidates for MIPS, while VQ focuses on data compression and computation acceleration. Actually, VQ can be used as a component of these algorithms for compression and fast computation as in~\cite{douze:linkcode}.

When using VQ for similarity search, the primary performance indicator is the quality of the similarity value calculated with the codebook-based approximation $\tilde{x}$. Existing VQ techniques were primarily designed for Euclidean nearest neighbor search (Euclidean NNS) instead of MIPS. They minimize the quantization error ($\Vert x-\tilde{x} \Vert$) explicitly or implicitly because it provides an upper bound for the error of the codebook based approximate Euclidean distance, i.e., $\left \vert \Vert x- q \Vert -  \Vert \tilde{x} -q \Vert \right \vert \le \Vert x- \tilde{x} \Vert$. However, inner product is different from Euclidean distance in several important aspects. In particular, inner product does not satisfy the triangle inequality and non-negativity. The inner product between an item and itself (i.e., $x^{\top}x$) is not guaranteed to be the largest, while self-distance (i.e., $\Vert x- x\Vert$) is guaranteed to be the smallest for Euclidean distance. These differences prompt us to ask the following two questions: \textit{Does minimizing quantization error necessarily lead to good performance for MIPS? Do we need a different design principle for VQ techniques when used for MIPS (than for Euclidean NNS)?}

To answer these questions, we start by analyzing the quantization errors of VQ techniques from a new angle. Instead of treating the quantization error $\Vert x-\tilde{x} \Vert$ as a whole, we decompose it into two parts: \textit{norm error} ($\Vert x\Vert - \Vert \tilde{x} \Vert$) and \textit{angular error} ($1-\frac{x^{\top}\tilde{x}}{\Vert x\Vert \Vert \tilde{x} \Vert}$). We found that norm error has a more significant influence on inner product than angular error. Based on this observation, we propose \textit{norm-explicit quantization} (\textit{NEQ}), which quantizes  the norm $\Vert x \Vert$ and the unit-norm direction vector $x/\Vert x\Vert$ separately. Quantizing norm explicitly using dedicated codebooks allows to reduce errors in norm, which is beneficial for MIPS. The direction vector can be quantized using existing VQ techniques without modification. NEQ is simple in that the complexity of both codebook learning and approximate inner product computation is not increased compared with the baseline VQ technique used for direction vector quantization. More importantly, NEQ is general and powerful in that it can significantly boost the performance of many existing VQ techniques for MIPS.

We evaluated NEQ on four popular benchmark datasets, where the cardinalities of the datasets range from 17K to 100M and their norm distributions are significantly different. The experimental results show that NEQ improves the performance of PQ~\cite{h:pq}, OPQ~\cite{kaiming:opq}, RQ~\cite{chen:rq} and AQ~\cite{babenko:aq} for MIPS consistently on all datasets and parameter configurations (e.g., the number of codebooks and the required top-$k$ items). NEQ also significantly outperforms the state-of-the-art LSH-based MIPS methods and provides better time-recall performance than the graph-based ip-NSW algorithm.

\textbf{Contributions.} Our contributions are three-folds. First, we challenge the common wisdom of minimizing the quantization error in existing VQ techniques and questioned whether it is a suitable design principle for MIPS. Second, we show that norm error has more significant influence on inner product than angular error, which leads to a more suitable design principle for MIPS. Third, we propose NEQ, a general framework that can be seamlessly combined with existing VQ techniques and consistently improves their performance for MIPS, which is beneficial to applications that involve MIPS.

\section{Related Work}\label{sec:background}

In this section, we introduce some popular VQ techniques to facilitate further discussion and discuss the relation between NEQ and some related work.

\textbf{PQ and OPQ}. PQ~\cite{h:pq} first generates $M$ sub-datasets $\mathcal{X}^1,\mathcal{X}^2,...,\mathcal{X}^M$ for the original dataset, each containing $d'=d/M$ features from all items. K-means is used to learn a codebook on each sub-dataset independently and each codeword is a $d'$-dimensional vector. An item $x$ is approximated by the concatenation of its corresponding codewords from each of the codebooks, i.e., $\tilde{x}=[c^1[i^1_x],c^2[i^2_x],...,c^M[i^M_x]]$. OPQ~\cite{kaiming:opq} uses an orthonormal matrix $R$ to rotate the items by $Rx$ before applying PQ. OPQ achieves lower quantization error when the features are correlated or some features have larger variance than others. However, codebook learning is more complex for OPQ as it involves multiple rounds of alternating optimization of the codebooks and the rotation matrix $R$.   

\textbf{RQ and AQ}. Different from PQ and OPQ, in RQ~\cite{chen:rq} every codebook covers all features and each codeword is a $d$-dimensional vector. The original data are used to train the first codebook with K-means and the residues ($x-c^1[i^1_x]$) are used to train the second codebook. This process is recursive in that the $m$-th codebook is trained with the residues from the previous ($m-1$) codebooks. Similar to RQ, each codebook in AQ~\cite{babenko:aq} also covers all features. AQ improves RQ by jointly optimizing all the $M$ codebooks. Beam search is used for encoding (finding the optimal codeword indexes of an item in the codebooks) with given codebooks and a least-square formulation is used to optimize the codebooks under given encoding. 

In addition to the VQ techniques introduced above, there are many other VQ techniques, such as CQ~\cite{zhang:cq}, TQ~\cite{babenko:treeq} LOPQ~\cite{kalantidis:lopq} and LSQ~\cite{martinez2016:LSQ}. Although these VQ techniques differ in their quantization strategies (e.g., partitioning the features or not) and the codebook learning algorithms (e.g., K-means or alternating minimization), all of them explicitly or implicitly minimize the quantization error $\Vert x- \tilde{x} \Vert$, which is believed to provide good performance for Euclidean NNS. In the next section, we show that this principle does not apply for MIPS.   

\textbf{Existing work}. Similar to some other VQ algorithms used for similarity search (e.g., PQ and RQ), the prototype of NEQ can also be found in earlier researches on signal compression. The shape-gain algorithm~\cite{gersho2012vector} separately quantizes the magnitude and direction of a signal to achieve efficiency with some loss in accuracy. Instead of hurting accuracy, NEQ shows that the separate quantization of norm and direction actually improves performance for MIPS. A recent work, multi-scale quantization~\cite{wu2017} also explicitly quantizes the norm and the motivation is to better reduce the quantization error when the dynamic range (i.e., spread of the norm distribution) is large. In contrast, NEQ does not try to minimize the quantization error and is not limited to the case that the data have large dynamic range. In fact, NEQ still provides significant performance improvement even if items in the dataset have almost identical norm.

\section{Analysis of Quantization Error for MIPS}\label{sec:analysis}

\begin{figure}[!t]	
	\centering
	\includegraphics[width=0.48\columnwidth]{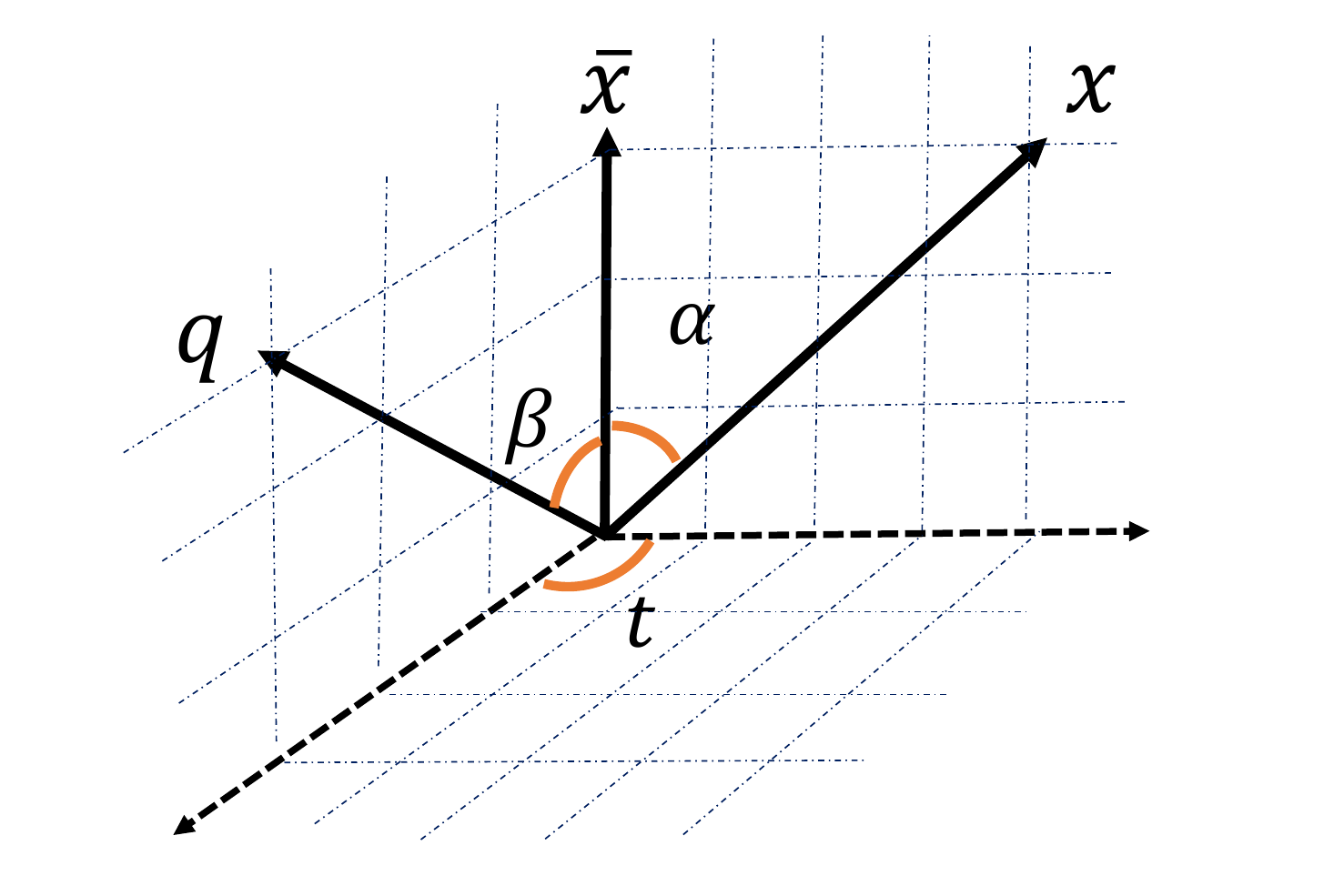}
	\includegraphics[width=0.48\columnwidth]{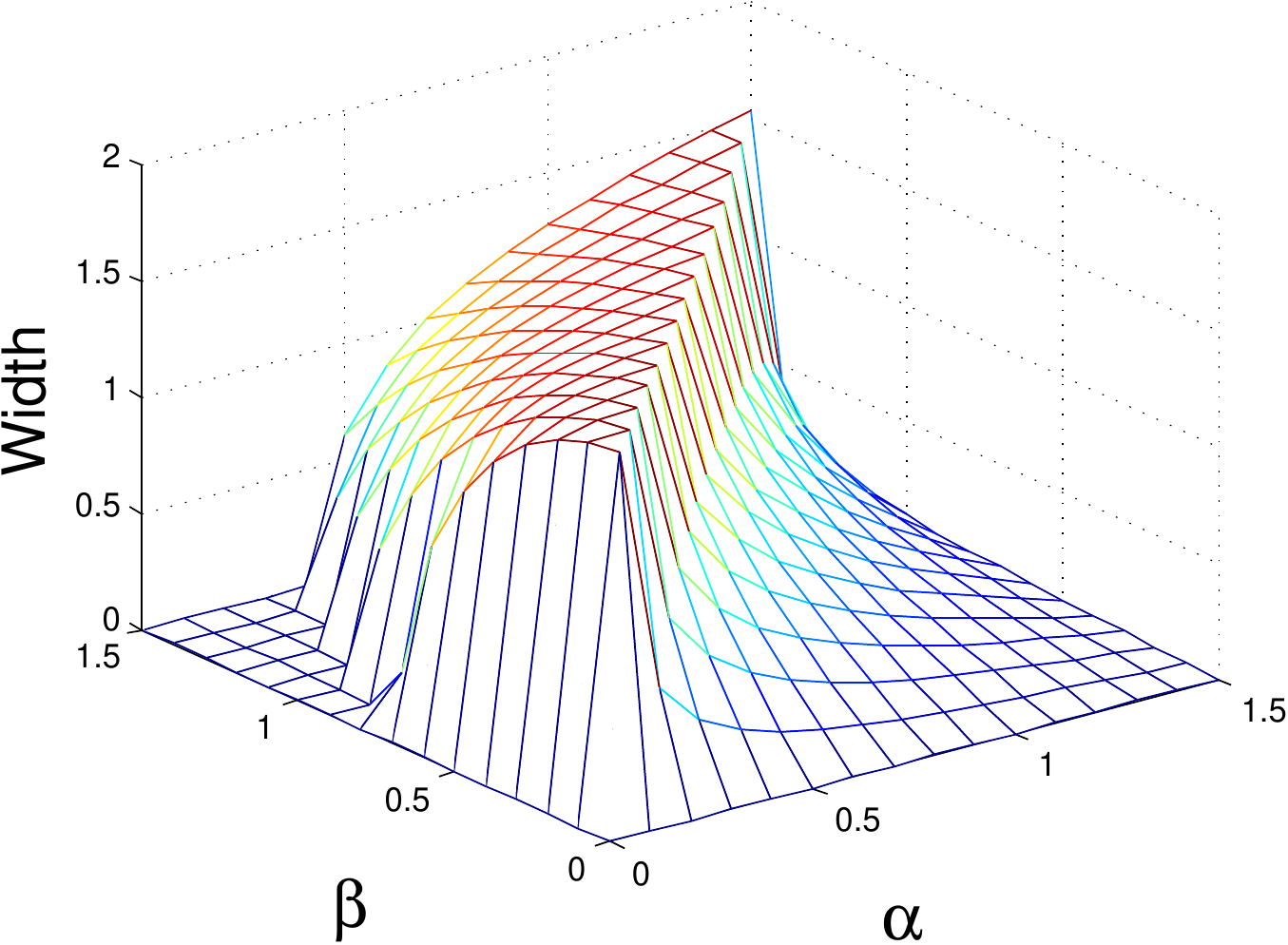}
	\caption{Illustration of Theorem~\ref{theorem:error}}\label{fig:example}
\end{figure}

For Euclidean distance, quantization error provides an upper bound on the error in the approximate Euclidean distance due to the triangle inequality, i.e., $\left \vert \Vert x- q \Vert -  \Vert \tilde{x} -q \Vert \right \vert \le \Vert x- \tilde{x} \Vert$. Therefore, almost all VQ techniques try to minimize the quantization error when learning the codebooks. For approximate inner product, $\Vert x- \tilde{x} \Vert$ provides a trivial error bound because $\left \vert q^{\top}x - q^{\top}\tilde{x} \right \vert \le \Vert q \Vert \Vert x- \tilde{x} \Vert$. As high-dimensional vectors tend to be orthogonal to each other~\cite{cai:angle}, the bound is loose and $q^{\top}(x- \tilde{x})$ can be significantly smaller than $\Vert q \Vert \Vert x- \tilde{x} \Vert$. Thus, we need to understand the influence of quantization error on inner product from a new angle. The exact inner product and its codebook-based approximation can be expressed as,
\begin{equation}\label{equ:error}
\begin{aligned}
&q^{\top}x=\Vert x \Vert \cdot \left(q^{\top} \frac{x}{\Vert x \Vert} \right)  \ \ \! \! \! \! and \! \! \!
&q^{\top}\tilde{x}=\Vert \tilde{x} \Vert \cdot \left(q^{\top} \frac{\tilde{x}}{\Vert \tilde{x} \Vert} \right)
\end{aligned}
\end{equation}                 
in which $x/\Vert x \Vert$ and $\tilde{x}/\Vert \tilde{x} \Vert$ are the unit-norm direction vectors of $x$ and $\tilde{x}$, respectively. It can be observed from~\eqref{equ:error} that the accuracy of the approximate inner product depends on two factors, i.e., the quality of norm approximation ($\Vert \tilde{x} \Vert$ for $\Vert x \Vert$) and the quality of direction vector approximation ($\tilde{x}/\Vert \tilde{x} \Vert$ for $x/\Vert x \Vert$).~\textit{But how do the two factors affect the quality of approximate inner product? Does one have greater influence than the other?} To facilitate further analysis, we formally define inner product error, norm error, and angular error as follows. 

\begin{definition}
	For an item $x$ and its codebook-based approximation $\tilde{x}$, given a query $q$, the inner product error $u$, norm error $\gamma$, and angular error $\eta$ are given as:
	\begin{align}\nonumber
	&u = \left \vert \frac{q^{\top}x - q^{\top}\tilde{x}}{q^{\top}x} \right \vert,
	&\gamma = \left \vert \frac{\Vert x \Vert- \Vert \tilde{x} \Vert}{\Vert x \Vert} \right \vert,
	&&\eta = 1-\frac{x^{\top}\tilde{x}}{\Vert x\Vert \Vert \tilde{x} \Vert}.
	\end{align}
\end{definition}

We define the inner product error and norm error as ratios over the actual values to exclude the scaling effect of $q$ and $\Vert x\Vert$. For angular error, $\eta=0$ if $x$ and $\tilde{x}$ are perfectly aligned in direction. 

To analyze the influence of norm error and angular error individually, we need to exclude the influence of the other. Therefore, we used the approximation $\hat{x}=\Vert \tilde{x} \Vert \cdot \frac{x}{\Vert x \Vert}$, which is accurate in direction, to calculate inner product error caused by norm approximation. Similarly, we used $\bar{x}=\Vert x \Vert \cdot \frac{\tilde{x}}{\Vert \tilde{x} \Vert}$, which is accurate in norm, to calculate the inner product error caused by direction approximation. A norm error of $\gamma$ will cause an inner product error $u\!=\!\gamma$ when there is no angular error as $u = \left \vert \frac{q^{\top}x - q^{\top}\hat{x}}{q^{\top}x} \right \vert = \left \vert \frac{\Vert x \Vert- \Vert \tilde{x} \Vert}{\Vert x \Vert} \right \vert$. Theorem~\ref{theorem:error} formally establishes that there are cases that an angular error $\eta$ results in an inner product error $u<\eta$.

\begin{theorem}\label{theorem:error}
	For an item $x$, its approximation $\bar{x}$ which is accurate in norm but inaccurate in direction, and a query $q$, denote the angle between $x$ and $\bar{x}$ as $\alpha$ and assume $\alpha\in(0, \pi/2)$, the angle between $\bar{x}$ and $q$ as $\beta$ and assume $\beta\in (0, \pi/2)$, the angle between the two planes defined by ($x$, $\bar{x}$) and ($\bar{x}$, $q$) as $t$. The inner product error $\left \vert \frac{q^{\top}x - q^{\top}\bar{x}}{q^{\top}x} \right \vert$ is not larger than the angular error $1-\frac{x^{\top}\bar{x}}{\Vert x\Vert \Vert \bar{x} \Vert}$ if angle $t$ satisfies $\frac{\cos(\beta)}{\sin(\alpha)\sin(\beta)}\left[\frac{1}{2-\cos(\alpha)}-\cos(\alpha)\right]\leq\cos(t)\leq\frac{\cos(\beta)}{\sin(\alpha)\sin(\beta)}\left[\frac{1}{\cos(\alpha)}-\cos(\alpha)\right]$. 	
\end{theorem}

We provide an illustration of the vectors in Theorem~\ref{theorem:error} in Figure~\ref{fig:example} and the proof can found in the supplementary material. We also plot the width of the feasible region of $t$ in the range of $(0, \pi/2)$, i.e., the difference between the maximum value and minimum value for Theorem~\ref{theorem:error} to hold, under different configurations of $\alpha$ and $\beta$ in Figure~\ref{fig:example}. The results show that when both $\alpha$ and $\beta$ are small and $\alpha<\beta$, for almost all $t\in (0, \pi/2)$, the the inner product error is smaller than the angular error. The required conditions are not very restrictive as we analyze below.

We consider an item $x$ having large inner product with $q$ as it is easy to distinguish items having large inner products with the query from those having small inner products. To achieve good performance, a VQ method should be able to distinguish items having large but similar inner products with the query. Firstly, the conditions that $\alpha\in(0, \pi/2)$ and $\alpha$ is small are easy to satisfy as $\bar{x}$ is the codebook based approximation of $x$ and it should have a small angle with $x$. Secondly, as $x$ has a large inner product with query $q$, its approximation $\bar{x}$ should also have a small angle with $q$, therefore the condition that $\beta\in (0, \pi/2)$ and $\beta$ is small is likely to hold. Finally, as $\bar{x}$ is trained to approximate $x$ and $q$ is not, $\alpha<\beta$ is again easy to satisfy. As $q$, $x$ and $\bar{x}$ have small angles with each other, $t$ is likely to fall in $(0, \pi/2)$.

\begin{figure}[!t]	
	\centering
	\includegraphics[width=0.49\columnwidth]{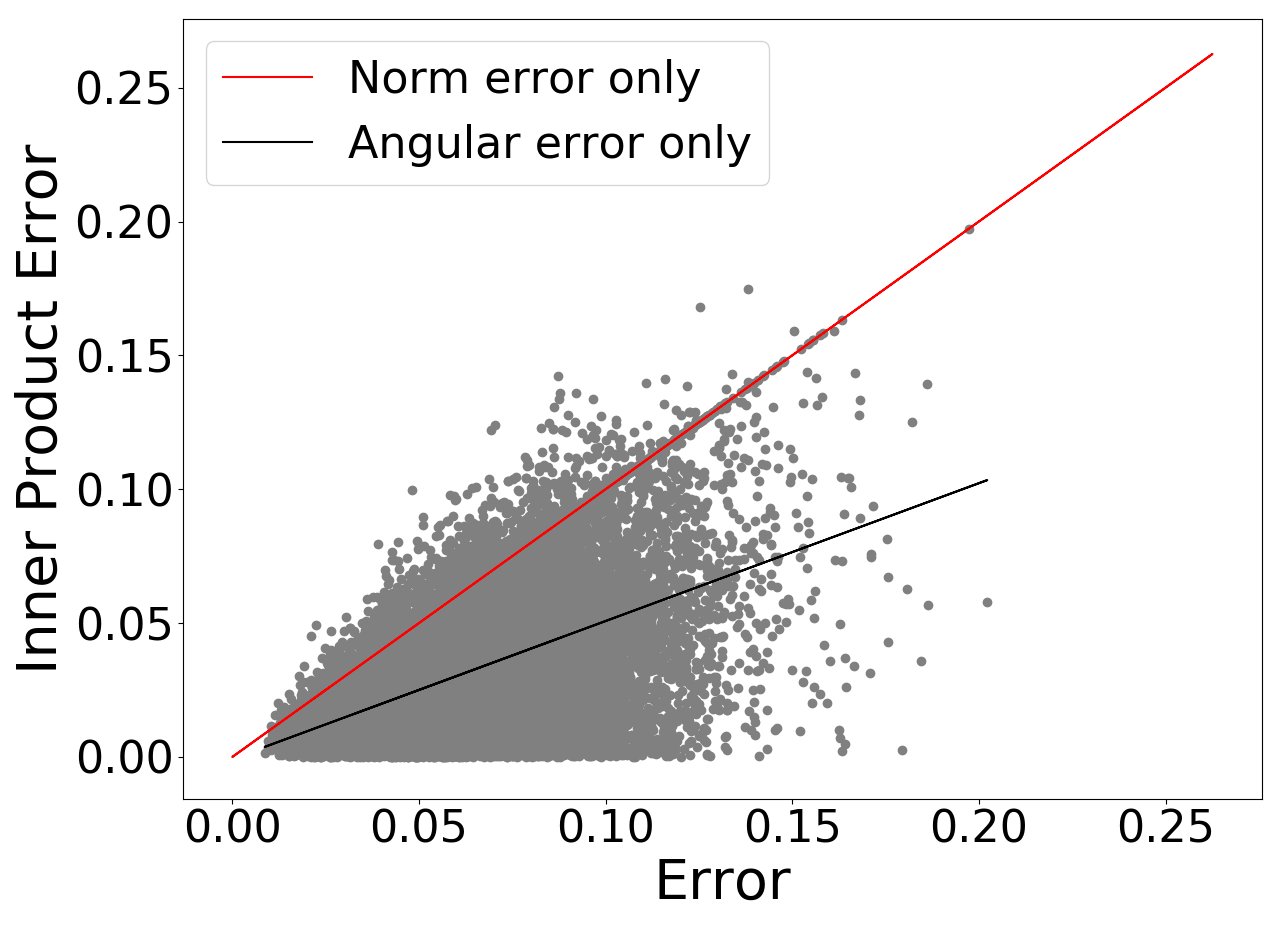}
	\includegraphics[width=0.49\columnwidth]{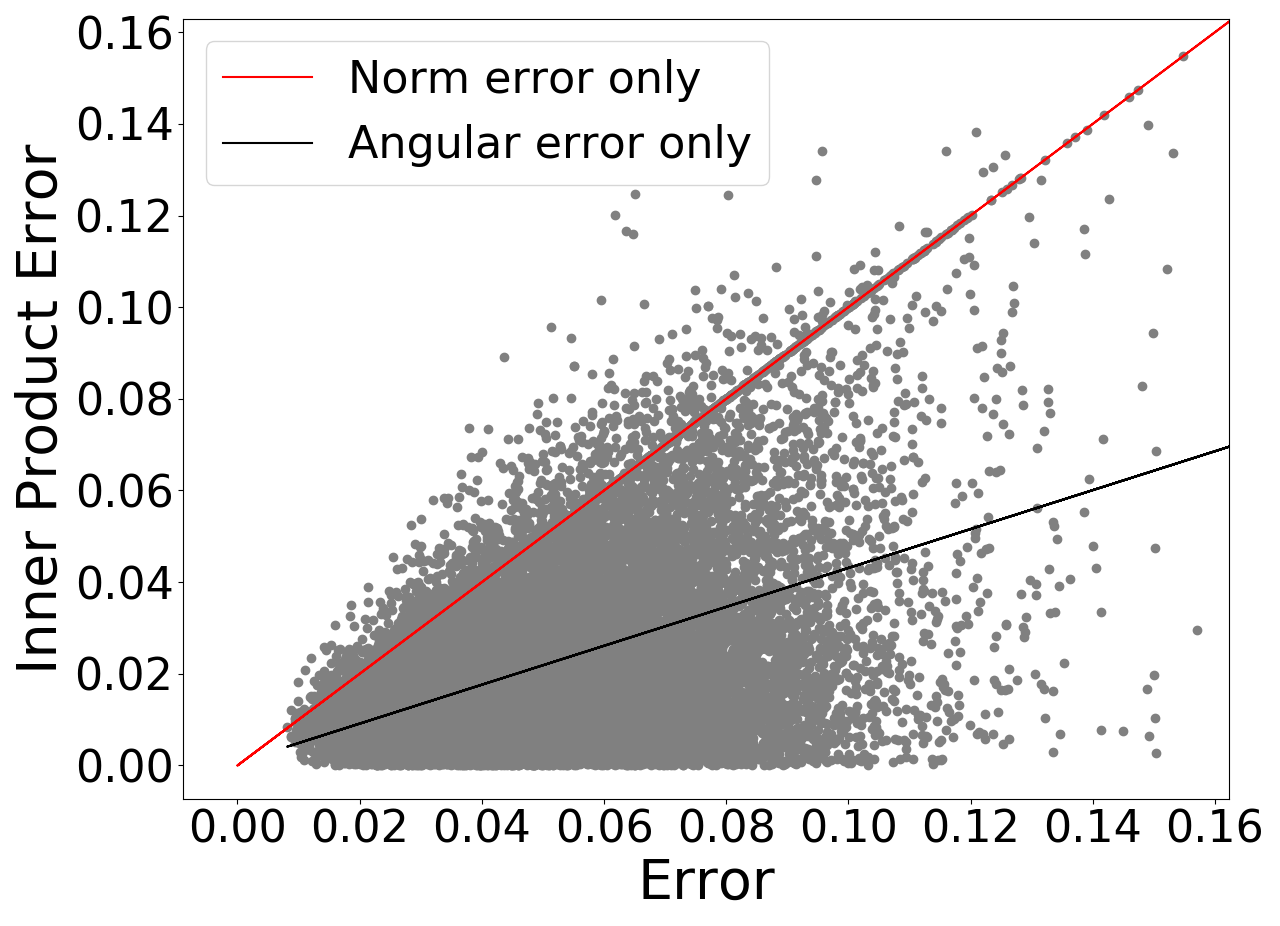}
	\caption{Influence of norm error and angular error on inner product for PQ (left) and RQ (right), all red points reside on the red line}\label{fig:error}
\end{figure}

Theorem~\ref{theorem:error} is also supported by the following experiment on the SIFT1M dataset~\footnote{SIFT1M is sampled from the SIFT100M dataset used in the experiments in Section~\ref{sec:exp}.}. We used 10,000 randomly selected queries and the errors are calculated on their ground-truth top-20 MIPS results~\footnote{Researches~\cite{neyshabur:simple-lsh,shrivastava:alsh,guo2016quantization} on MIPS usually use a value of $k$ ranging from 1 to 50, 20 is the middle of this range.} in the dataset. We experimented with PQ and RQ using 8 codebooks each containing 256 codewords. For each item-query pair ($x$, $q$), we plot two points in Figure~\ref{fig:error}. One (in red) shows the norm error and the inner product error caused by inaccurate norm (using $\hat{x}$). The other (in gray) shows the angular error and the inner product error caused by inaccurate direction vector (using $\bar{x}$). The results show that all red points reside on the line with a slope of 1, which verifies that a norm error of $\gamma$ will cause an inner product error $u=\gamma$. In contrast, most of the gray points are below the red line, which means that an angular error $\eta$ usually results in an inner product error $u<\eta$. We fitted a line for the gray points and the slopes for PQ and RQ are 0.510 and 0.426, respectively. The Pearson's correlation coefficients between norm error and inner product error are 1 for both PQ and RQ. While the Pearson's correlation coefficients between angular error and inner product error are 0.475 and 0.382 for PQ and RQ, respectively. We also plot the influence of norm error and angular error on Euclidean distance in the supplementary material~\footnote{See https://arxiv.org/pdf/1911.04654.pdf for the supplementary material.}, which shows that angular error has larger influence than norm error on Euclidean distance.

In conclusion, the results in this section show that norm error has more significant influence on inner product than angular error in most cases. Therefore, to improve the performance of VQ techniques for MIPS, we should reduce quantization errors in norm. To achieve this goal, we can modify the formulations of the codebook learning problem in existing VQ algorithms to consider norm error (e.g., incorporating norm error into the cost function or constraints). However, this methodology has a problem in generality as we need to modify each VQ algorithm individually. In contrast, norm-explicit quantization (NEQ) uses the fact that norm is a scalar summary of the vector and explicitly quantizes it to reduce error. As a result, NEQ can be naturally combined with any VQ algorithm by using it to quantize the direction vector.

\section{Norm-Explicit Quantization}\label{sec:method}

Existing VQ techniques try to minimize the quantization error and do not allow explicit control of norm error and angular error. However, MIPS could benefit from methods that explicitly reduce the error in norm because accurate norm is important for MIPS. Therefore, the core idea of NEQ is to quantize the norm $\Vert x \Vert$ and the direction vector $\frac{x}{\Vert x \Vert}$ of the items separately. The norm is encoded explicitly using separate codebooks to achieve a small error, while the direction vector can be quantized using an existing VQ quantization technique without modification. To be more specific, the $M$ codebooks in NEQ are divided into two parts. The first $M'$ codebooks $\mathcal{L}^1, \mathcal{L}^2,...,\mathcal{L}^{M'}$ are norm codebooks, in which each codeword $l^m[k] \in \mathbb{R}$ for $1 \le m \le M' $ and $1 \le k \le K$. The other $M-M'$ codebooks $\mathcal{C}^{M'+1}, \mathcal{C}^{M'+2},..,\mathcal{C}^M$ are vector codebooks for the direction vector. In NEQ, the codebook based approximation $\tilde{x}$ of $x$ can be expressed as,
\begin{equation}\label{equ:NEQ approximation}
\tilde{x}=\left(\sum_{m=1}^{M'}l^m[i^m_x]\right) \cdot \left(\sum_{m=M'+1}^{M}c^m[i^m_x] \right),
\end{equation}      
in which $i^1_x,i^2_x,...,i^M_x$ are the codeword indexes of $x$ in the codebooks. According to~\eqref{equ:NEQ approximation}, NEQ-based approximate inner product $q^{\top}\tilde{x}$ can be calculated using Algorithm~\ref{alg:calculation}. Lines 4-6 reconstruct the approximate norm of $x$ and Lines 7-9 compute the inner product between $q$ and the approximate direction vector of $x$. Note that the inner product computation $q^{\top}c^m[i^m_x]$ in Line 8 can be replaced by table lookup when the inner products between $q$ and the codewords are precomputed.

\begin{algorithm}
	\caption{NEQ: Approximate Inner Product Calculation}
	\label{alg:calculation}
	\begin{algorithmic}[1]
		\STATE {\bfseries Input:} Query $q$, $M$ codeword indexes $i^1_x, i^2_x, ..., i^M_x$ of item $x$
		\STATE {\bfseries Output:} An approximation of $q^{\top}x$
		\STATE $l=0, p=0$;
		\FOR{$m$ from $1$ to $M'$}
		\STATE $l=l+l^m[i^m_x]$;
		\ENDFOR
		\FOR{$m$ from $M'+1$ to $M$}
		\STATE $p=p+q^{\top}c^m[i^m_x]$;
		\ENDFOR
		\STATE return $l \cdot p$;
	\end{algorithmic}
\end{algorithm} 

The remaining problem is how to train the norm and vector codebooks. A straightforward solution, which trains the norm codebooks with $\Vert x \Vert$ and the vector codebooks with $x/\Vert x \Vert$, does not work. This is because the codebook based approximation $\bar{x}=\sum_{m=M'+1}^{M}c^m[i^m_{x'}]$ of the direction vector is not guaranteed to be unit norm due to the intrinsic norm errors of vector quantization. Therefore, even if we quantize $\Vert x \Vert$ accurately with the norm codebooks, $\tilde{x}$ in~~\eqref{equ:NEQ approximation} can still have large norm error. NEQ solves this problem with the codebook learning process in Algorithm~\ref{alg:learning}.  

\begin{algorithm}
	\caption{NEQ: Codebook Learning}
	\label{alg:learning}
	\begin{algorithmic}[1]
		\STATE {\bfseries Input:} Dataset $\mathcal{X}$, \# codebook $M$, \# norm codebook $M'$ 
		\STATE {\bfseries Output:} $M'$ norm codebooks, $M-M'$ vector codebooks
		\STATE Extract the direction vector $x'=\frac{x}{\Vert x \Vert}$;  
		\STATE Train $M-M'$ vector codebooks on $x'$ using a VQ method;
		\STATE Encode $x'$ with the vector codebooks, obtain the codebook based approximation $\bar{x}$ of $x'$; 
		\STATE Get the \textit{relative norm} $l_x$ of item $x$ as $\frac{\Vert x \Vert}{\Vert \bar{x} \Vert}$;  
		\STATE Train $M'$ norm codebooks to quantize $l_x$;
		\STATE Return the $M$ codebooks;
	\end{algorithmic}
\end{algorithm}                   

Line 4 trains the vector codebooks using an exiting VQ method, such as PQ or RQ. Instead of quantizing the actual norm $\Vert x \Vert$, NEQ quantizes the relative norm $l_x=\Vert x \Vert/\Vert \bar{x} \Vert$ in Line 7 of Algorithm~\ref{alg:learning}. This design absorbs the norm error of VQ into the relative norm $l_x$ and ensures that the codebook based approximation $\tilde{x}$ in~\eqref{equ:NEQ approximation} has the same norm as $x$ if $l_x$ is quantized accurately. As we will show in the experiments, NEQ also works for datasets in which items have almost identical norms thanks to this design. The norm codebooks are learned in a recursive manner similar to RQ. The norm is used to train the first codebook $\mathcal{L}^1$ with K-means. The residuals ($\Vert x \Vert-l^1[i^1_x]$) are used to train $\mathcal{L}^2$ and this process is conducted iteratively. The normalization in Line 3 may look unnecessary as we can quantize the original item $x$ directly using the vector codebooks and define the relative norm as $\Vert x \Vert/\Vert \tilde{x} \Vert$. However, we observed that this alternative does not perform as well as Algorithm~\ref{alg:learning}. One possible reason is that unit vectors may be easier to quantize for VQ techniques.

As a demonstration of the effectiveness of NEQ in reducing the quantization error in norm, we report some statistics of the Yahoo!Music dataset. For the original RQ, a norm error of $1.51\times 10^{-2}$ and $6.47\times 10^{-3}$ are achieved with 8 and 16 codebooks, respectively. Keeping the total number of codebooks the same and using only one codebook for norm, norm explicit quantization based RQ reaches a norm error of $1.1\times 10^{-3}$ under both 8 and 16 codebooks. We will show that the lower norm error of NEQ translates into better performance for MIPS in the experiments in Section~\ref{sec:exp}.

\textbf{Setting the number of norm codebooks.} Generally, a good $M'$ can be chosen by testing the recall-item performance of all $M-1$~\footnote{There should be at least 1 and at most $M-1$ norm codebooks.} configurations on a set of sample queries. When the number of codewords in each codebook is 256 (i.e., $K=256$), we found empirically that using one codebook for norm provides the best performance in most cases. This is because the norm error is already small with one norm codebook. Using more codebooks for norm provides limited reduction in norm error but increases angular error as the number of angular codebooks is reduced.

\textbf{Why not storing the norm?} As the relative norm $l_x$ is a scalar, one may wonder why not storing its exact value to completely eliminate norm error. This is because storing $l_x$ with a 4-byte floating point number costs too much space and VQ algorithms are usually evaluated with a fixed per-item space budget (especially when used for data compression). With the usual setting $K=256$, using $M$ codebooks results in a per-item index size of $M$ bytes. If $l_x$ is stored exactly, the direction vector can only use $M-4$ codebooks. Empirically, we found that using 1 norm codebook already makes the norm error very small, which leaves direction vector $M-1$ codebooks and achieves better overall performance.

\textbf{Complexity analysis.} For index building, NEQ learns $M-M'$ vector codebooks and the original VQ method learns $M$ vector codebooks. Although NEQ needs to conduct normalization twice (Line 3 and Line 6 of Algorithm~\ref{alg:learning}) and learn the norm codebooks, the complexity of these operations is generally low compared with learning vector codebooks. For inner product computation with lookup table, the original VQ method needs $M$ lookups and $M-1$ additions. NEQ needs $M'$ lookups and $M'-1$ additions to reconstruct the relative norm, and $M-M'$ lookups and $M-M'-1$ additions to add the inner product. Then one more multiplication is needed to assemble the final result. Thus, approximate inner product computation in NEQ costs $M$ lookups and $M-1$ additions, which is exactly the same as the original VQ method. Therefore, NEQ does not increase the complexity of codebook learning and approximate inner product computation.

We would like to emphasize that the strength of NEQ lies in its simplicity and generality. NEQ is simple in that it uses existing VQ methods to quantize the direction vector without modifying their formulations of the codebook learning problem. This makes NEQ easy to implement as off-the-shelf VQ libraries can be reused. NEQ is also general in that it can be combined with any VQ methods, including PQ, OPQ, RQ and AQ. In the supplementary material, we show that NEQ with two codebooks can adopt the multi-index algorithm~\cite{babenkol:imi} for candidate generation in MIPS. We will also show in Section~\ref{sec:exp} that NEQ boosts the performance of many VQ methods for MIPS.

\section{Experiments}\label{sec:exp}

\begin{table*}
	\centering
	\caption{Dataset statistics}
	\label{tab:datasets}
	\begin{center}
		\begin{sc}
			\fontsize{8}{9}\selectfont
			\begin{tabular}{ccccl}
				\toprule		
				Dataset & Netflix & Yahoo!Music & ImageNet & SIFT100M\\ 
				\midrule
				\# items & 17,770 & 136,736 & 2,340,373 &  100,000,000\\
				\# dimensions& 300 & 300 & 150 & 128\\  
				\bottomrule
			\end{tabular}
		\end{sc}
	\end{center}
\end{table*}

\textbf{Experiment setting.} We used four popular datasets, Netflix, Yahoo!Music, ImageNet and SIFT100M, whose statistics are summarized in Table~\ref{tab:datasets}. Netflix and Yahoo!Music record user ratings for items. We obtained item and user embeddings from these two datasets using alternating least square (ALS)~\cite{yun:als} based matrix factorization. The item embeddings were used as dataset items, while the user embeddings were used as queries. ImageNet and SIFT100M contain descriptors of images. The four datasets vary significantly in norm distribution (see details in the supplementary material) and we deliberately chose them to test NEQ's robustness to different norm distributions. ImageNet has a long tail in its norm distribution, while items in SIFT100M have almost the same norm. For Netflix and Yahoo!Music, most items have a norm close to the maximum~\footnote{See https://github.com/xinyandai/product-quantization for all experiment codes.}.

\begin{figure*}[h]
	\centering
	\includegraphics[width=\textwidth]{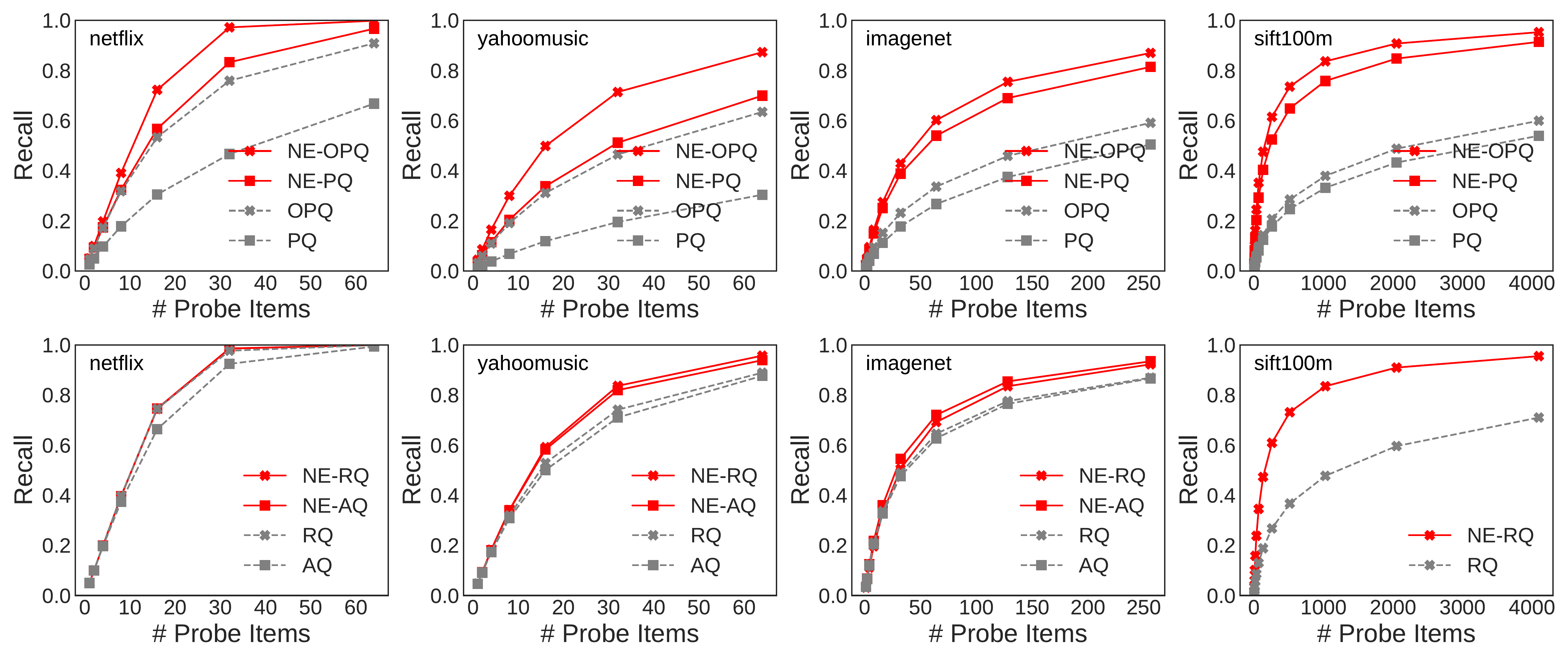}
	\caption{Item-recall performance of the VQ methods and their NEQ-based variants}\label{fig:general performance}
\end{figure*}

Following the standard protocol for evaluating VQ techniques~\cite{babenko:aq,babenko:treeq,zhang:cq}, we used the recall-item curve as the main performance metric and it measures the ability of a VQ method to preserve the similarity ranking of the items. To obtain the recall-item curve, all items in a dataset are first sorted according to the codebook based approximate inner products. For a query, denote the set of items ranking top $T$ as $\mathcal{S}'$ and the set of ground truth top-$k$ MIPS results as $\mathcal{S}$, the recall is $|\mathcal{S}'\cap \mathcal{S}|/|\mathcal{S}|$. At each value of $T$, we report the average recall of 10,000 randomly selected queries. We do not report the running time as the VQ methods have almost identical running time~\footnote{AQ and RQ have more expensive inner product table computation than PQ and OPQ. However, this difference has negligible impact on the running time when the dataset is large.} given the same number of codebooks $M$.

For a VQ method X (e.g., RQ), its NEQ version is denoted as NE-X (e.g., NE-RQ). The NEQ variants use the same number of codebooks (norm codebooks plus direction codebooks) as the original VQ methods. Each codebook has $K=256$ codewords and only one codebook is used for norm in NEQ. The default value of $k$ (the number of target top inner product items) is 20~\footnote{The performance of MIPS is usually evaluated by setting $k$ as 1, 10, 20 or 50 and the results are usually consistent under different configurations of $k$. Due to space limit, we provide the results under more configurations of $k$ in the supplementary material.}. For Neflix, the codebooks were trained using the entire dataset. For the other datasets, the codebooks were trained using a sample of size 100,000.

\textbf{Improvements over existing VQ methods.} We report the performance of the original VQ methods (in dotted lines) and their NEQ-based variants (in solid lines) in Figure~\ref{fig:general performance}. The number of codebooks is 8. We do not report the performance of AQ and NE-AQ on SIFT100M as the encoding process of AQ did not finish in 72 hours. The results show that the NEQ-based variants consistently outperform their counterparts on all the four datasets. The performance improvements of NEQ on PQ and OPQ are much more significant than on AQ and RQ. Moreover, there is a trend that the performance benefit increases with the dataset cardinality. These two phenomenons can be explained by the fact that reducing the error in norm is more helpful when the quantization error is large. With 8 codebooks, the small Netflix dataset is already quantized accurately, while the SIFT100M dataset is not well quantized. With the same number of codebooks, PQ and OPQ generally have larger quantization errors than RQ and AQ and thus the performance gain of NEQ is more significant.

\begin{figure}[!t]	
	\centering 
	\begin{minipage}[b]{0.49\textwidth}
		\includegraphics[width=0.49\textwidth]{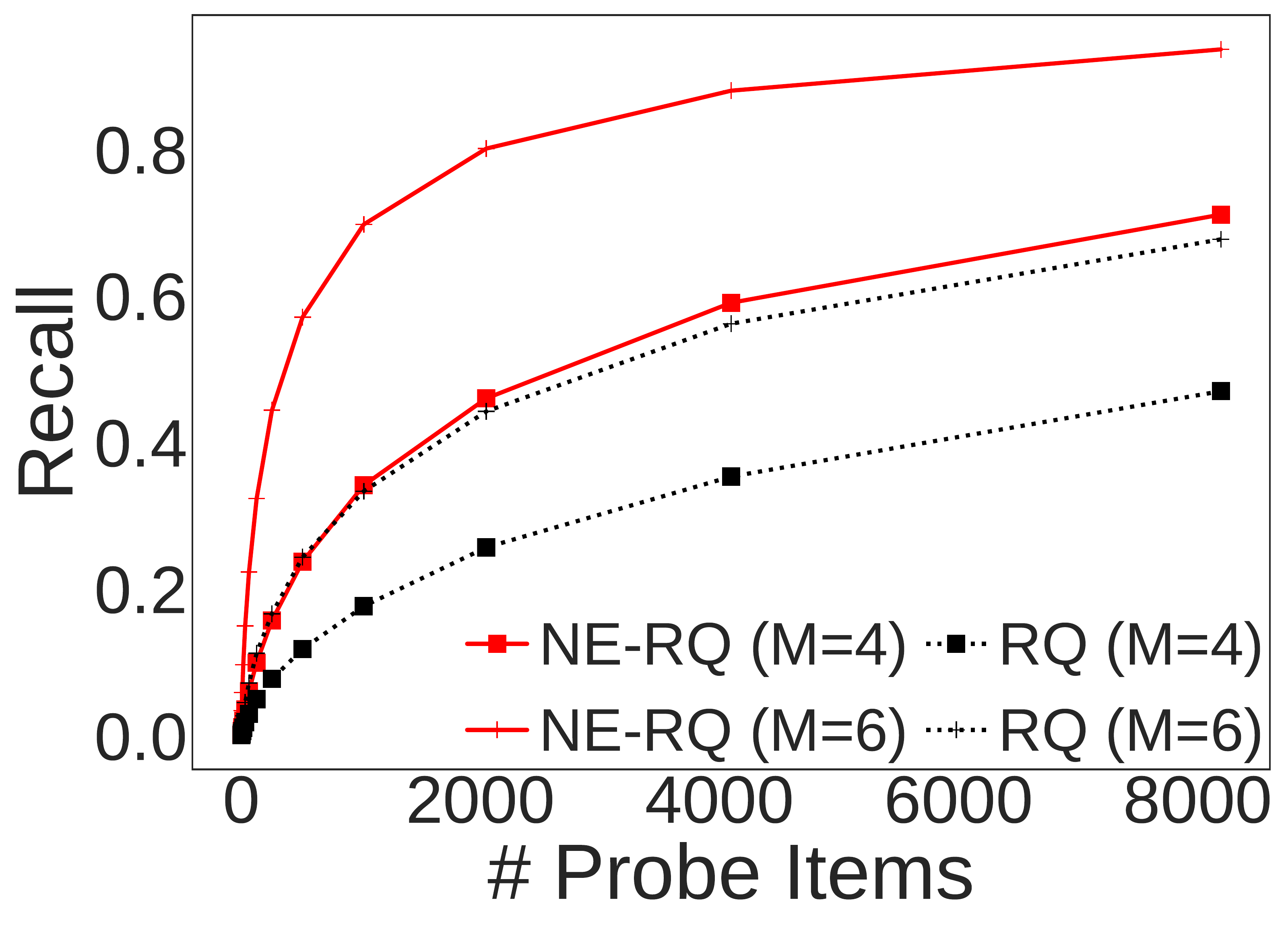}
		\includegraphics[width=0.49\textwidth]{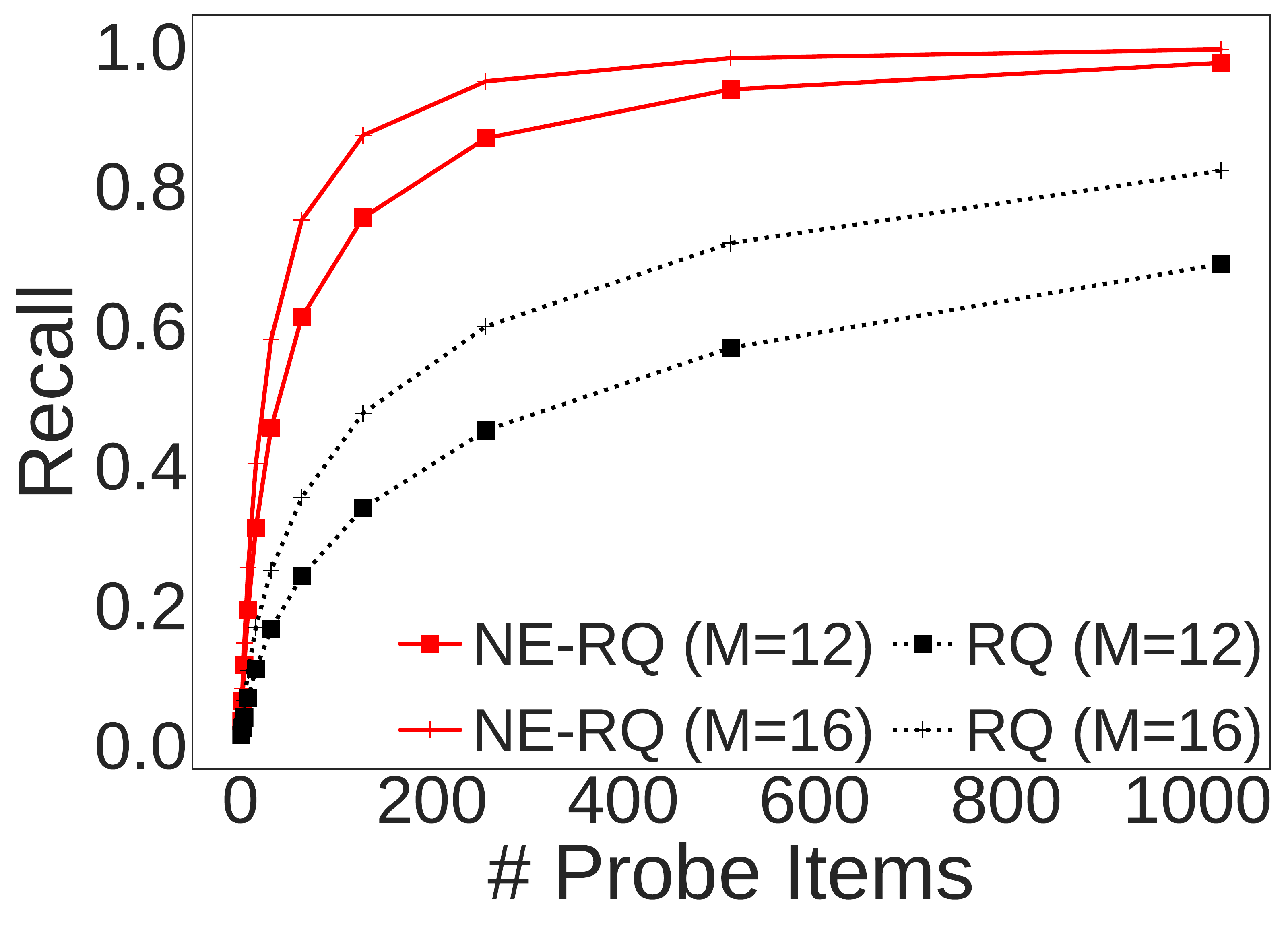}
		\caption{Different number of codebooks}\label{fig:number of codebooks}
	\end{minipage}
	\begin{minipage}[b]{0.49\textwidth}
		\includegraphics[width=0.49\textwidth]{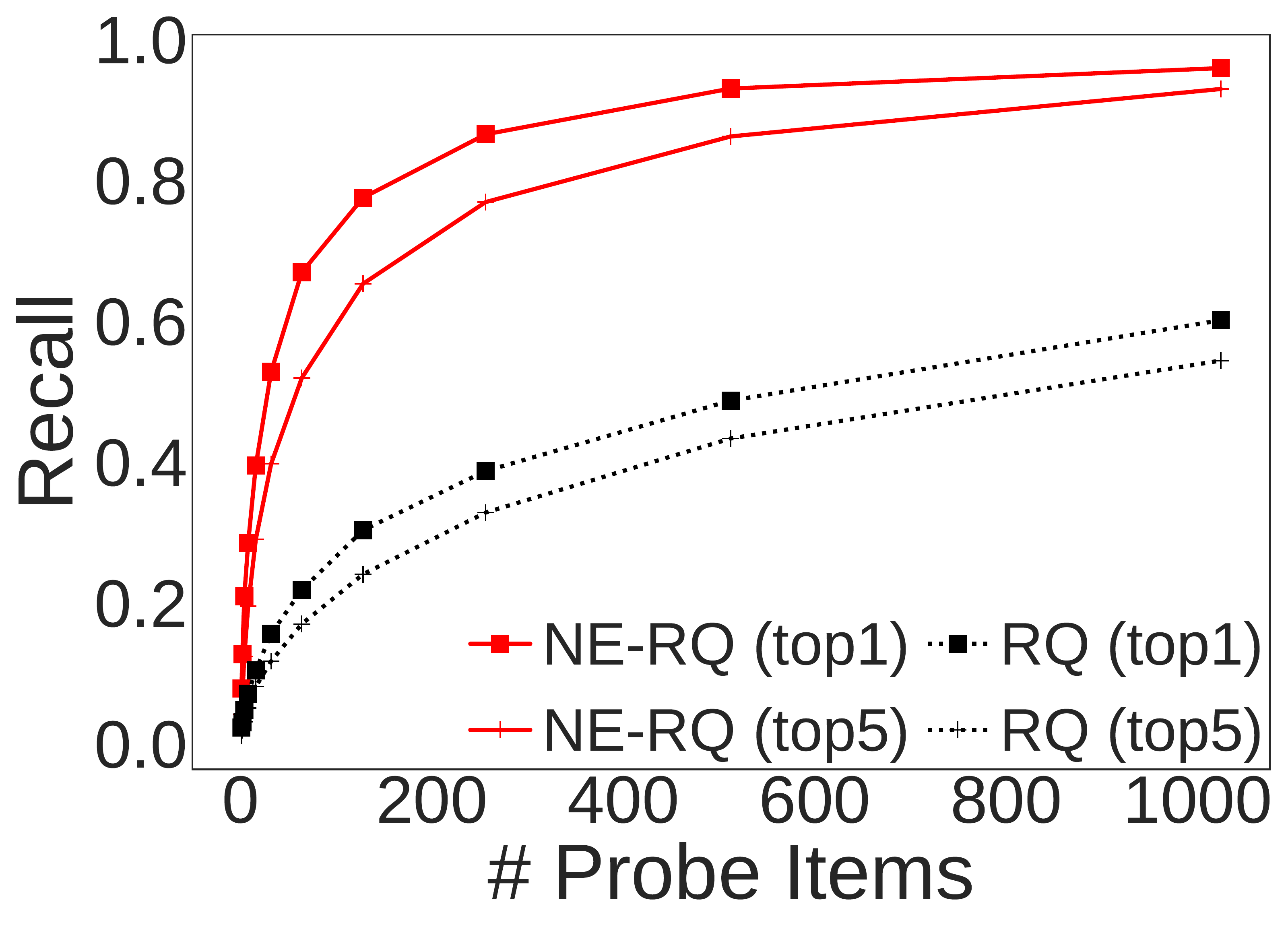}
		\includegraphics[width=0.49\textwidth]{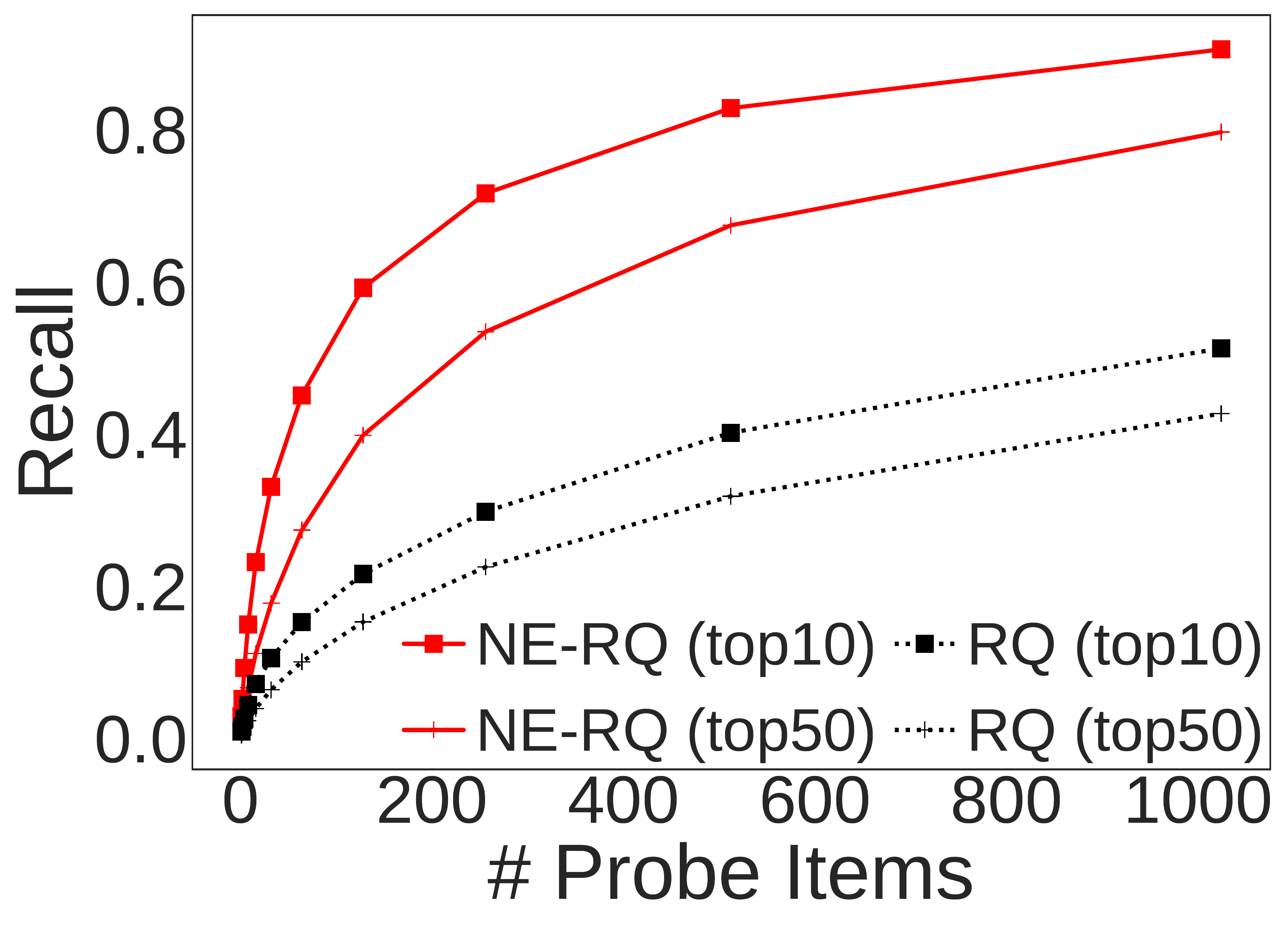}
		\caption{Different values of $k$}\label{fig:value of k}
	\end{minipage}
\end{figure}

Next, we test the robustness of NEQ to the parameter configurations, i.e., the number of codebooks $M$ and the value of $k$. We report the performance of RQ and NE-RQ on the SIFT100M dataset in Figure~\ref{fig:number of codebooks} and Figure~\ref{fig:value of k} (the results of other VQ methods and datasets can be found in the supplementary material). Figure~\ref{fig:number of codebooks} shows that NE-RQ outperforms RQ across different number of codebooks. Figure~\ref{fig:value of k} shows that NE-RQ consistently outperforms RQ for different values of $k$ with 8 codebooks and the performance gap is similar for different values of $k$. The results in the supplementary material show that the robustness of NEQ to the parameter configurations also holds for PQ, OPQ and AQ. In addition, we examine the robustness of the VQ methods and their NEQ variants across different runs of the codebook learning algorithm in the supplementary material. The results show that NEQ usually provides smaller standard deviation in recall across different runs.

\begin{figure}[!t]	
	\centering 
	\begin{minipage}[b]{0.49\textwidth}
		\includegraphics[width=0.49\textwidth]{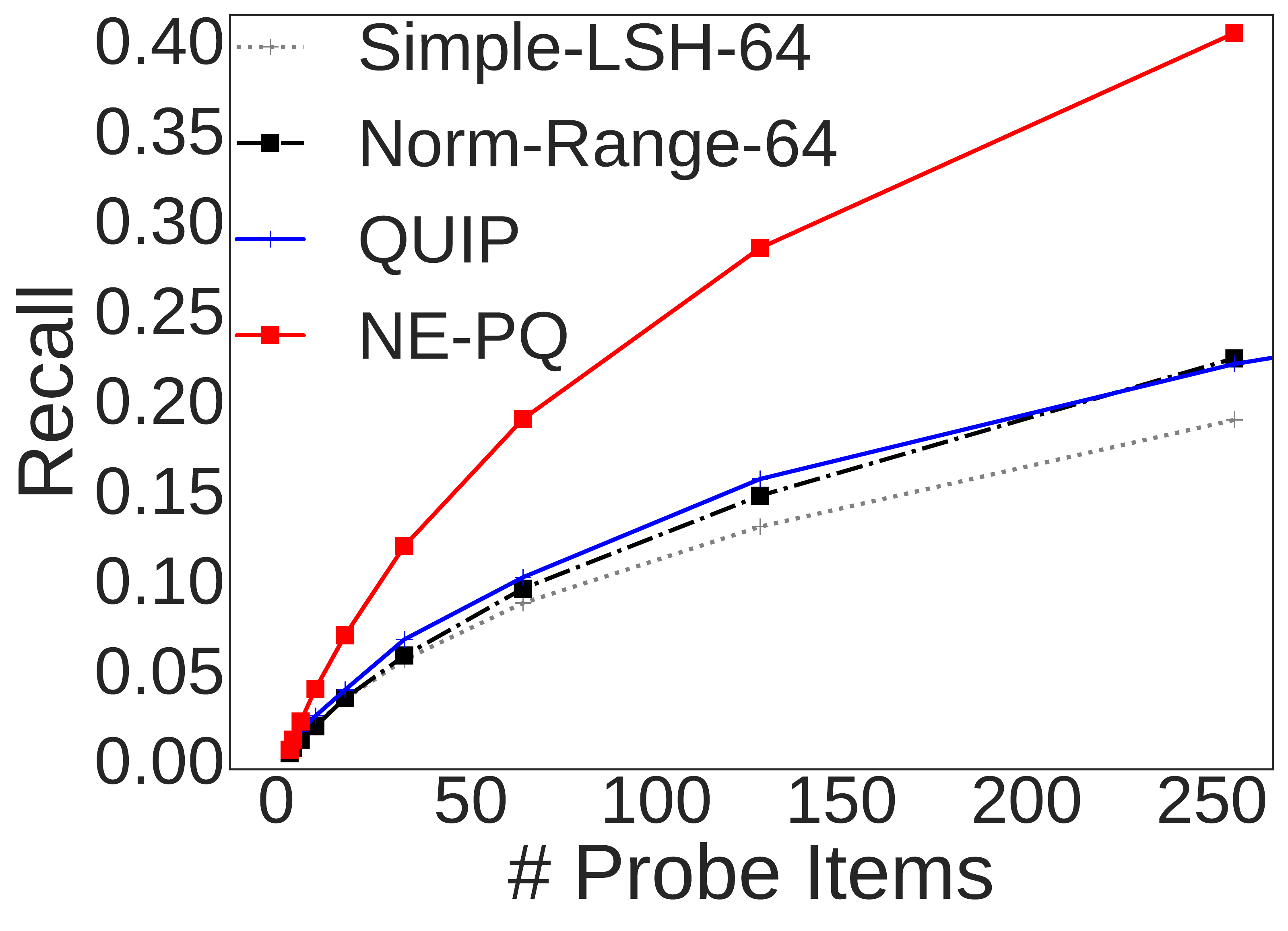}
		\includegraphics[width=0.49\textwidth]{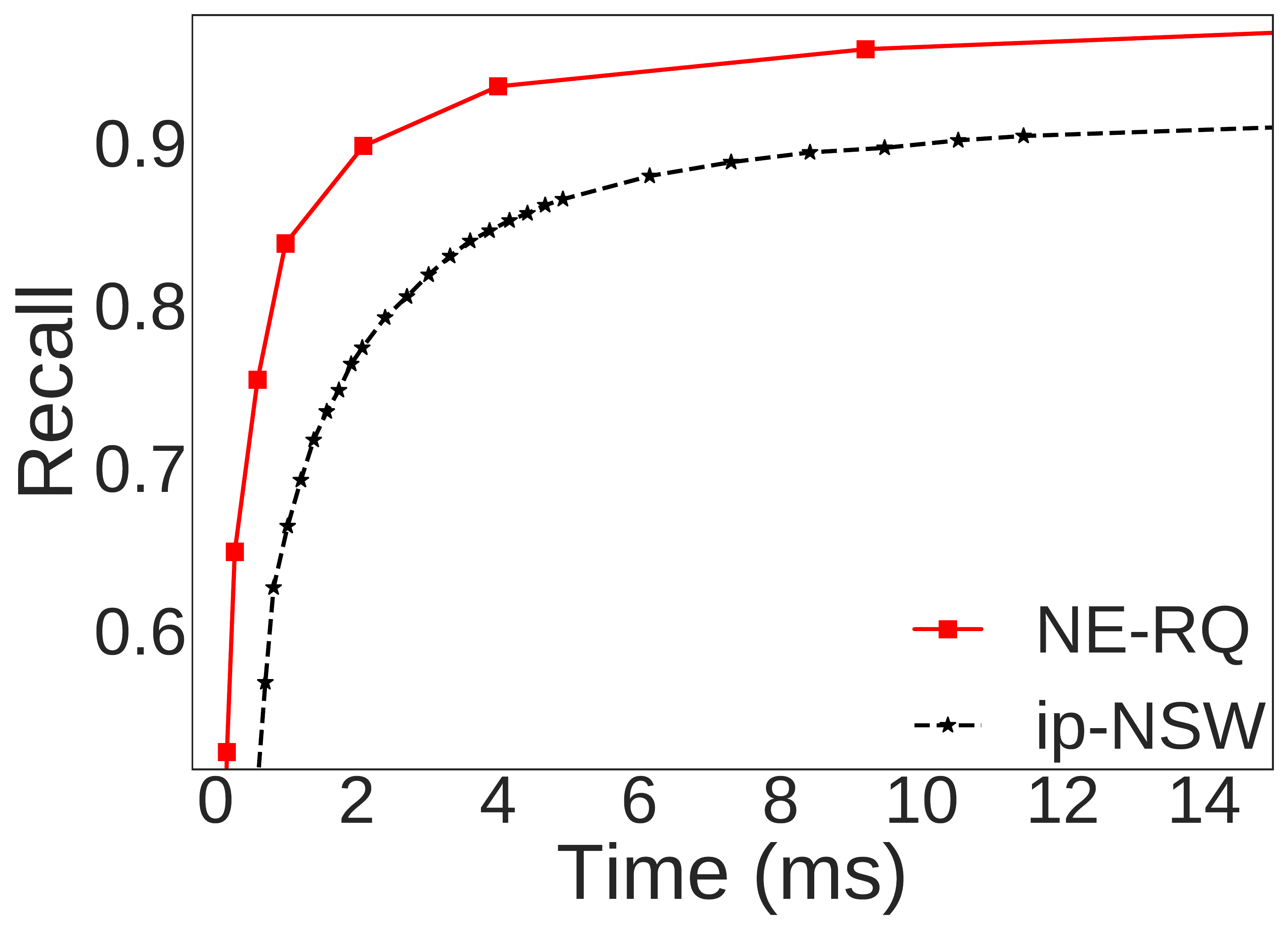}
		\caption{Comparison with LSH \& graph methods}\label{fig:comparsion with binary hashing}
	\end{minipage}
	\begin{minipage}[b]{0.49\textwidth}
		\includegraphics[width=0.49\textwidth]{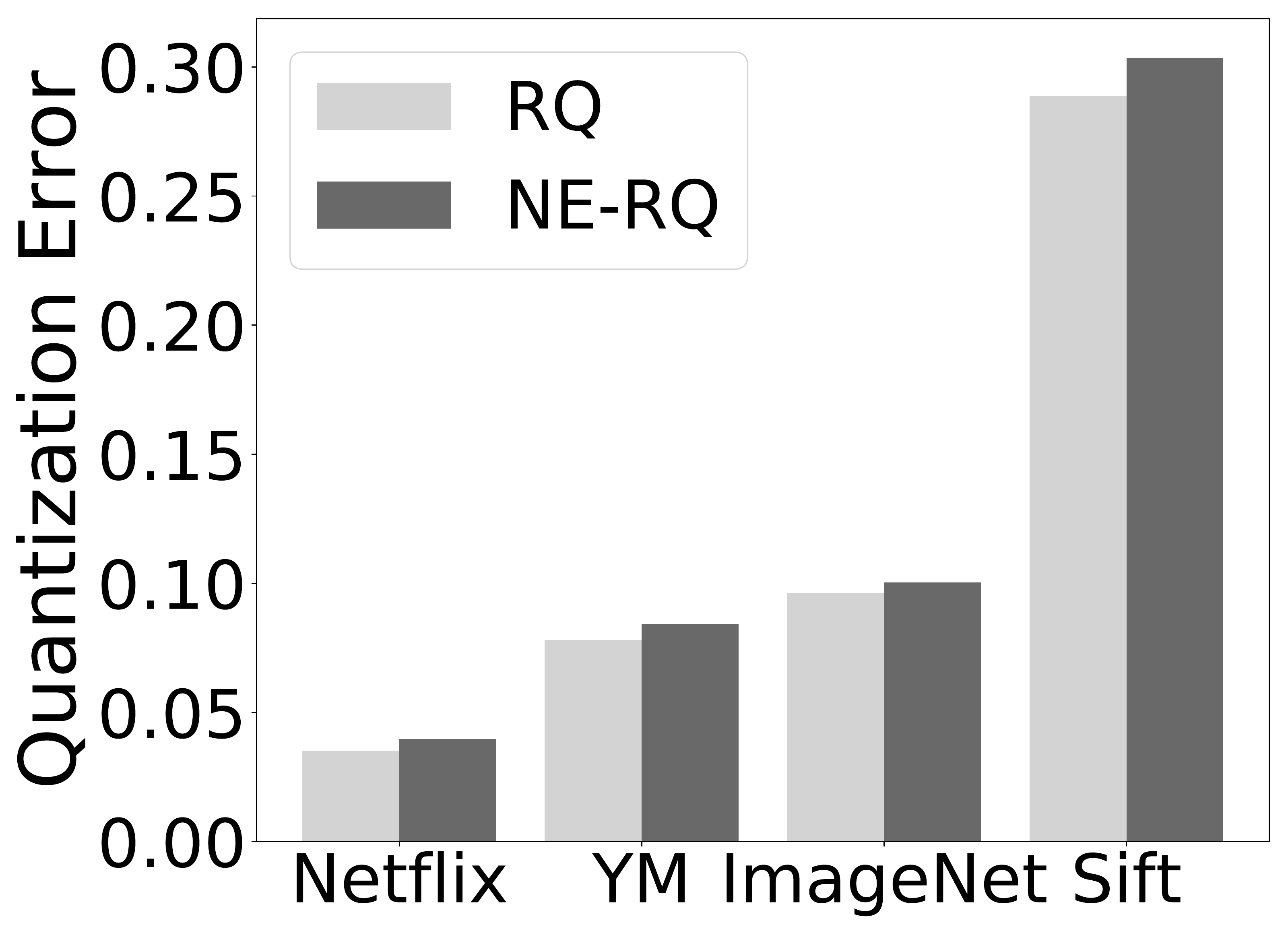}
		\includegraphics[width=0.49\textwidth]{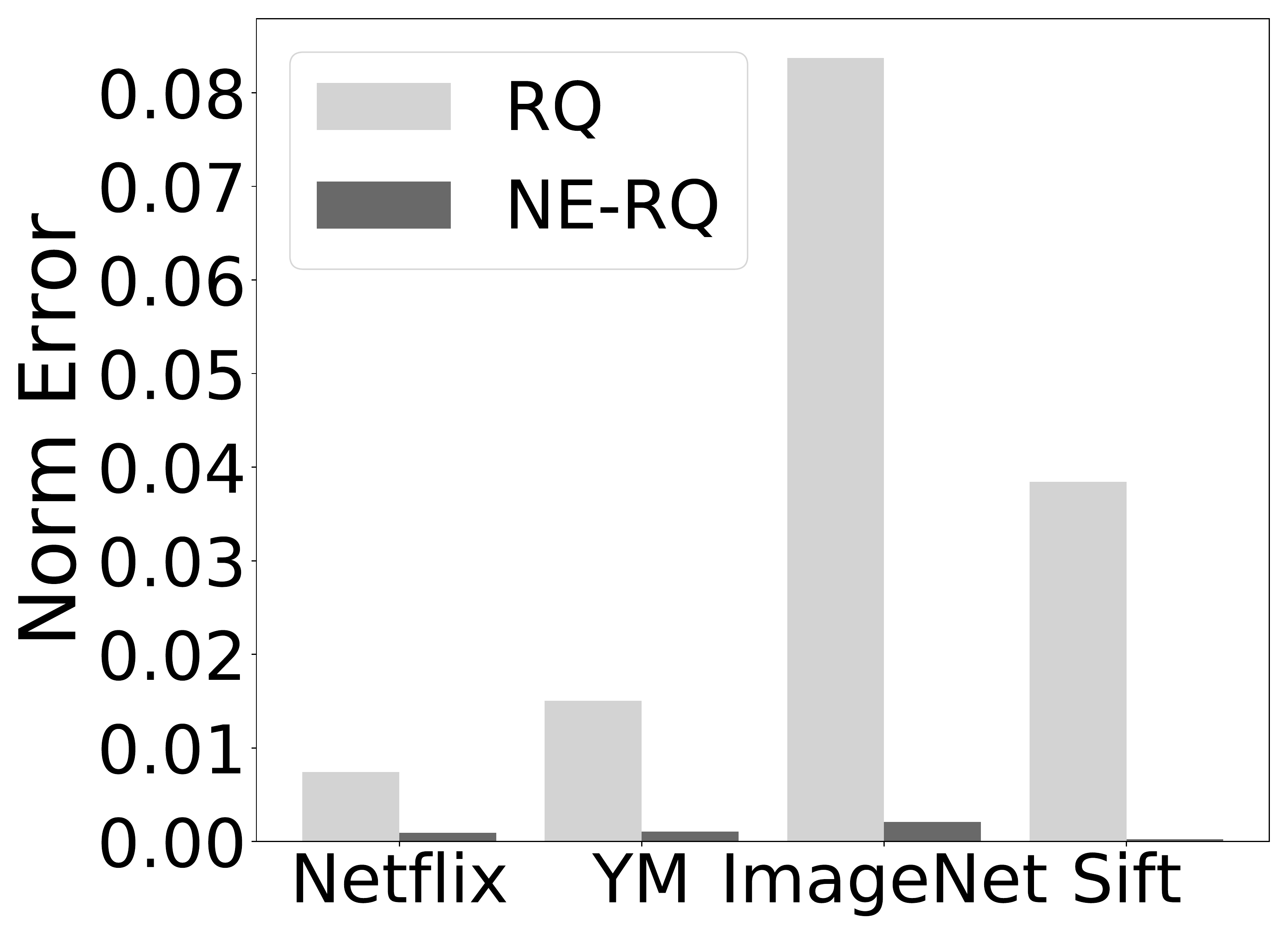}
		\caption{Quantization error of NE-RQ and RQ}
		\label{fig:quantization error}
	\end{minipage}
\end{figure}

\textbf{Comparison with other methods} Norm-Range LSH~\cite{yan:normrange} and Simple-LSH~\cite{neyshabur:simple-lsh} use binary hashing and are the state-of-the-art LSH-based algorithms for MIPS. QUIP~\cite{guo2016quantization} is a vector quantization method specialized for MIPS, which explicitly minimizes the squared inner product error ($(q{\top}x-q{\top}\tilde{x})^2$) to learn the codebooks. QUIP has several variants and we used~\textit{QUIP-cov(x)} for fair comparison as other variants use knowledge about the queries but NEQ does not. According to the QUIP paper, the performance gap between other variants and~\textit{QUIP-cov(x)} is small for the ImageNet dataset. For Norm-Range LSH, we partitioned the dataset into 64 sub-datasets as recommended in~\cite{yan:normrange}. We report the performance results on the ImageNet dataset in Figure~\ref{fig:comparsion with binary hashing}~(left). Simple-LSH and Norm-Range used 64 bit binary code. NE-PQ and QUIP use two codebooks each containing 256 codewords. This means that the per item index size of NE-PQ (and QUIP) is 16 bit and only a quarter of that of the LSH-based methods. The results show that the vector quantization based methods (NE-PQ and QUIP) outperform the LSH-based algorithms with smaller per-item index size. Moreover, NE-PQ significantly outperforms QUIP even if QUIP uses a more complex codebook learning strategy.

We also compared the recall-time performance of NE-RQ with the proximity graph-based ip-NSW algorithm~\cite{morozov:graphmips} on the ImageNet dataset in Figure~\ref{fig:comparsion with binary hashing}~(right). ip-NSW is shown to achieve the state of the art recall-time performance in existing MIPS algorithms in~\cite{morozov:graphmips}. NE-RQ with two codebooks was used for candidate generation (by combining with the multi-index algorithm~\cite{babenkol:imi}) and the candidates were verified by calculating the exact inner product in this experiment. The results show that NE-RQ achieves higher recall than ip-NSW given the same query processing time. As the implementation may affect the running time, we also plot recall vs. inner product calculation in the supplementary material, which shows that NE-RQ requires fewer inner product computation at the same recall. However, we found ip-NSW provides better recall-time performance than NEQ on the SIFT1M dataset. Although the main design goal of NEQ is good recall-item performance instead of recall-time performance, this experiment shows that using NEQ to generate candidate is beneficial to some datasets.

\textbf{Insights.} A natural question arises after observing the good performance of NEQ:~\textit{Does NEQ only reduce the error in norm? Or it reduces the quantization error as a by-product of its design?} To answer this question, we compared the quantization error ($\Vert x-\tilde{x} \Vert$ normalized by the maximum norm in the dataset) and the norm error of RQ and NE-RQ in Figure~\ref{fig:quantization error}. The number of codebooks is 8 and the reported errors are averaged over all items in the dataset. The results show that NE-RQ indeed reduces norm error significantly but its quantization error is slightly larger than RQ on all the four datasets. This can be explained by the fact that NE-RQ uses 1 codebook to encode the norm and has fewer vector codebooks than RQ. This result shows that a smaller quantization error does not necessarily result in better performance for MIPS. Originally designed for Euclidean distance, existing VQ methods minimize the quantization error. With NEQ, we have shown that the minimizing quantization error is not a suitable design principle for inner product due to its unique properties.

\section{Conclusions}\label{sec:conclusions}

In this paper, we questioned whether minimizing the quantization error is a suitable design principle of VQ techniques for MIPS. We found that the quantization error in norm have great influence on inner product and can be significantly reduced by explicitly encoding it using separate codebooks. Based on this observation, we proposed NEQ --- a general paradigm that specializes existing VQ techniques for MIPS. NEQ is simple as it does not modify the codebook learning process of existing VQ methods. NEQ is also general as it can be easily combined with existing VQ methods. Experimental results show that NEQ provides good performance consistently on various datasets and parameter configurations. Our work shows that inner product requires different design principles from Euclidean distance for VQ techniques and we hope to inspire more researches in this direction.  

~\\
\textbf{Acknowledgment}: This work was supported by ITF 6904945, and GRF 14208318 \& 14222816, and the National Natural Science Foundation of China (NSFC) (Grant No. 61672552).

\bibliographystyle{aaai}
\bibliography{3229.neq}

\newpage
\onecolumn
\appendix
	\vbox{
	\hsize\textwidth
	\linewidth\hsize
	\vskip 0.1in
	\hrule height 4pt
	\vskip 0.25in
	\vskip -\parskip
	\centering
	{\LARGE\bf Supplementary Material for Norm-Explicit Quantization\par}
	\vskip 0.29in
	\vskip -\parskip
	\hrule height 1pt                  
	\vskip 0.09in
}

\section{Proof of Theorem 1}
In this part, we re-state Theorem~\ref{theorem:error} in the main paper and provide its proof.
\begin{theorem}\label{theorem:error}
	For an item $x$, its approximation $\bar{x}$ which is accurate in norm but inaccurate in direction, and a query $q$, denote the angle between $x$ and $\bar{x}$ as $\alpha$ and assume $\alpha\in(0, \pi/2)$, the angle between $\bar{x}$ and $q$ as $\beta$ and assume $\beta\in (0, \pi/2)$, the angle between the two planes defined by ($x$, $\bar{x}$) and ($\bar{x}$, $q$) as $t$. The inner product error $\left \vert \frac{q^{\top}x - q^{\top}\bar{x}}{q^{\top}x} \right \vert$ is not larger than the angular error $1-\frac{x^{\top}\bar{x}}{\Vert x\Vert \Vert \bar{x} \Vert}$ if angle $t$ satisfies $\frac{\cos(\beta)}{\sin(\alpha)\sin(\beta)}\left[\frac{1}{2-\cos(\alpha)}-\cos(\alpha)\right]\leq\cos(t)\leq\frac{\cos(\beta)}{\sin(\alpha)\sin(\beta)}\left[\frac{1}{\cos(\alpha)}-\cos(\alpha)\right]$. 	
\end{theorem}

\begin{figure}[!h]	
	\centering
	\includegraphics[width=0.40\columnwidth]{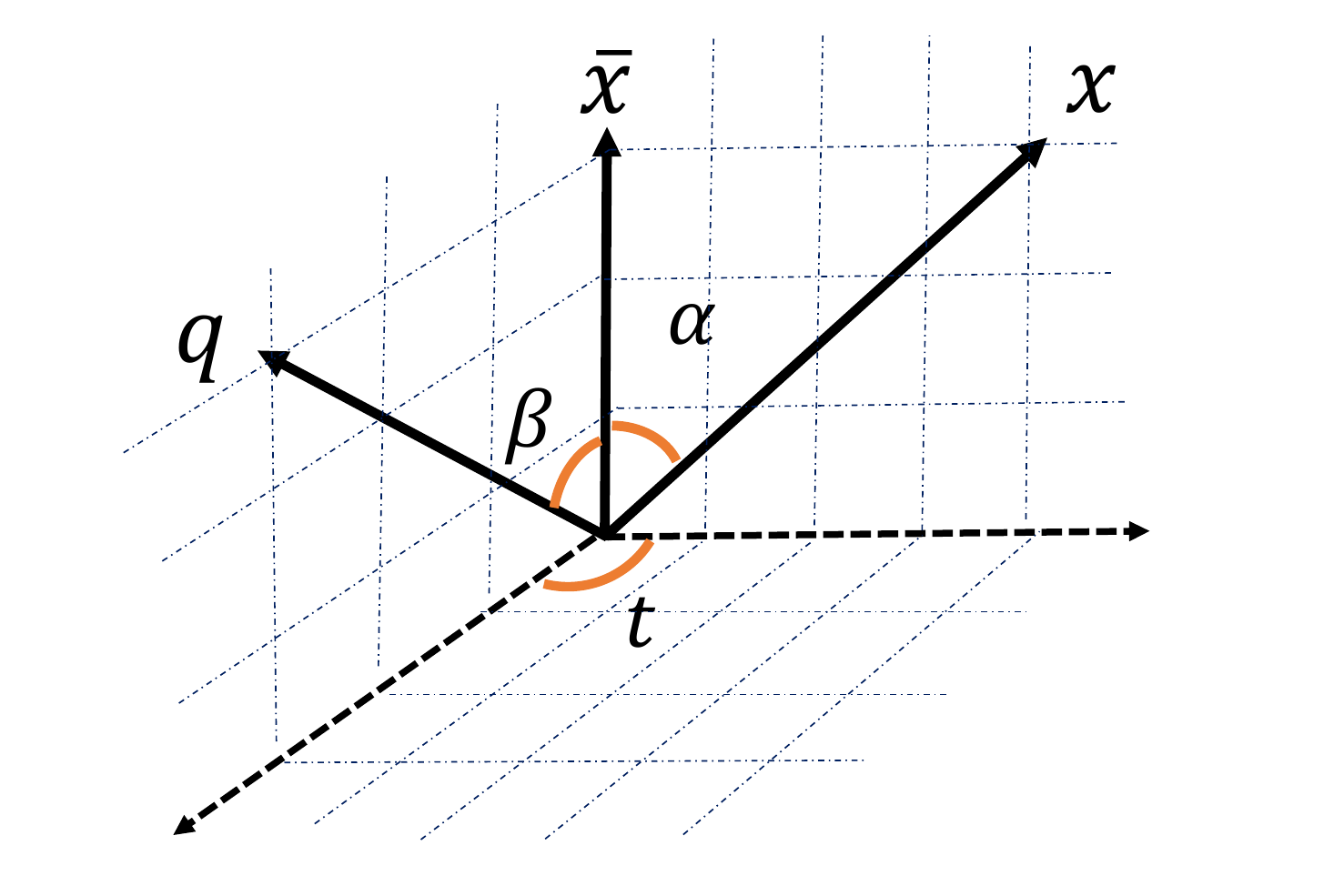}
	\caption{Illustration of the vectors in Theorem~\ref{theorem:error}}\label{fig:example}
\end{figure}

\begin{proof}
	Define the angle between $x$ and $q$ as $\gamma$. As $\bar{x}$ is accurate in norm, the inner product error can be simplified as $\left \vert \frac{q^{\top}x - q^{\top}\bar{x}}{q^{\top}x} \right \vert=\left \vert \frac{\Vert x \Vert \Vert q \Vert\cos(\gamma) - \Vert x \Vert \Vert q \Vert\cos(\beta)}{\Vert x \Vert \Vert q \Vert\cos(\gamma)}\right \vert=\left \vert \frac{\cos(\gamma) -\cos(\beta)}{\cos(\gamma)} \right \vert$. The angular error is $1-\cos(\alpha)$. Therefore, to find the feasible region of $t$ for Theorem~\ref{theorem:error} to hold, we need to express $\gamma$ as a function of $t$ and compare the inner product error and angular error. 
	
	With some simple derivation, we have $\cos(\gamma)=\sin(\alpha)\sin(\beta)\cos(t)+\cos(\alpha)\cos(\beta)$. Note that we assume $\alpha$, $\beta$ and $t$ are all in $(0, \pi/2)$, thus $\cos(\alpha)>0$, $\cos(\beta)>0$ and $\cos(t)>0$ and we also have $\cos(\gamma)>0$.
	
	The inner product error ($\left \vert \frac{\cos(\gamma) -\cos(\beta)}{\cos(\gamma)} \right \vert$) is not larger than angular error ($1-\cos(\alpha)$) can happen in two cases. 
	
	Case 1: $\cos(\gamma)\ge \cos(\beta)$ and $1-\cos(\alpha)\ge 1-\frac{\cos(\beta)}{\cos(\gamma)}$, solving this case, we have $\frac{\cos(\beta)}{\sin(\alpha)\sin(\beta)}(1-\cos(\alpha))\le \cos(t)\le\frac{\cos(\beta)}{\sin(\alpha)\sin(\beta)}(\frac{1}{\cos(\alpha)}-\cos(\alpha))$.
	
	Case 2: $\cos(\gamma)\le \cos(\beta)$ and $1-\cos(\alpha)\ge \frac{\cos(\beta)}{\cos(\gamma)}-1$, solving this case, we have $\frac{\cos(\beta)}{\sin(\alpha)\sin(\beta)}(\frac{1}{2-\cos(\alpha)}-\cos(\alpha))\le \cos(t)\le\frac{\cos(\beta)}{\sin(\alpha)\sin(\beta)}(1-\cos(\alpha))$.
	\
	We can merge the two ranges and obtain the final feasible region of $t$ as $\frac{\cos(\beta)}{\sin(\alpha)\sin(\beta)}\left[\frac{1}{2-\cos(\alpha)}-\cos(\alpha)\right]\leq\cos(t)\leq\frac{\cos(\beta)}{\sin(\alpha)\sin(\beta)}\left[\frac{1}{\cos(\alpha)}-\cos(\alpha)\right]$.

\end{proof}

\section{Adaptation of NEQ to IMI}  
The inverted multi-index (IMI) is an efficient algorithm to generate candidates for Euclidean distance NNS using PQ with two codebooks. Assume all items that are encoded as $(c^1[i], c^2[j])$ are collected in list $\mathcal{W}_{ij}$ and there are $K^2$ such lists. IMI probes these lists in an ascending order of their Euclidean distances ($\Vert q- (c^1[i], c^2[j]) \Vert$) to the query. IMI only needs to sort the squared partial distances ($\Vert q_1 - c^1[i] \Vert^2$ and $\Vert q_2 - c^2[j] \Vert^2$) instead of computing the distances of all lists to $q$ explicitly. IMI uses a priority queue to generate the next list to probe on demand and is very efficient as the size of the priority queue is only $O(\sqrt{t})$ when $t$ lists are generated. Due to its efficiency and the ability to manage a large number of fine-grained lists, IMI significantly outperforms single inverted index based IVFADC for candidate generation in Euclidean distance NNS. We refer readers to~\cite{babenkol:imi} for a detailed discussion on IMI.                

\begin{figure}[!h]
	\centering
	\includegraphics[width=0.49\textwidth]{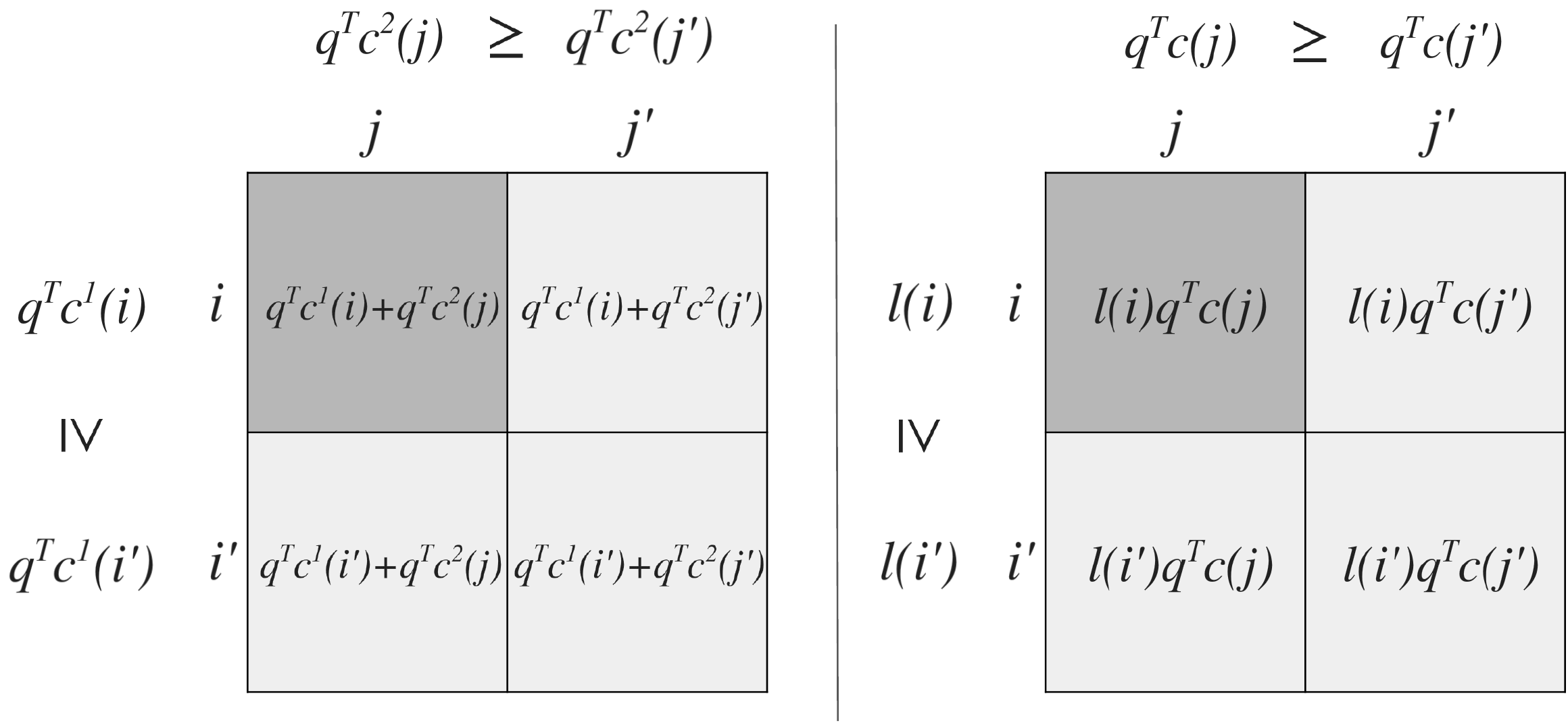}
	\caption{IMI for VQ (left) and NEQ (right)}\label{fig:imi}
\end{figure}

IMI builds on the following property of Euclidean distance,
\begin{equation}\label{equ:multi index Elucidean}
\Vert q- [c^1[i], c^2[j]] \Vert^2 = \Vert q_1 - c^1[i] \Vert^2 + \Vert q_2 - c^2[j] \Vert^2,
\end{equation}
which means that the overall squared distance is the sum of partial distances on the two codebooks. Similar property also holds when VQ with two codebooks are used for MIPS as     
\begin{equation}\label{equ:multi index inner product}
q^{\top}(c^1[i]+c^2[j]) = q^{\top}c^1[i] + q^{\top}c^2[j],  
\end{equation}
which suggests that the total inner product is the sum of the partial inner products from the two codebooks. Therefore, IMI can also be used for candidate generation for VQ. An illustration of VQ-based IMI is shown in Figure~\ref{fig:imi} (left). The codewords in a codebook are sorted in descending order of their inner products with $q$. The two codebooks are arranged in the horizontal direction and vertical direction of the table, respectively.~\eqref{equ:multi index inner product} ensures that the upper left list $\mathcal{W}_{ij}$ in Figure~\ref{fig:imi} (left) has the maximum inner product among the 4 lists, while the lower right list $\mathcal{W}_{i'j'}$ has the minimum inner product. In fact, this ranking relation is all that is required to prove the correctness and efficiency of IMI.       

For NEQ with two codebooks, i.e., one norm codebook $\mathcal{L}$ and one direction codebook $\mathcal{C}$, the approximate inner product between list $\mathcal{W}_{ij}$ (indexed by $l[i], c[j]$) and query $q$ can be expressed as,
\begin{equation}\label{equ:multi index neq}
q^{\top}(l[i] \cdot c[j]) = l[i] \cdot q^{\top}c[j].   
\end{equation}

Assume that $q^{\top}c[j] \ge q^{\top}c[j'] \ge 0$. The upper-left list $\mathcal{W}_{ij}$ in Figure~\ref{fig:imi} (right) has the largest inner product among the 4 lists, while the lower right list $\mathcal{W}_{i'j'}$ has the minimum inner product. Therefore, the ranking relation required by IMI also holds for NEQ, which means that NEQ can also use IMI for candidate generation. One subtlety is that for direction codewords with $q^{\top}c[j] < 0$, we have $l[i]q^{\top}c[j] < l[i']q^{\top}c[j]$ if $l[i] > l[i']$. This problem can be addressed by building a separate IMI for these direction codewords, in which the norm codewords are sorted in ascending order. However, this IMI will be activated only when the other IMI (for direction codewords with $q^{\top}c[j] \ge 0$) is exhausted. This is because lists in the other IMI have non-negative inner products with the query, while lists in the special IMI have negative inner products. As IMI is usually used to generate a small number of candidates (compared to the cardinality of the dataset), this IMI seldom needs to be activated. 

Using NEQ-based IMI for candidate generation has a couple of benefits compared with VQ-based IMI. First, the cost of initializing IMI is cheaper as the inner products of only one vector codebook need to be computed. Second, NEQ-based IMI provides an upper bound on $q^{\top}x$ without additional overhead, which can be used to terminate candidate generation. The sphere K-means in~\cite{auvolat:clustering} can be used to ensure that the codeword $c[j]$ has unit norm when there is only one vector codebook. In this case, the norm codebook quantizes the actual norm $\Vert x \Vert$ instead of the relative norm $l_x$. Moreover, in the final iteration of K-means for norm codeword learning, we simply set the center $l[i]$ as the maximum norm in its corresponding cluster. When probing the NEQ lists with IMI, we can keep the value of the $k$-th largest inner products found so far. If the norm codewords of all remaining lists are no larger than this value, candidate generation should be stopped. 

\section{Norm Distributions of Datasets Used in the Experiments} 

\begin{figure*}[h]
	\includegraphics[width=0.245\textwidth]{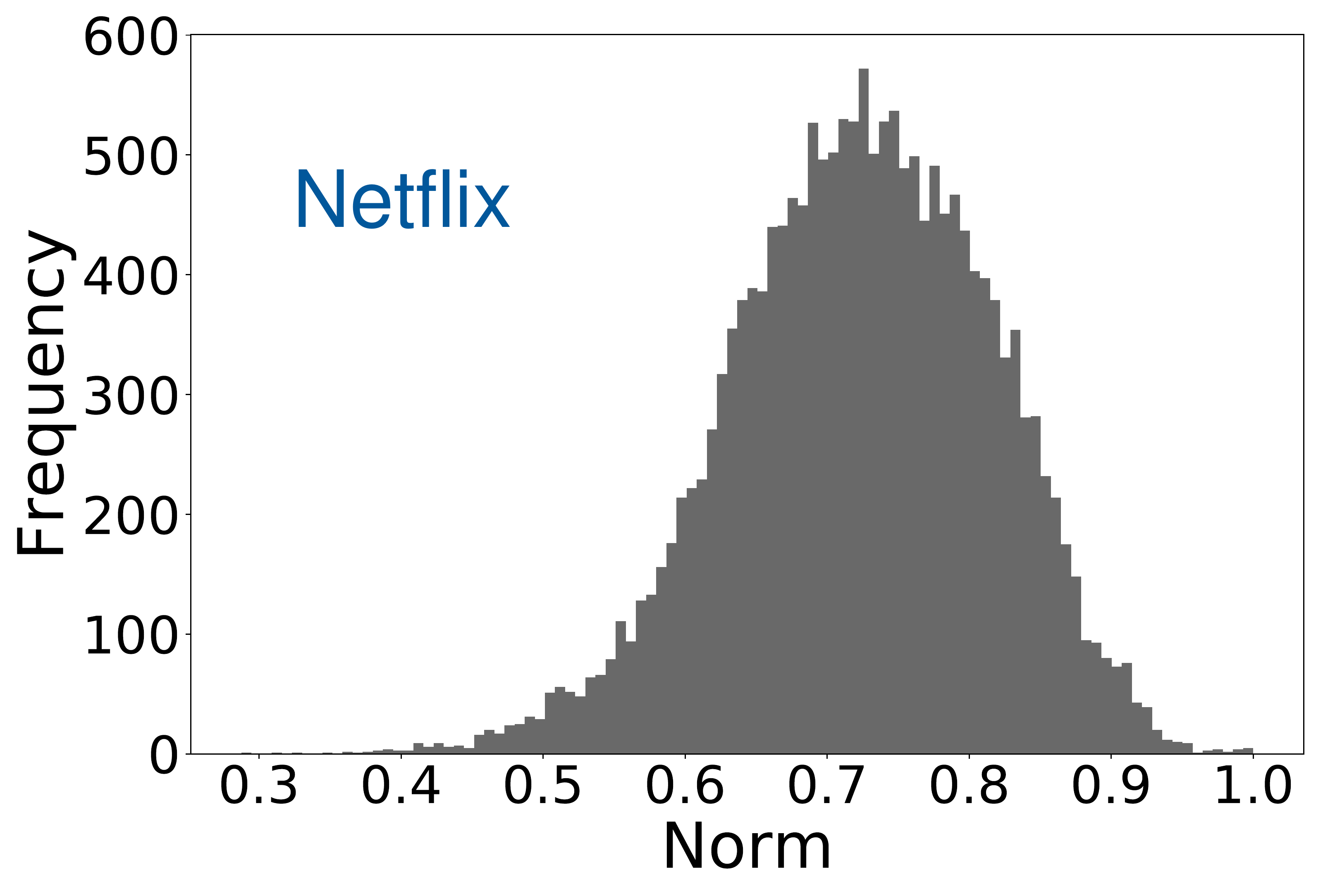}
	\includegraphics[width=0.245\textwidth]{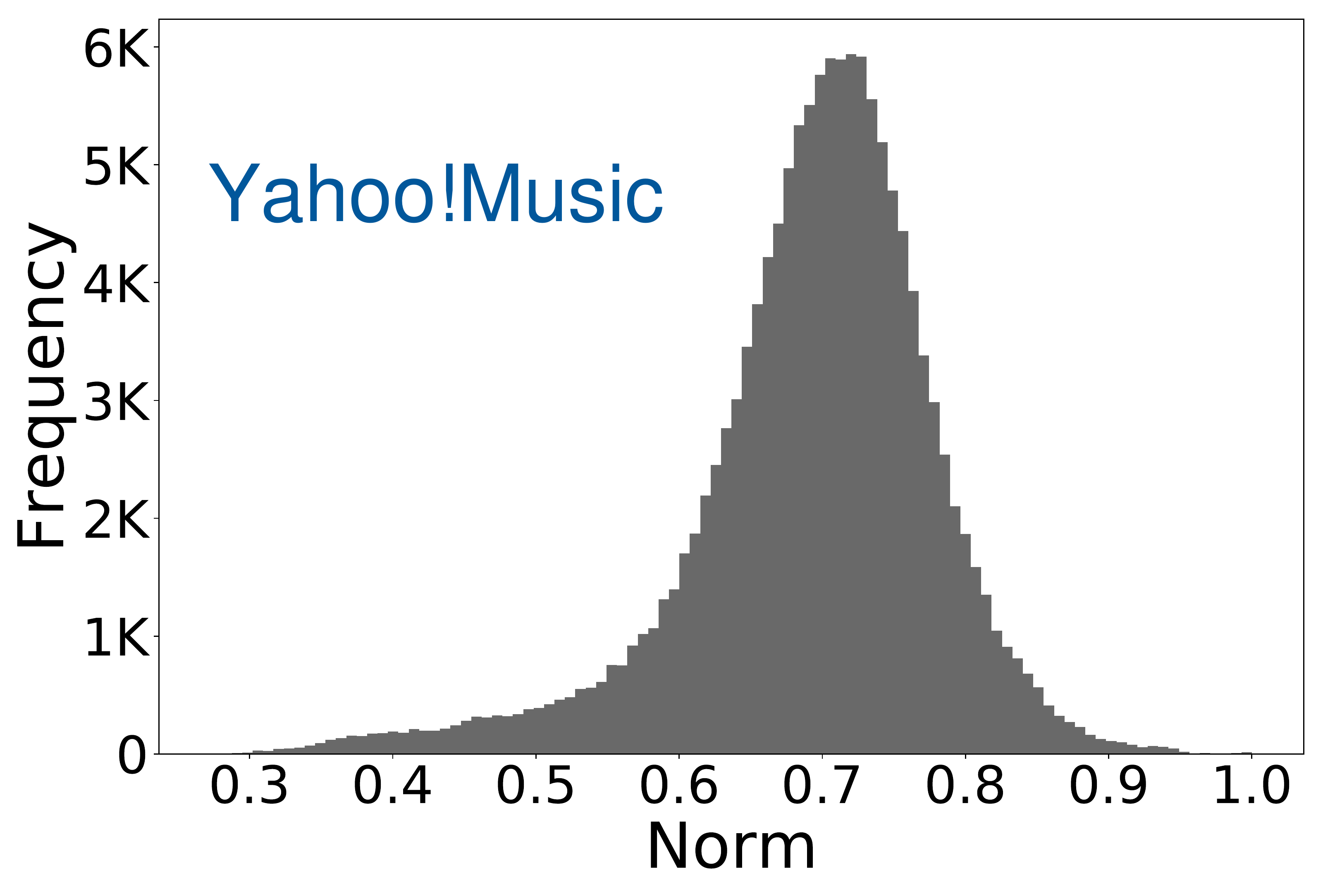}
	\includegraphics[width=0.245\textwidth]{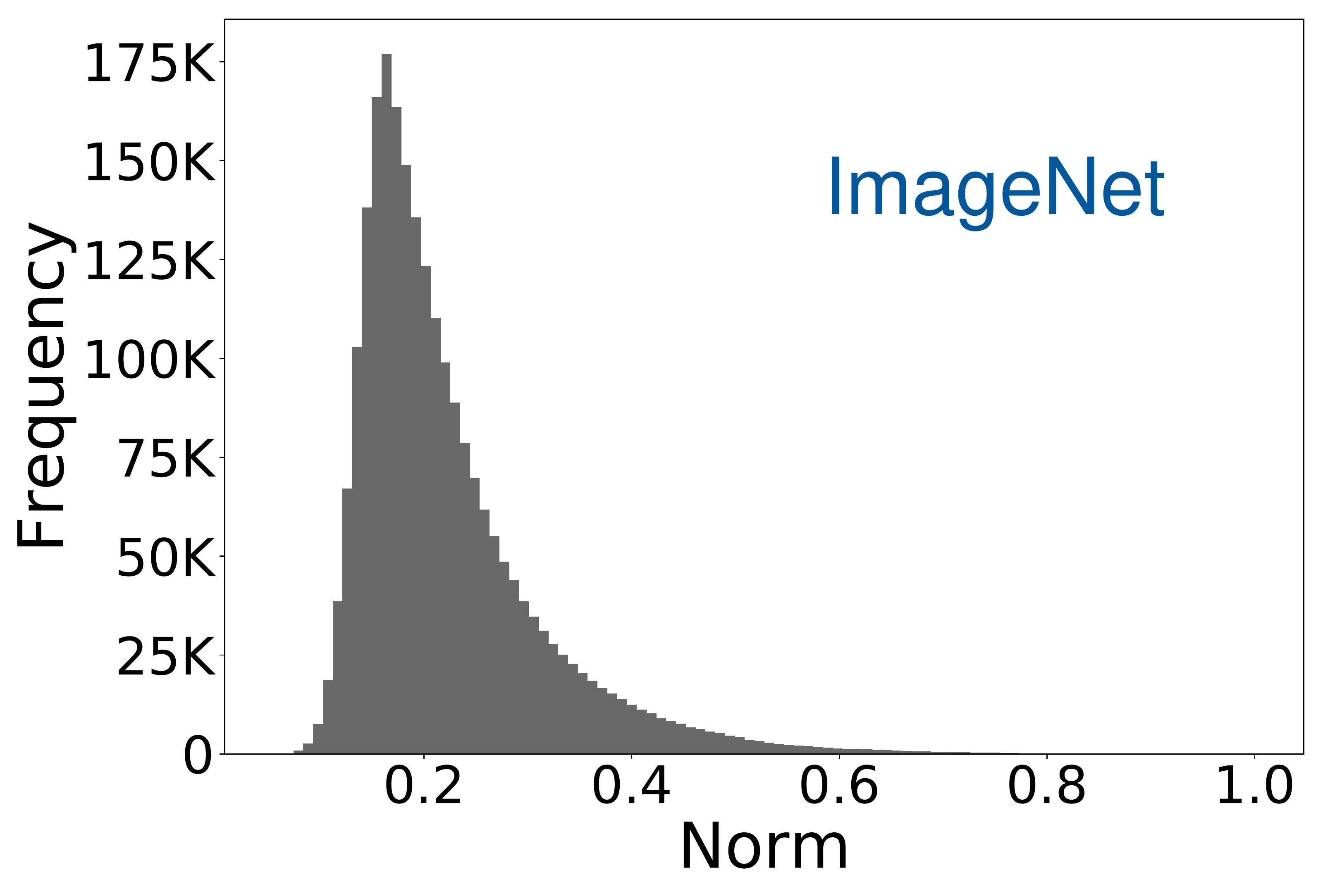}
	\includegraphics[width=0.245\textwidth]{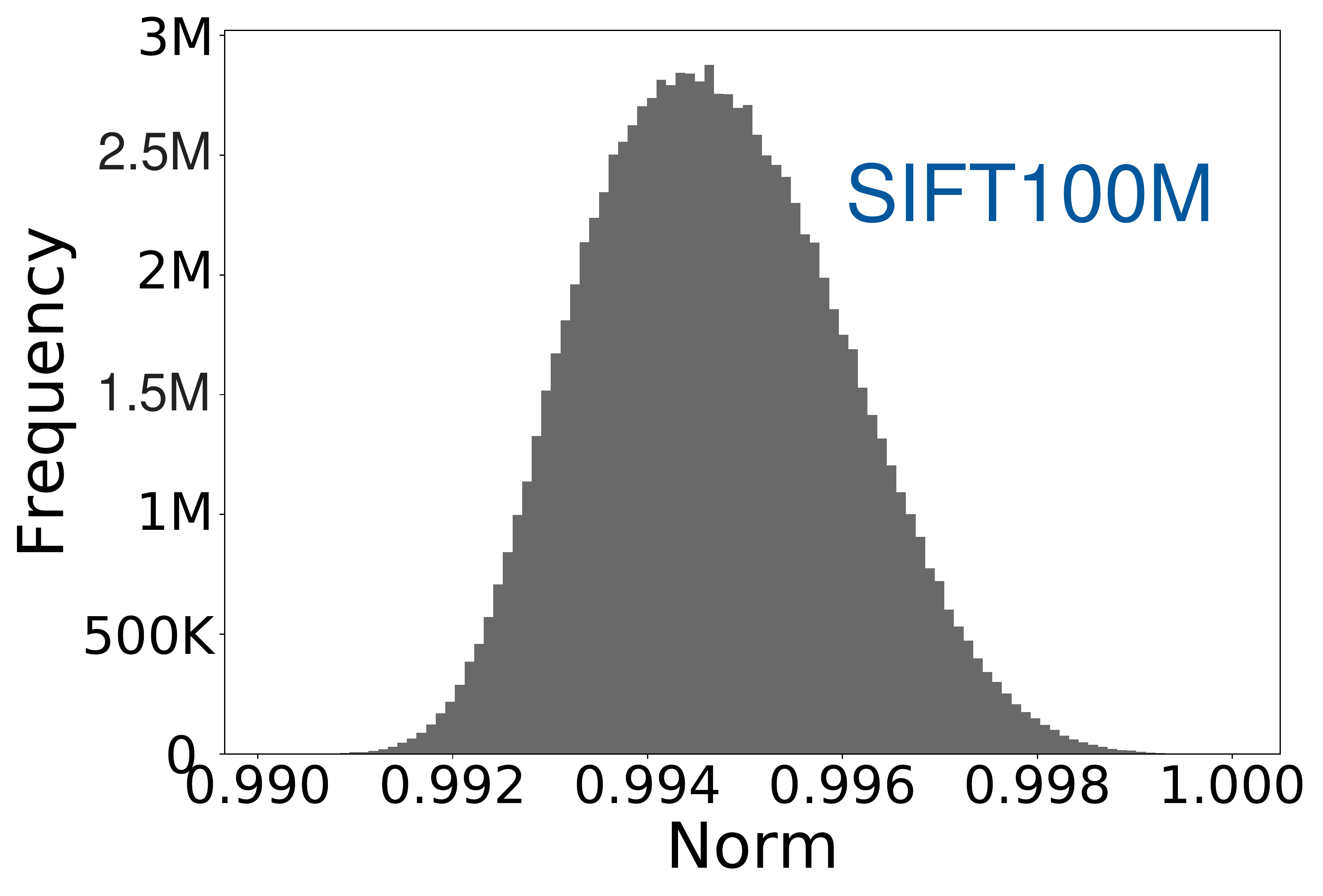}
	\caption{Norm distributions (maximum norm normalized to 1) of the datasets}\label{fig: norm distribution}
\end{figure*}

The norm distributions of the four datasets used in the experiments, i.e., Netflix, Yahoo!Music, ImageNet and SIFT100M, are plotted in Figure~\ref{fig: norm distribution}. Items in SIFT100M have almost identical norm and the value of norm only spans a very small range. ImageNet has a long tail in the norm distribution and a small portion of items have significantly larger norm than the majority. For Netflix and Yahoo!Music, the norms of most items are close to the maximum but a small portion of items have significantly smaller norm than the majority. We choose these datasets to test the robustness of NEQ to different norm distributions.               

\section{Influence of Quantization Error on Euclidean Distance}	

In this part, we analyze the influence of the norm error and angular error on  Euclidean distance. The norm error, angular error and Euclidean distance error are defined as follows.

\begin{definition}
	For an item $x$ and its codebook-based approximation $\tilde{x}$, given a query $q$, the norm error $\alpha$, angular error $\beta$ and the Euclidean distance error $v$ are given as:
	\begin{align}
	\nonumber
	&\alpha = \left \vert \frac{\Vert x \Vert- \Vert \tilde{x} \Vert}{\Vert x \Vert} \right \vert,
	&&\beta = 1-\frac{x^{\top}\tilde{x}}{\Vert x\Vert \Vert \tilde{x} \Vert},
	&v = \left \vert \frac{\Vert x -q \Vert - \Vert \tilde{x} -q \Vert  }{\Vert x -q \Vert} \right \vert.
	\end{align}
\end{definition}

Similar to Section 3 in the main paper, we plot the relation between norm error, angular error and Euclidean distance error on the SIFT1M dataset in Figure~\ref{fig:error}. We used 1,000 randomly selected queries and the errors were calculated on their top-20 Euclidean distance neighbors in the dataset. For each item-query pair ($x$, $q$), we plot two points in Figure~\ref{fig:error}. One (in red) shows the norm error and the Euclidean distance error caused by inaccurate norm (using the approximation $\hat{x}=\Vert \tilde{x} \Vert \cdot \frac{x}{\Vert x \Vert}$). The other (in gray) shows the angular error and the Euclidean distance error caused by inaccurate direction vector (using $\bar{x}=\Vert x \Vert \cdot \frac{\tilde{x}}{\Vert \tilde{x} \Vert}$). We also fit a line through the origin for each group of points in the figure. The results show that angular error has larger influence on Euclidean distance error than norm error, which is contrary to the case of inner product. This phenomenon is another evidence of the difference between inner product and Euclidean distance, which suggests that MIPS requires a design for VQ techniques different from Euclidean nearest neighbor search.         

\begin{figure}[h]	
	\centering 
	\includegraphics[width=0.3\columnwidth]{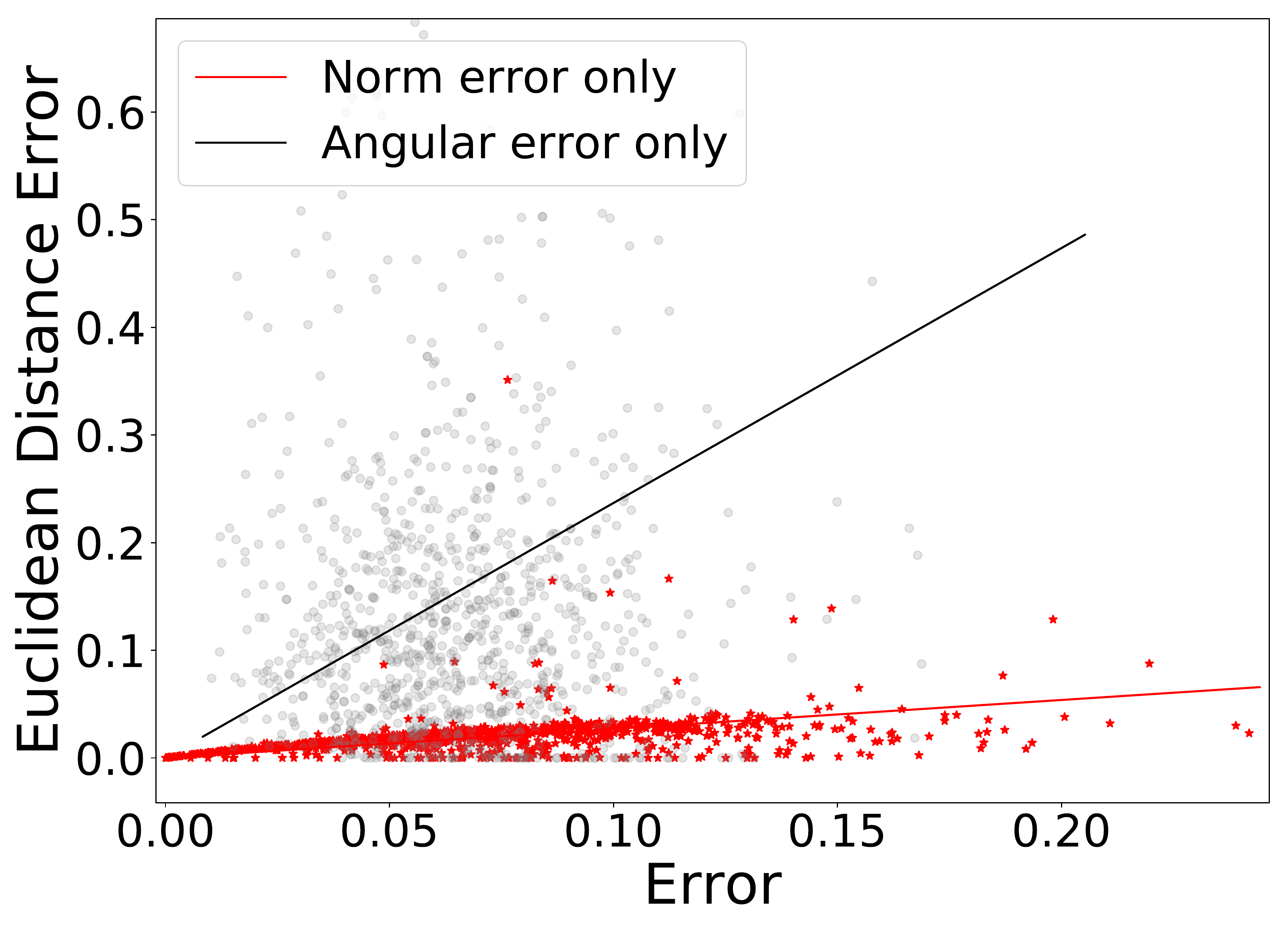}
	\includegraphics[width=0.3\columnwidth]{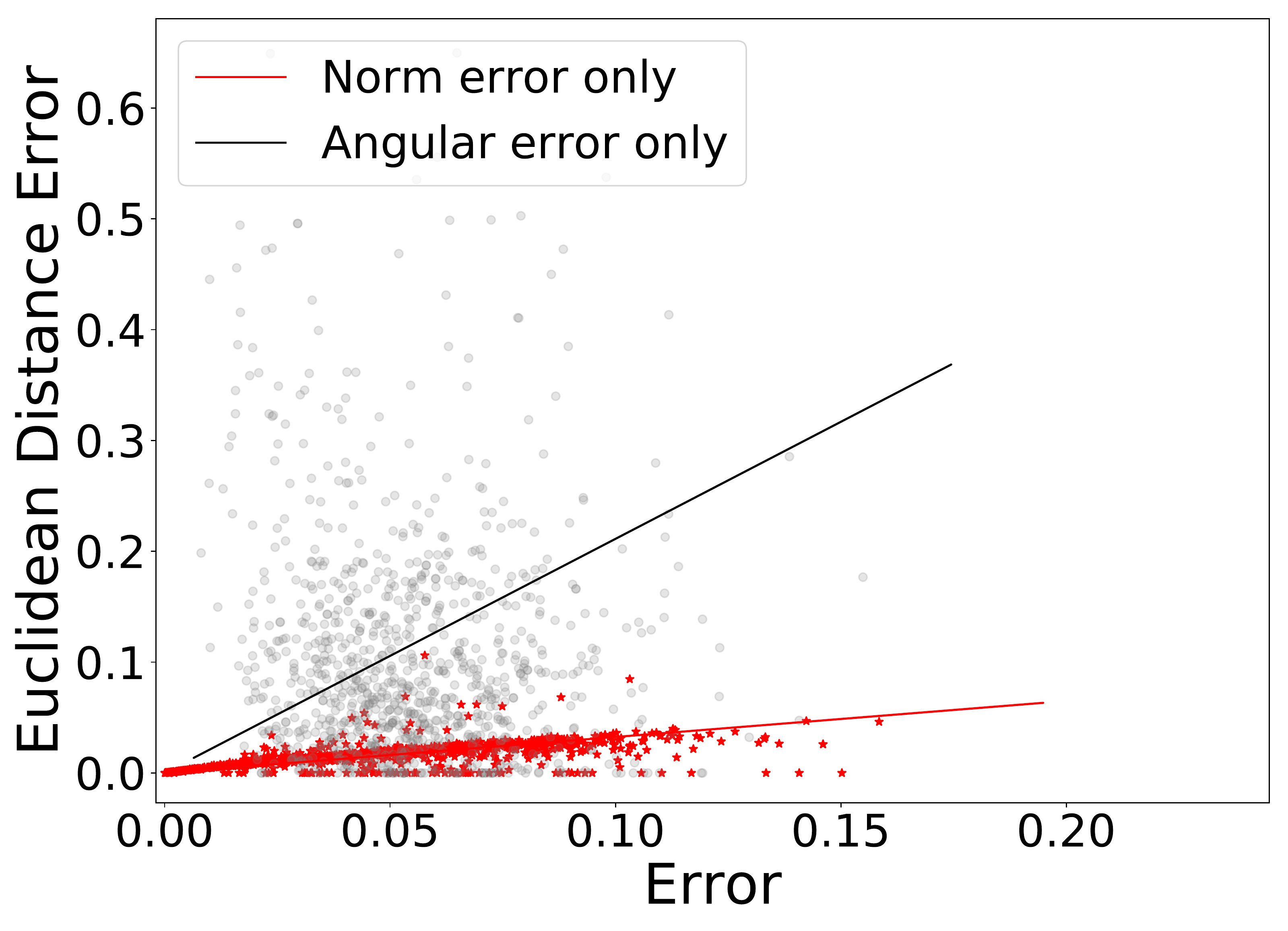}
	\caption{Influence of norm error and angular error on euclidean distance for PQ (left) and RQ (right)}\label{fig:error}
\end{figure}

\section{Additional Experimental Results}

In the main paper, we have shown that NE-RQ consistently improves the recall-item performance of RQ under different configurations of the number of codebooks ($M$) and the number of target inner product neighbors ($k$). In this section, we show that the robustness of NEQ to parameter configurations is consistent across different VQ techniques (e.g., PQ, OPQ and AQ). As the Netflix and Yahoo!Music datasets are small, we show the results on the larger SIFT100M and ImageNet datasets. As the encoding of AQ is complex and computationally heavy, we do not show the performance of AQ on SIFT100M. For ImageNet, we only show the performance of AQ and NE-AQ under part of the configurations. Note that the comparison for different values of $k$ was conducted using 8 codebooks for all figures.

\subsection{Results of PQ and OPQ on SIFT100M}

\begin{figure}[!h]	
	\centering 
	\begin{minipage}[b]{0.49\textwidth}
		\includegraphics[width=0.49\textwidth]{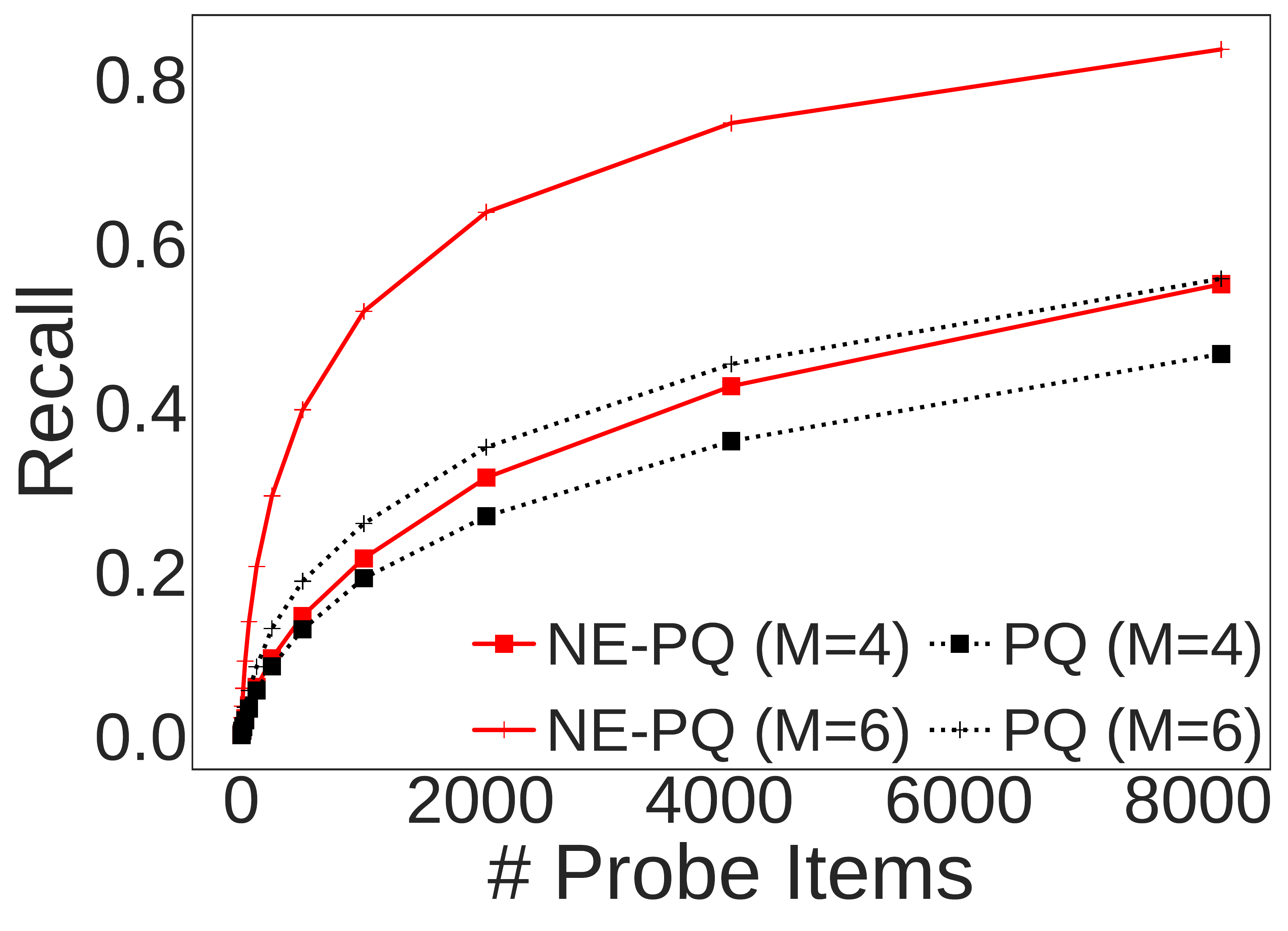}
		\includegraphics[width=0.49\textwidth]{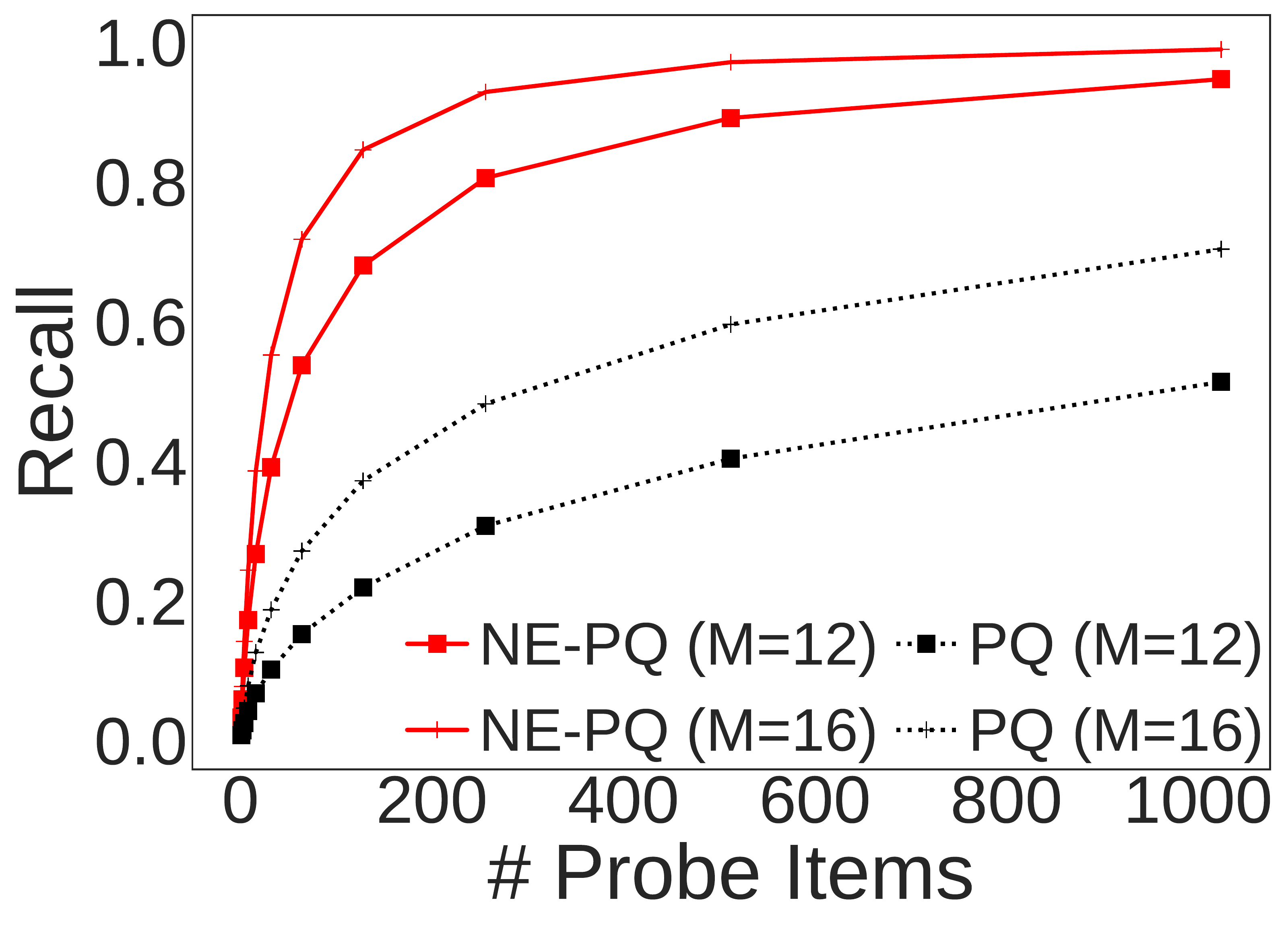}
		\caption{Different \# codebooks for PQ}\label{fig:number of codebooks for PQ on sift100m}
	\end{minipage}
	\begin{minipage}[b]{0.49\textwidth}
		\includegraphics[width=0.49\textwidth]{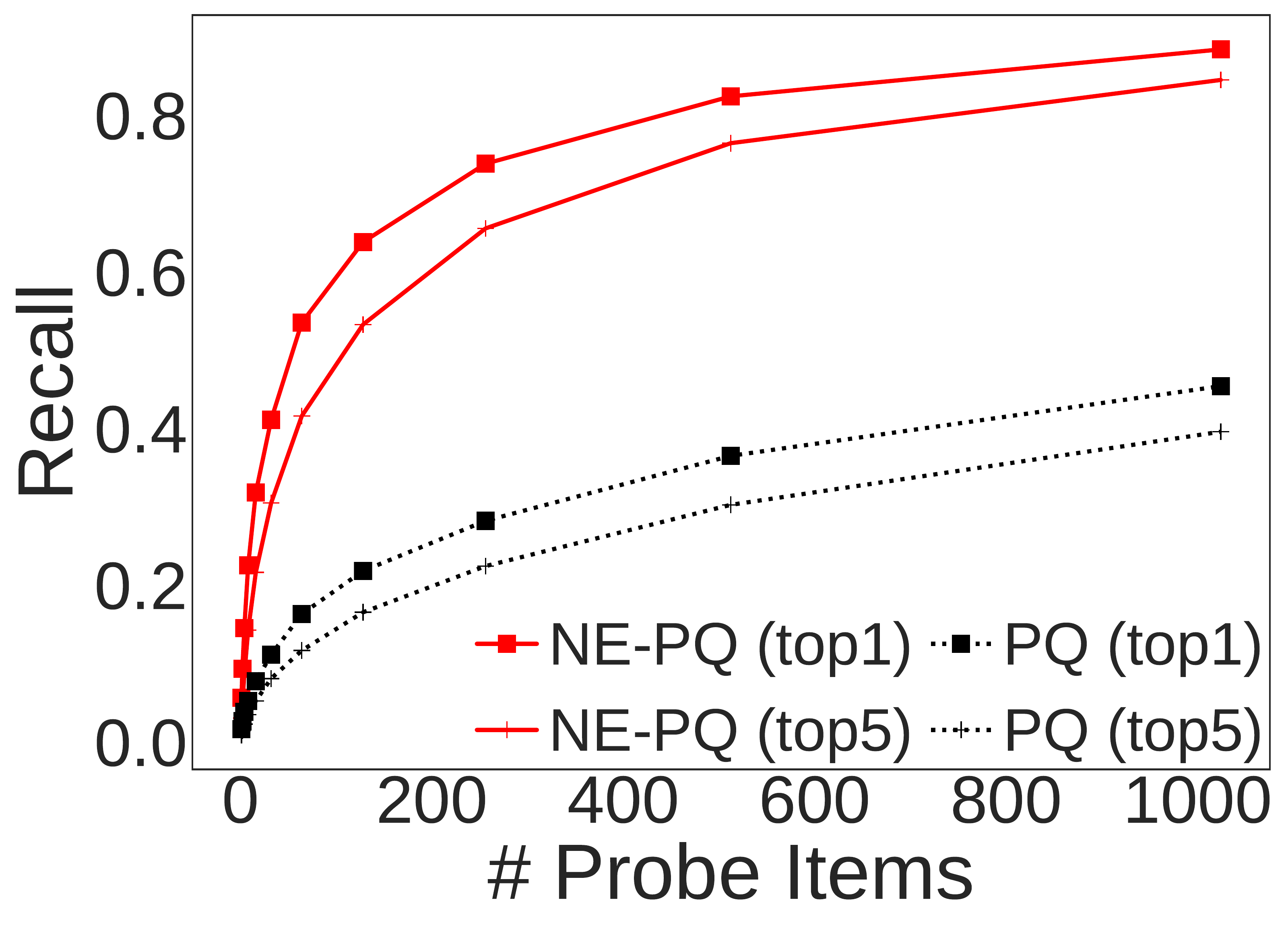}
		\includegraphics[width=0.49\textwidth]{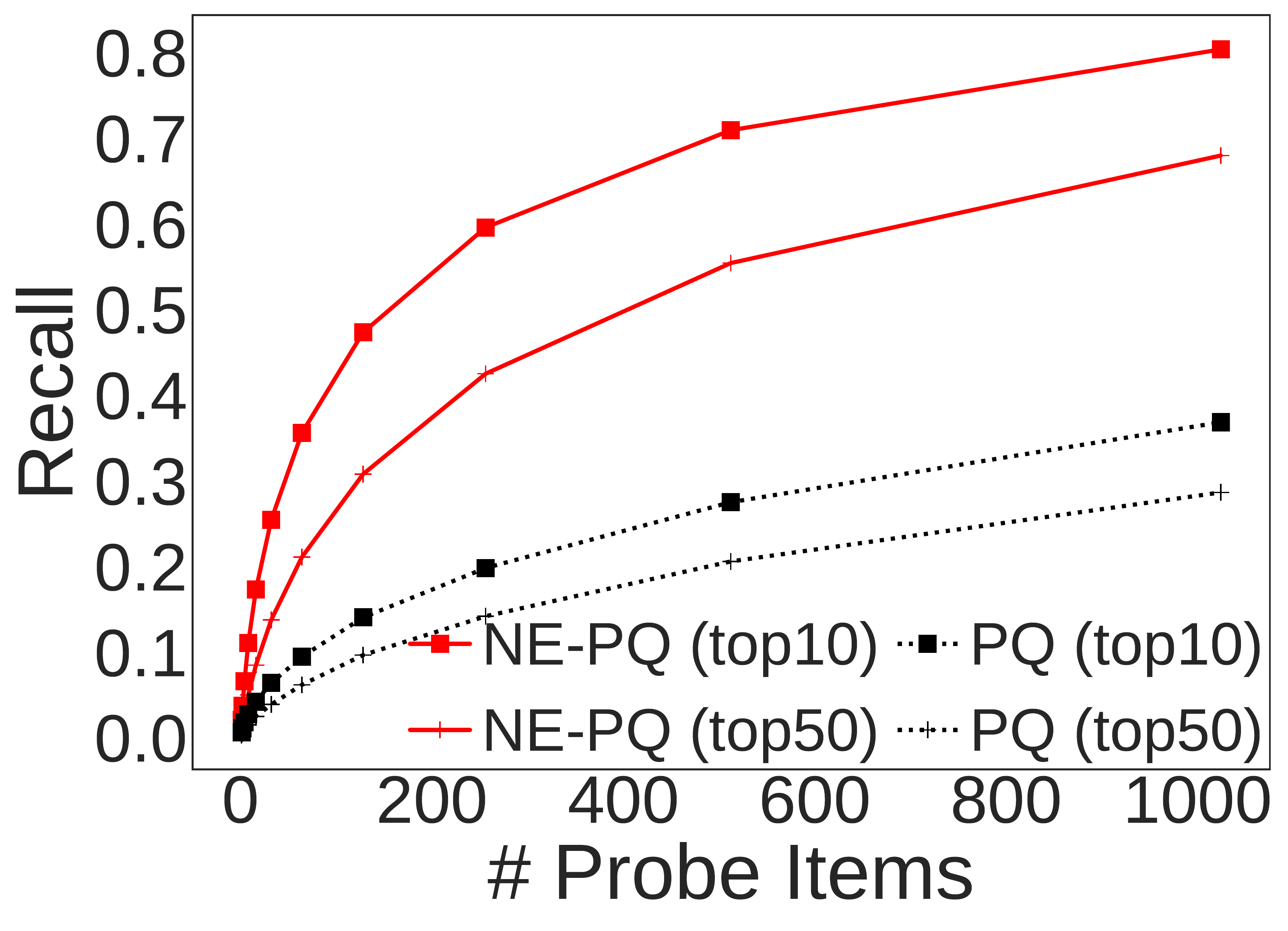}
		\caption{Different values of $k$ for PQ}\label{fig:value of k for PQ on sift100m}
	\end{minipage}
\end{figure}
\begin{figure}[!h]	
	\centering 
	\begin{minipage}[b]{0.49\textwidth}
		\includegraphics[width=0.49\textwidth]{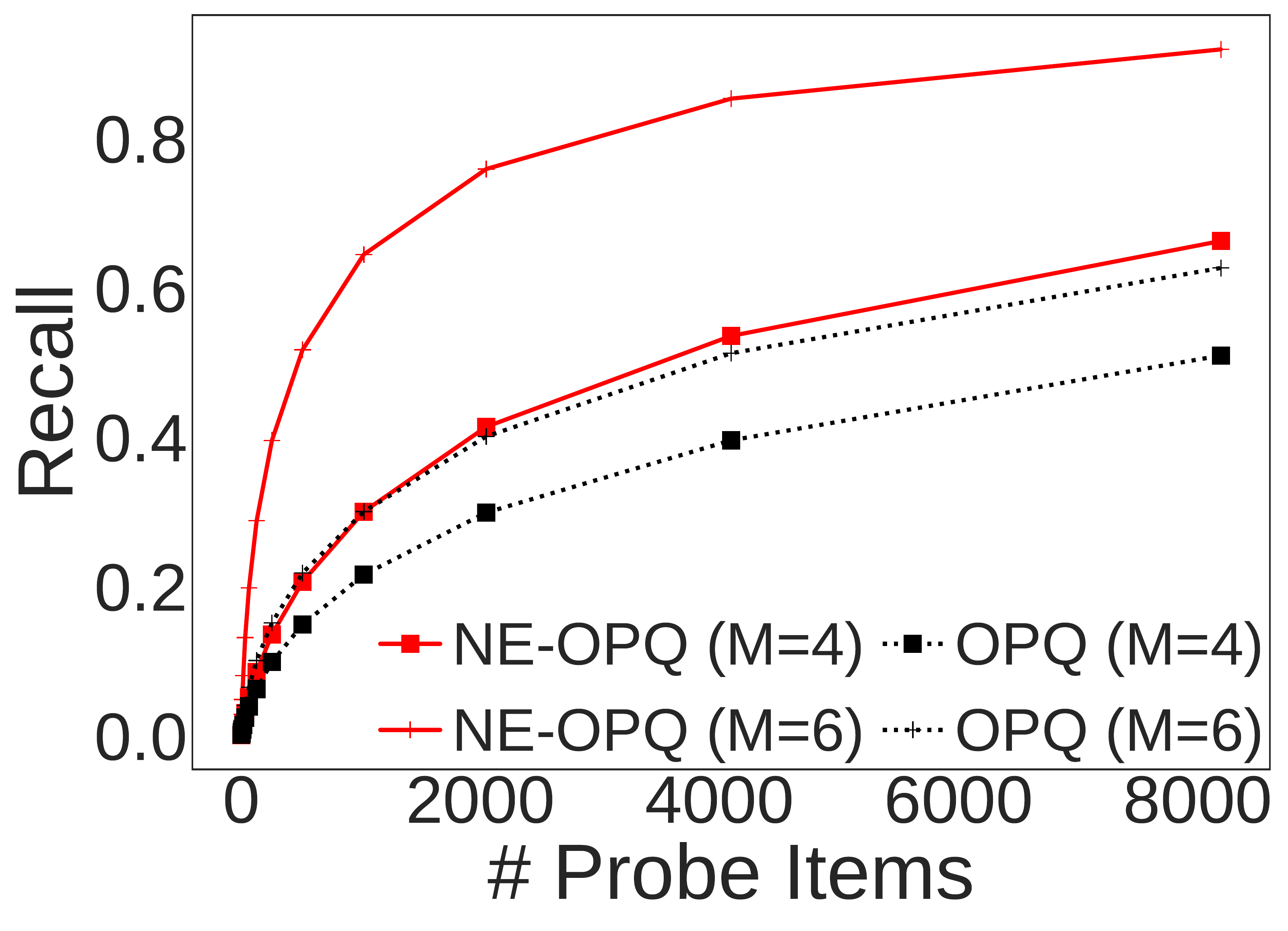}
		\includegraphics[width=0.49\textwidth]{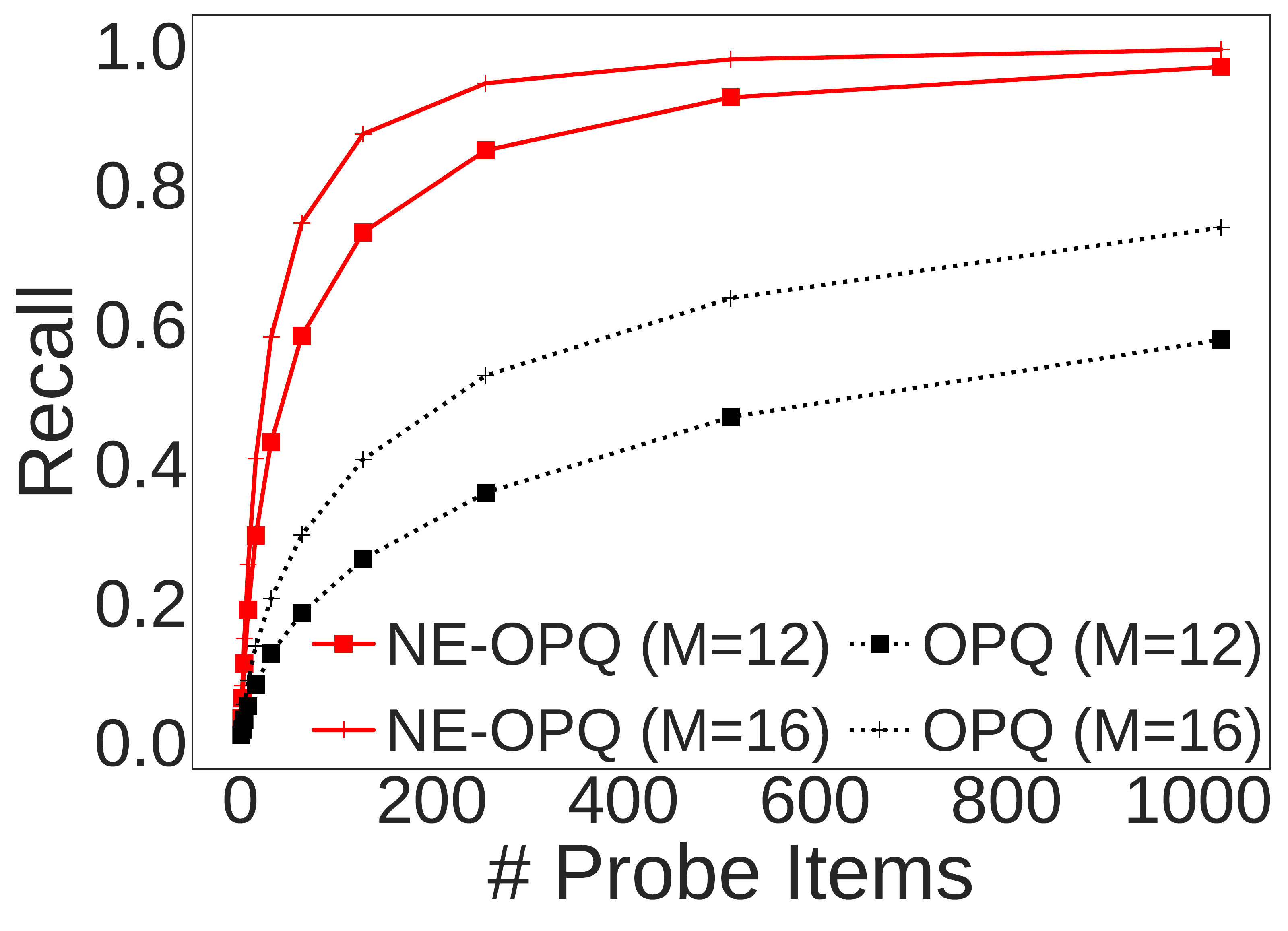}
		\caption{Different \# codebooks for OPQ}\label{fig:number of codebooks for OPQ on sift100m}
	\end{minipage}
	\begin{minipage}[b]{0.49\textwidth}
		\includegraphics[width=0.49\textwidth]{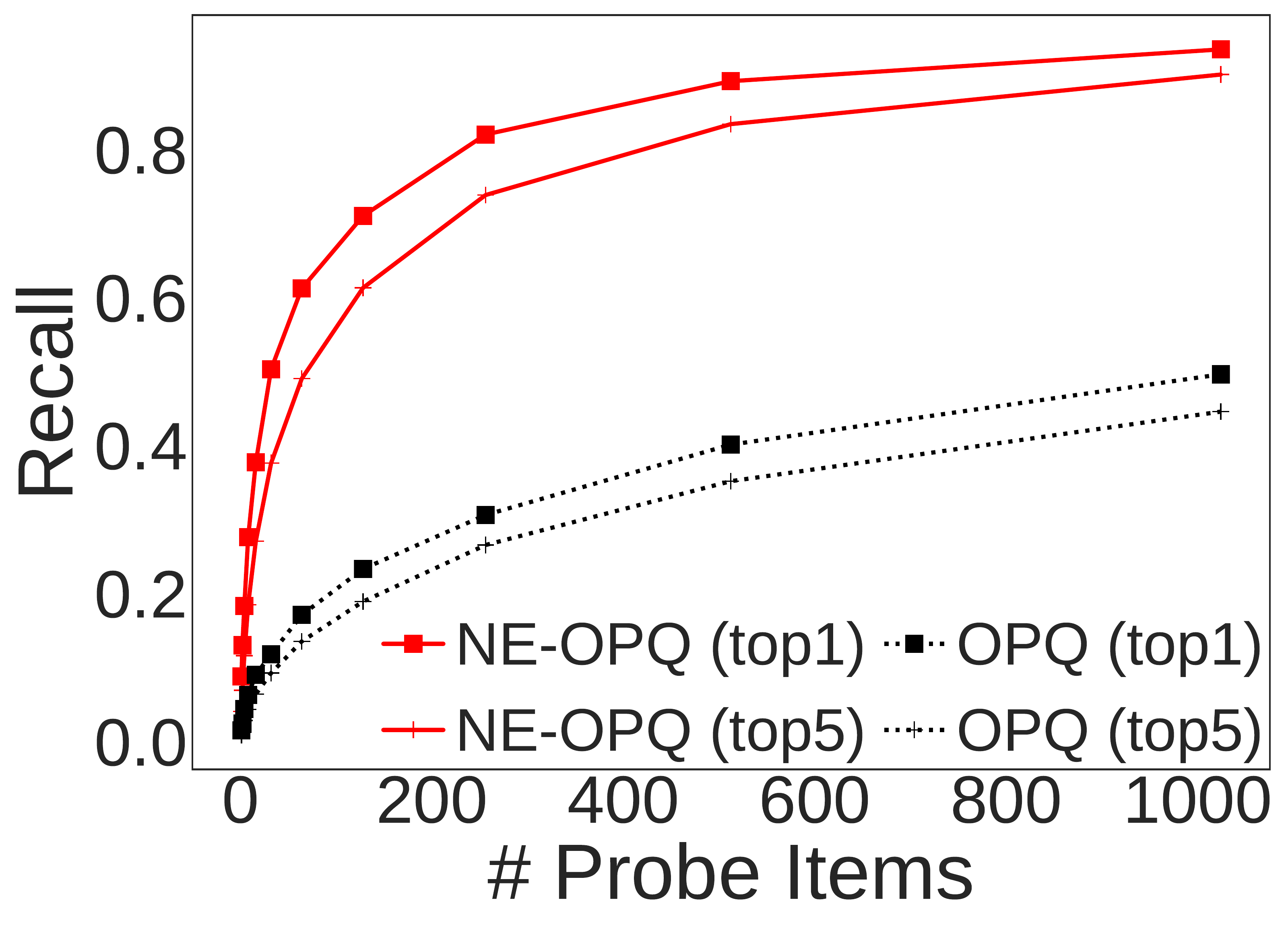}
		\includegraphics[width=0.49\textwidth]{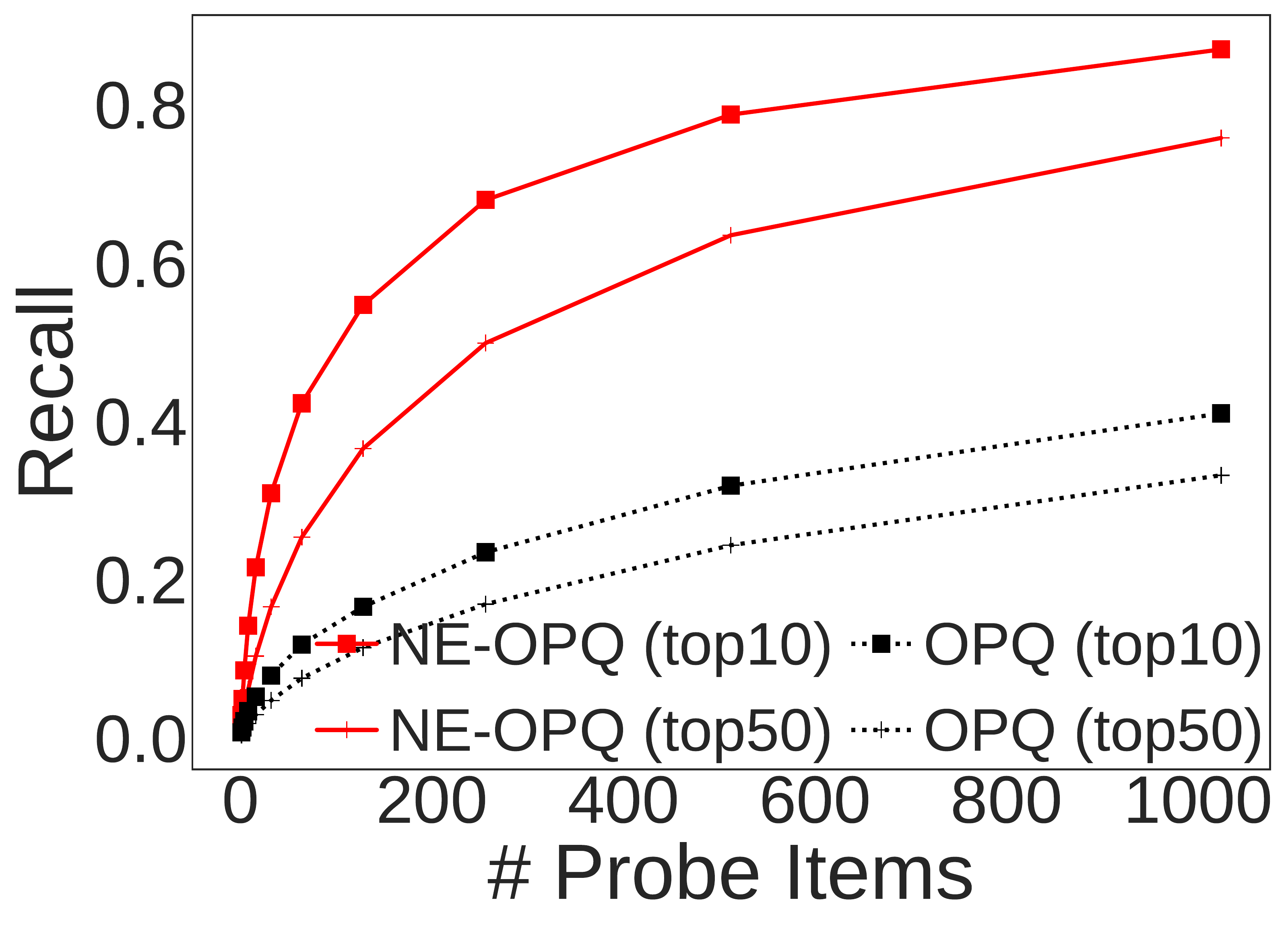}
		\caption{Different values of $k$ for OPQ}\label{fig:value of k for OPQ on sift100m}
	\end{minipage}
\end{figure}
%
%
%

\subsection{Results of PQ, OPQ, RQ and AQ on ImageNet}

\begin{figure}[!h]	
	\centering 
	\begin{minipage}[b]{0.49\textwidth}
		\includegraphics[width=0.49\textwidth]{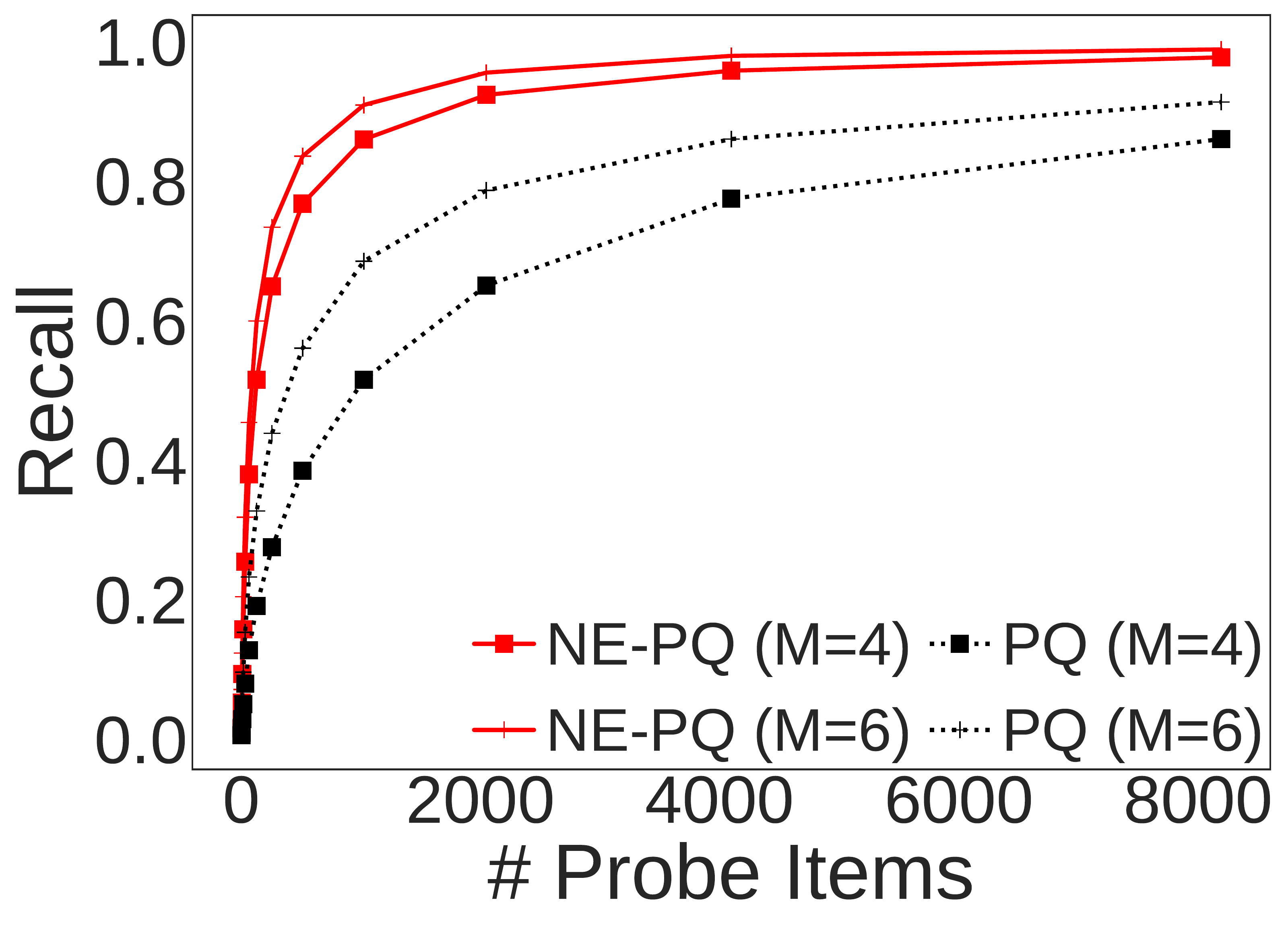}
		\includegraphics[width=0.49\textwidth]{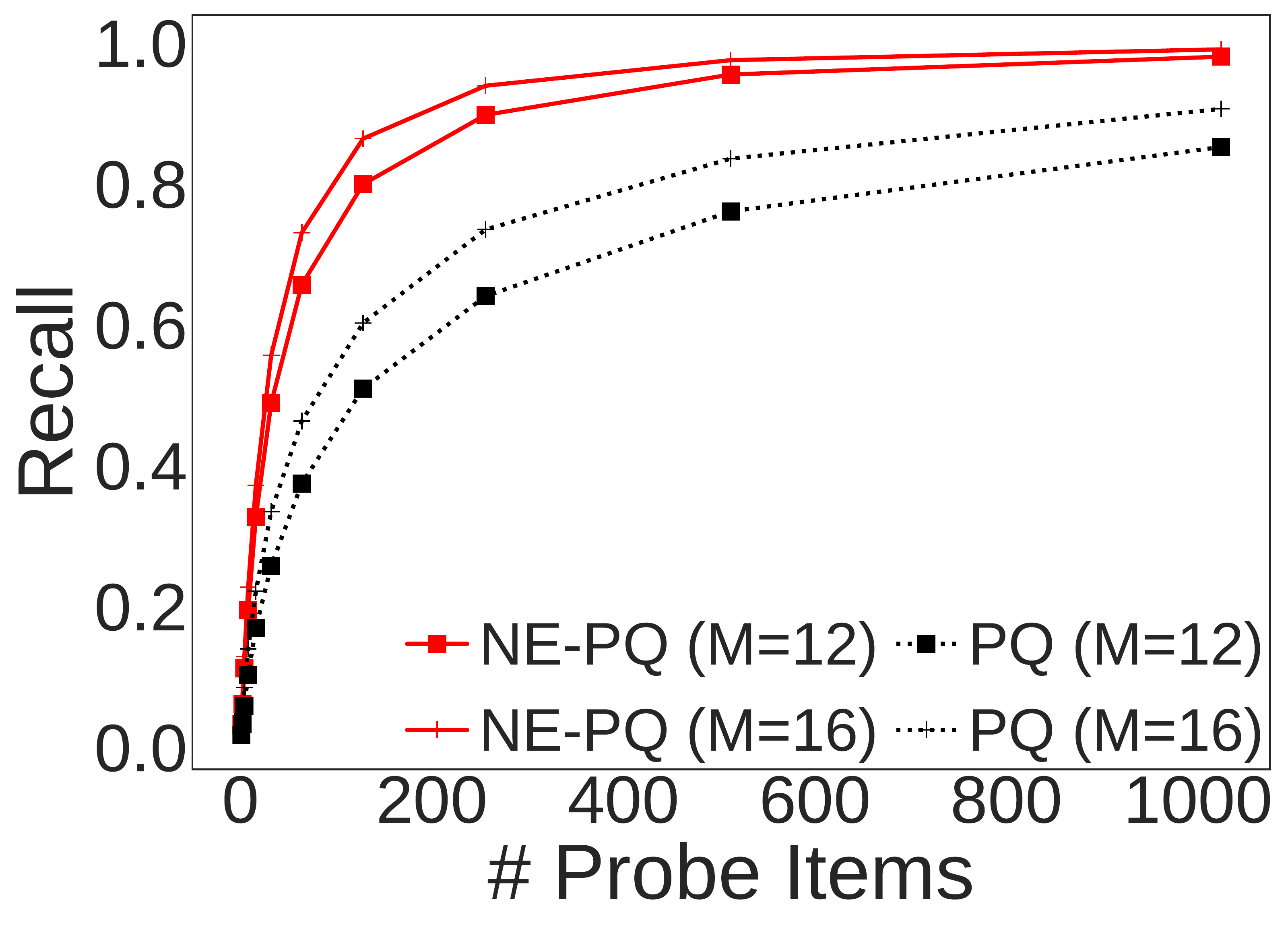}
		\caption{Different \# codebooks for PQ}\label{fig:number of codebooks for PQ on imagenet}
	\end{minipage}
	\begin{minipage}[b]{0.49\textwidth}
		\includegraphics[width=0.49\textwidth]{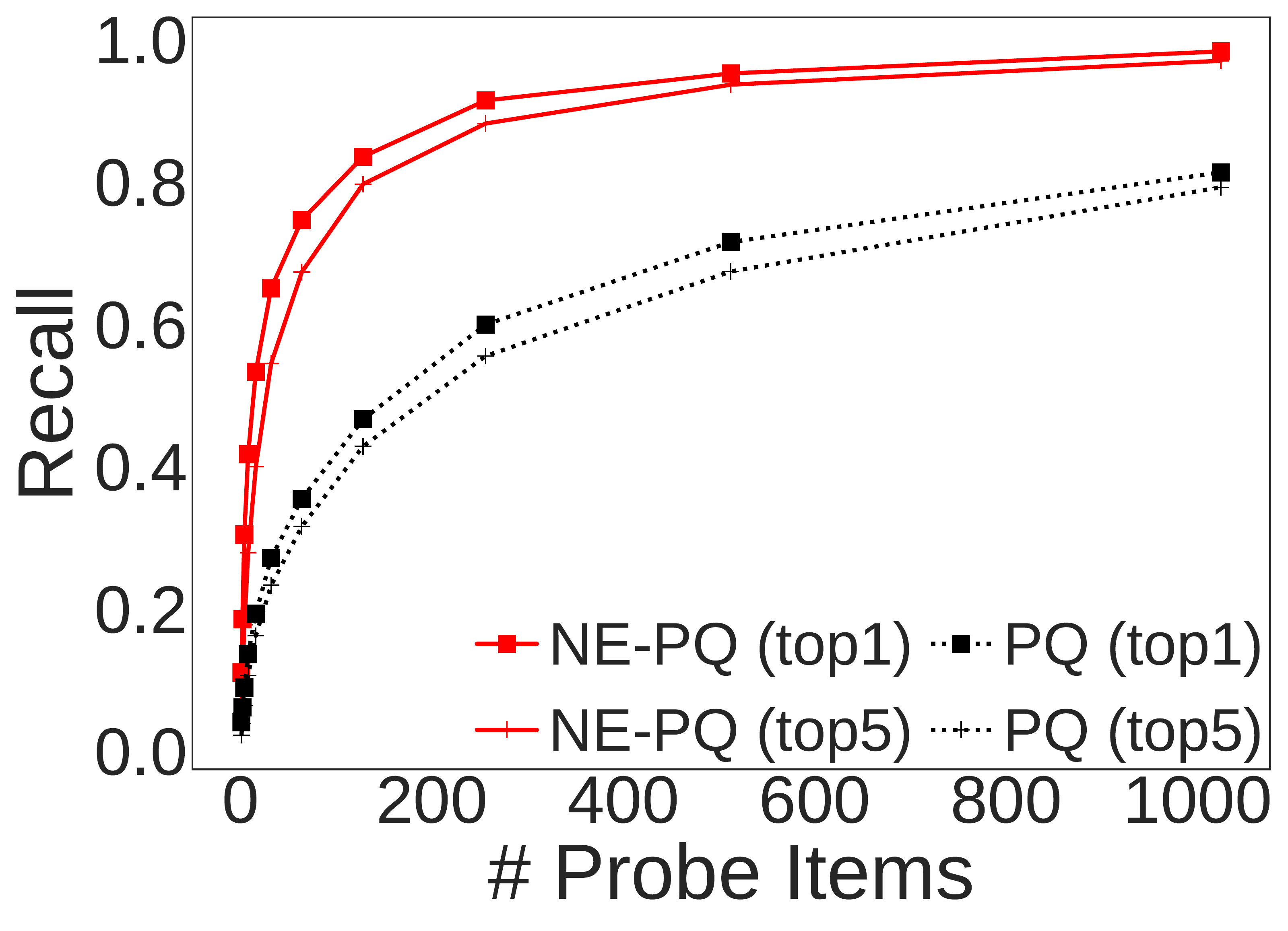}
		\includegraphics[width=0.49\textwidth]{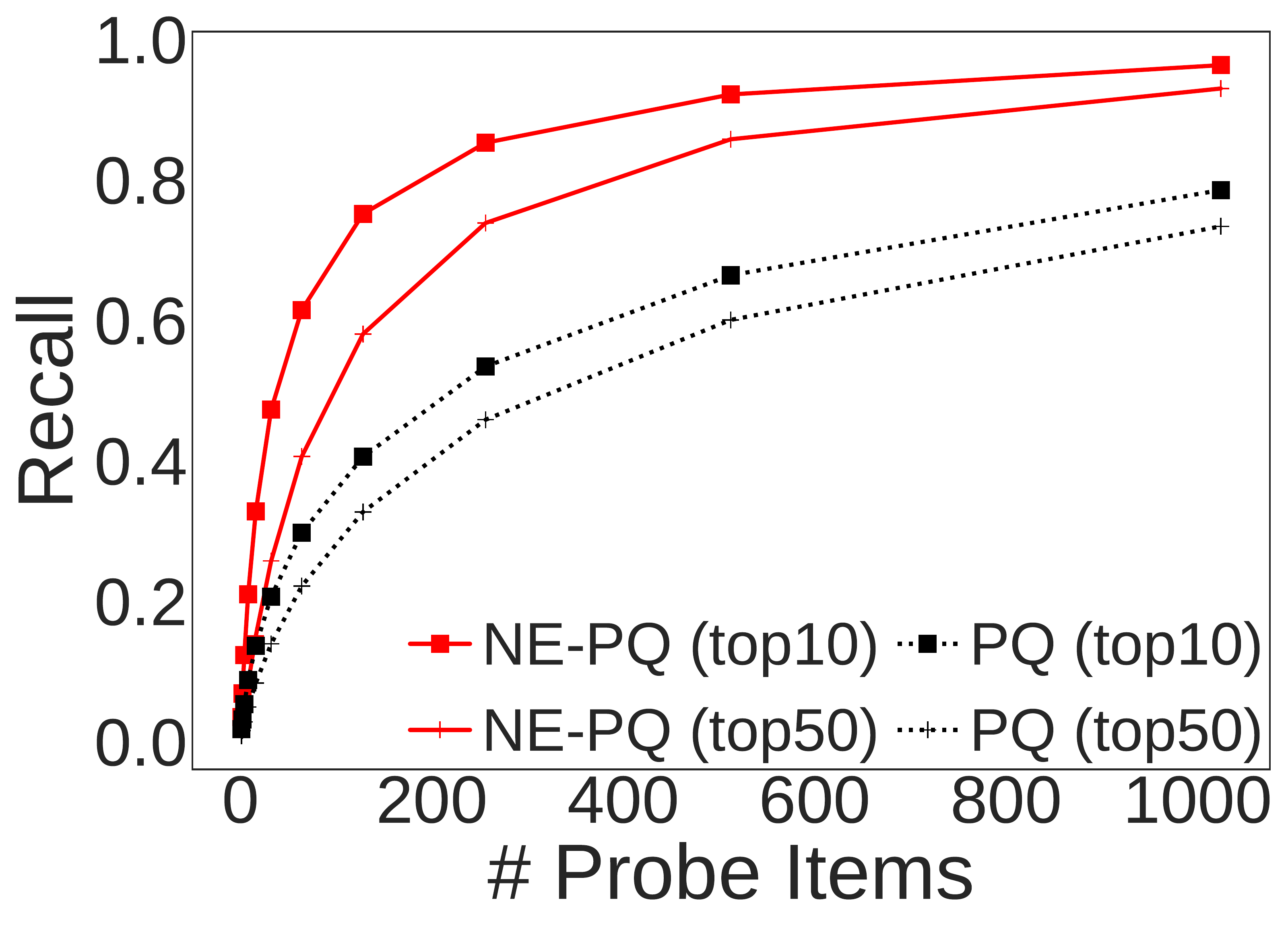}
		\caption{Different values of $k$ for PQ}\label{fig:value of k for PQ on imagenet}
	\end{minipage}
\end{figure}
\begin{figure}[!h]	
	\centering 
	\begin{minipage}[b]{0.49\textwidth}
		\includegraphics[width=0.49\textwidth]{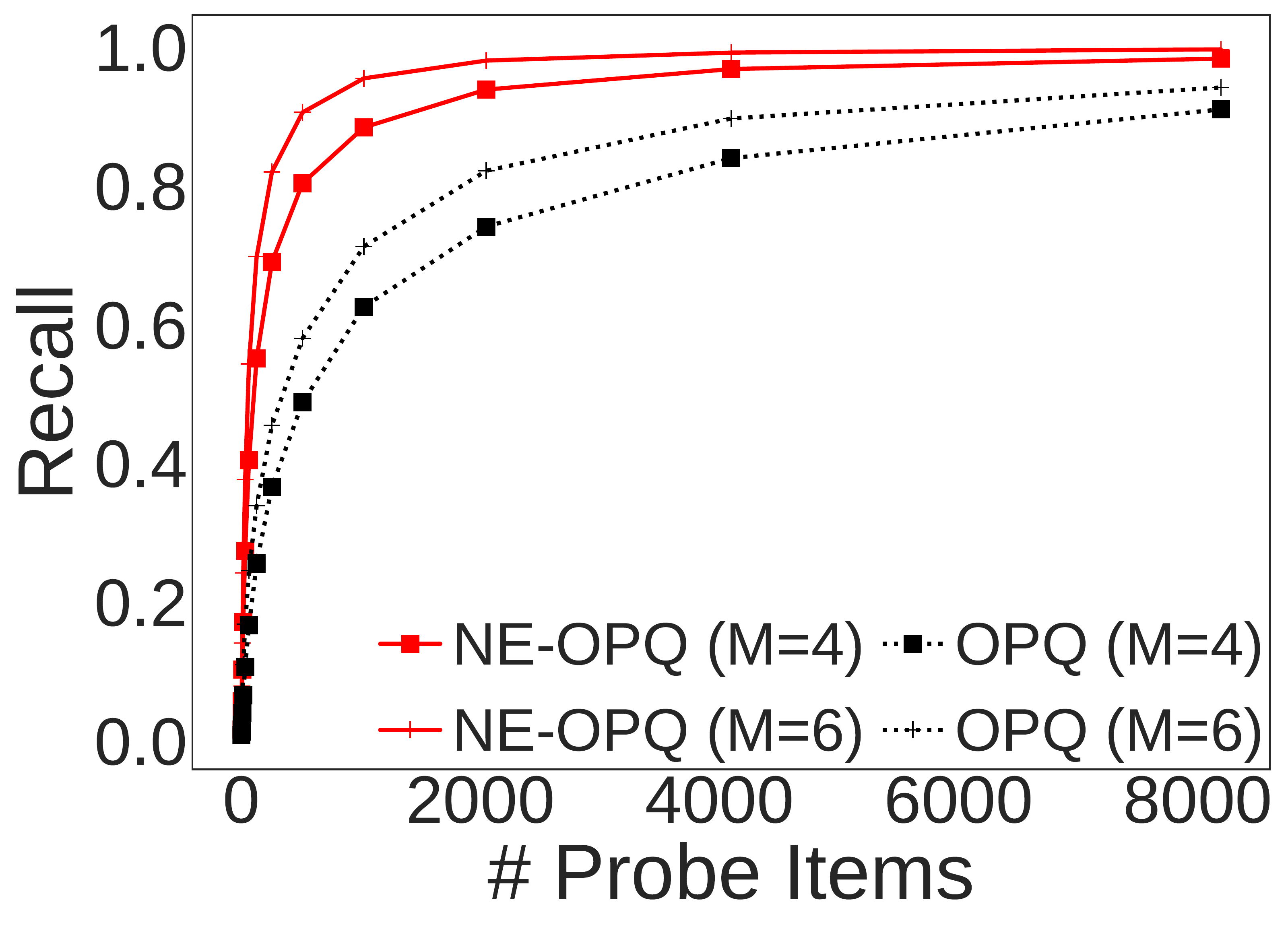}
		\includegraphics[width=0.49\textwidth]{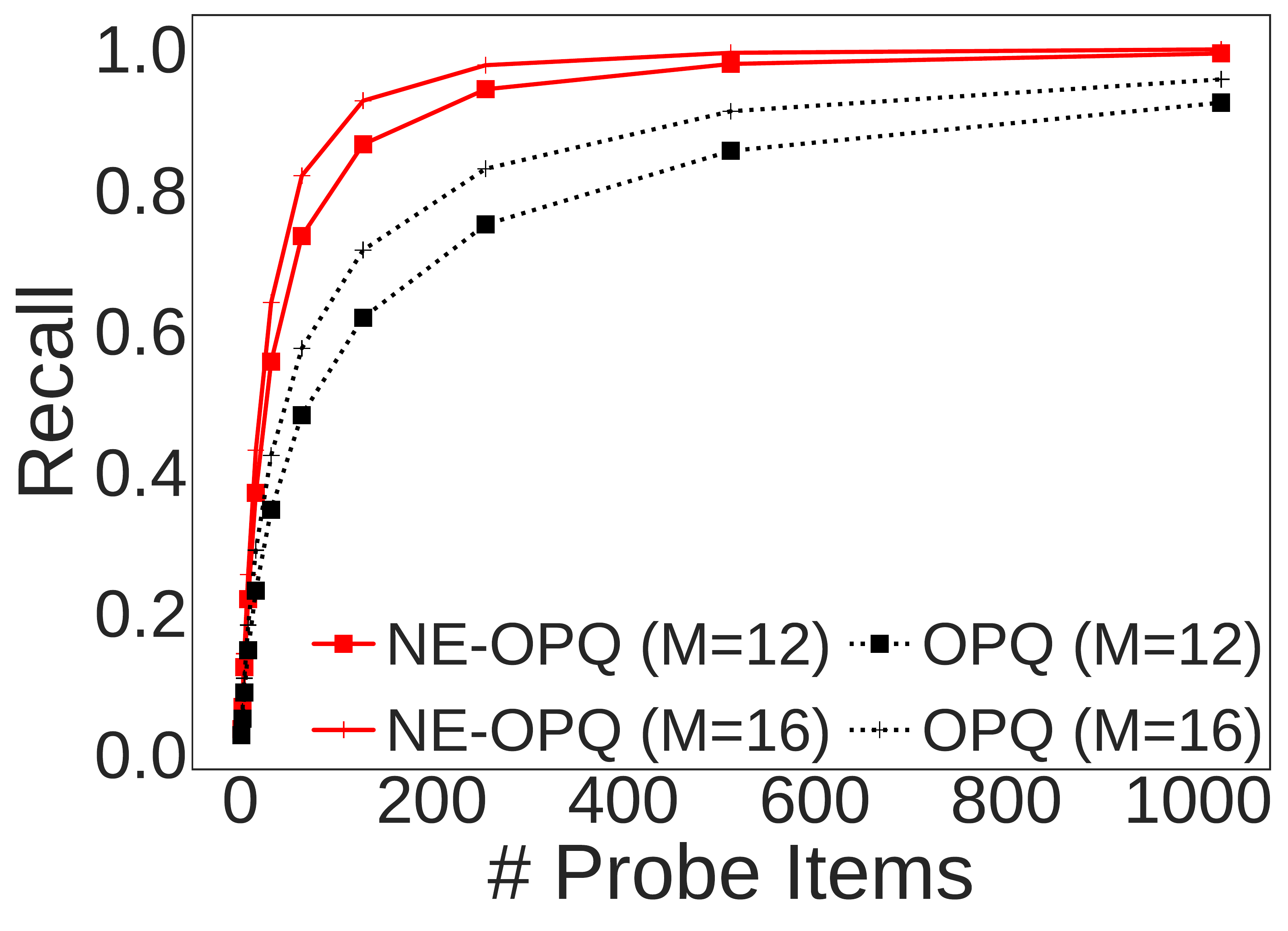}
		\caption{Different \# codebooks for OPQ}\label{fig:number of codebooks for OPQ on imagenet}
	\end{minipage}
	\begin{minipage}[b]{0.49\textwidth}
		\includegraphics[width=0.49\textwidth]{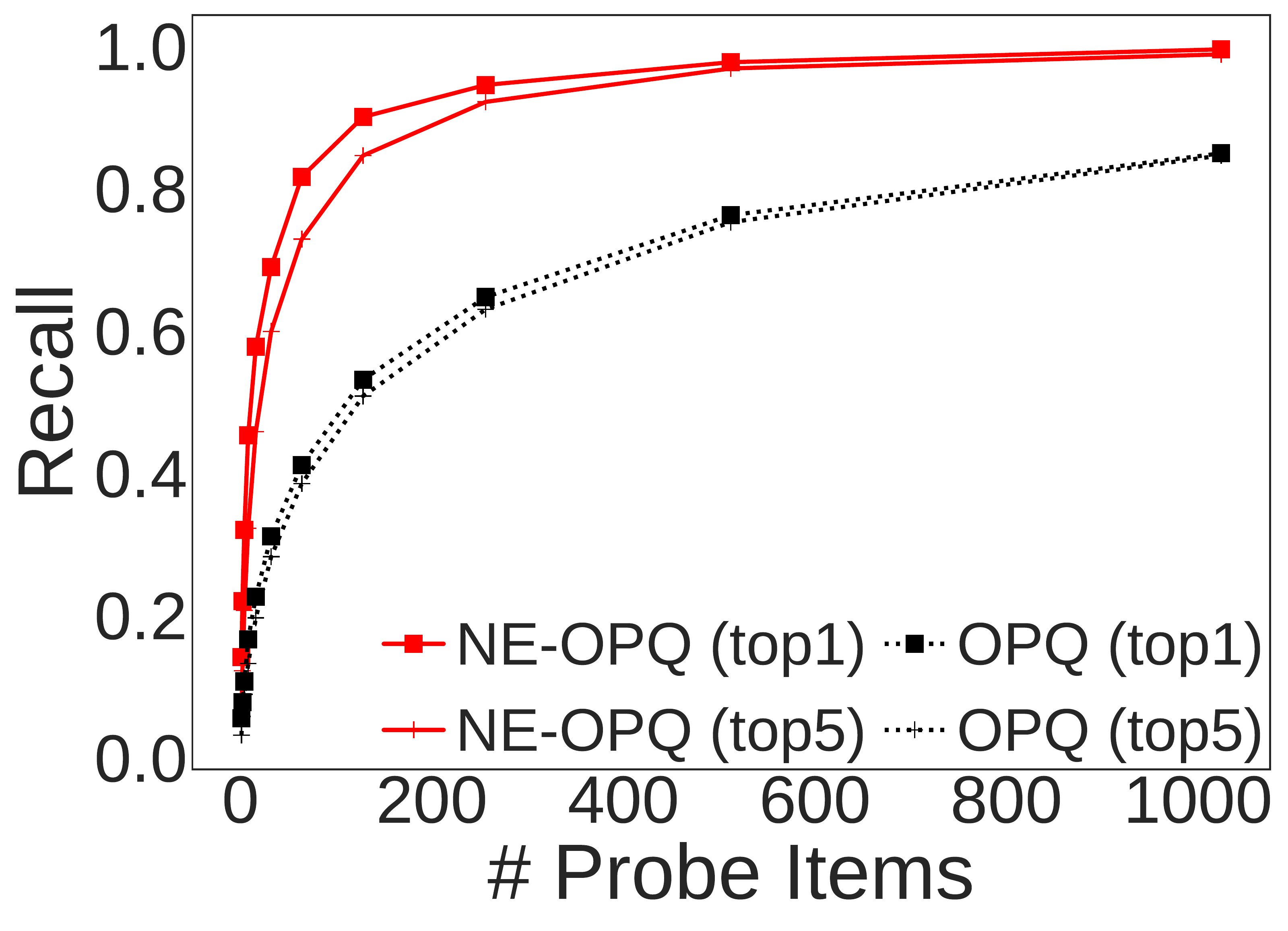}
		\includegraphics[width=0.49\textwidth]{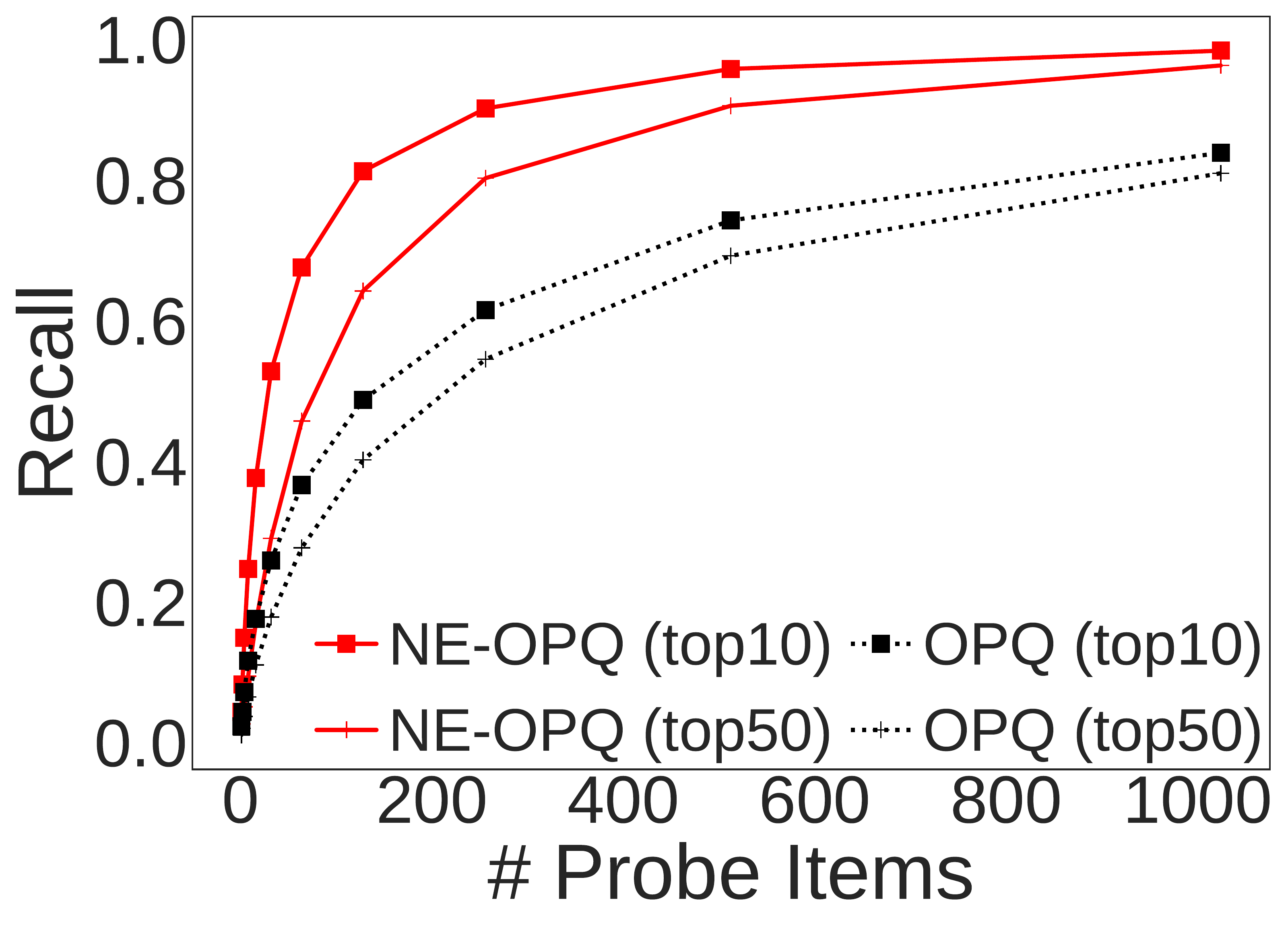}
		\caption{Different values of $k$ for OPQ}\label{fig:value of k for OPQ on imagenet}
	\end{minipage}
\end{figure}
\begin{figure}[!h]	
	\centering 
	\begin{minipage}[b]{0.49\textwidth}
		\includegraphics[width=0.49\textwidth]{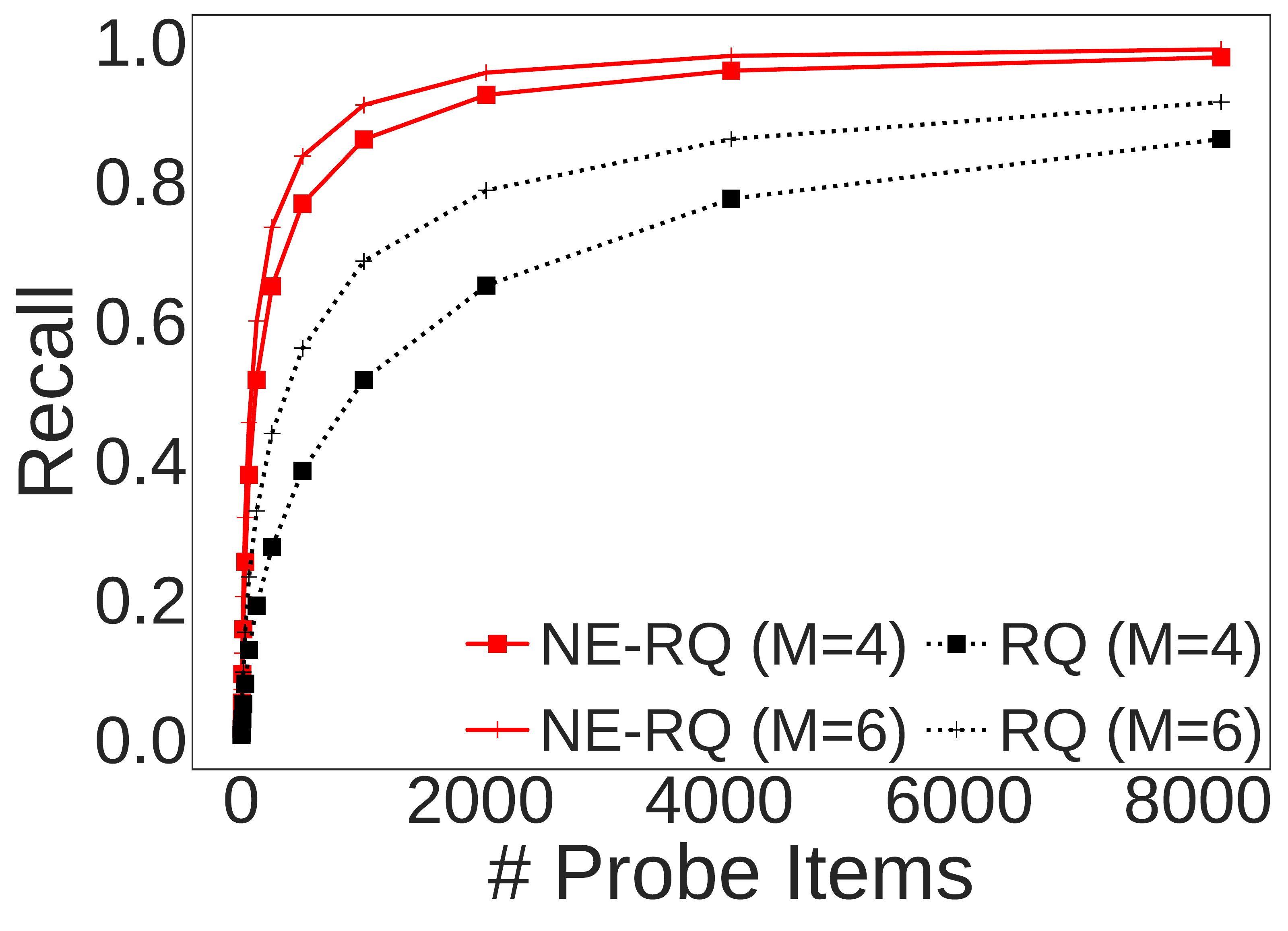}
		\includegraphics[width=0.49\textwidth]{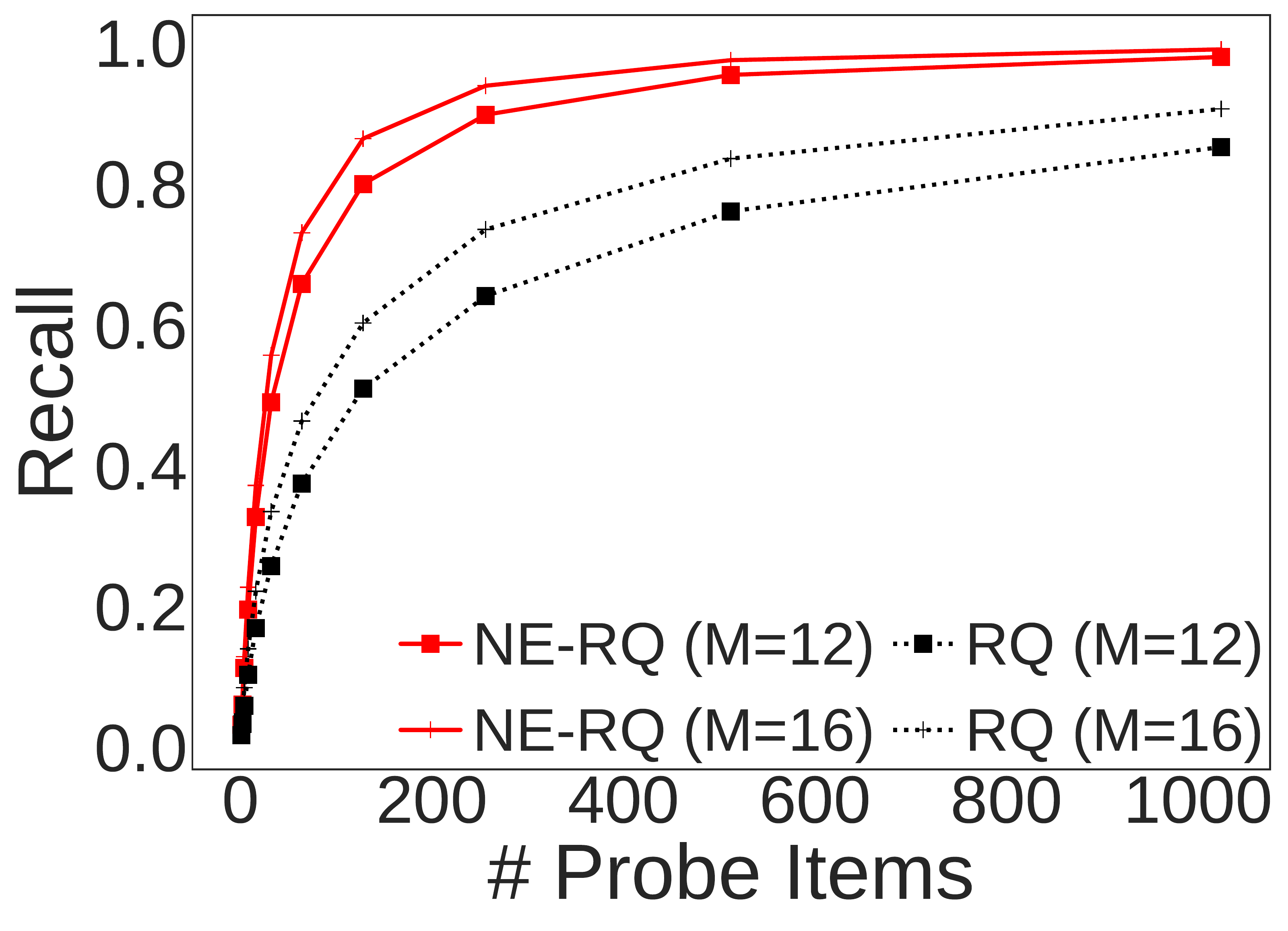}
		\caption{Different \# codebooks for RQ}\label{fig:number of codebooks for RQ on imagenet}
	\end{minipage}
	\begin{minipage}[b]{0.49\textwidth}
		\includegraphics[width=0.49\textwidth]{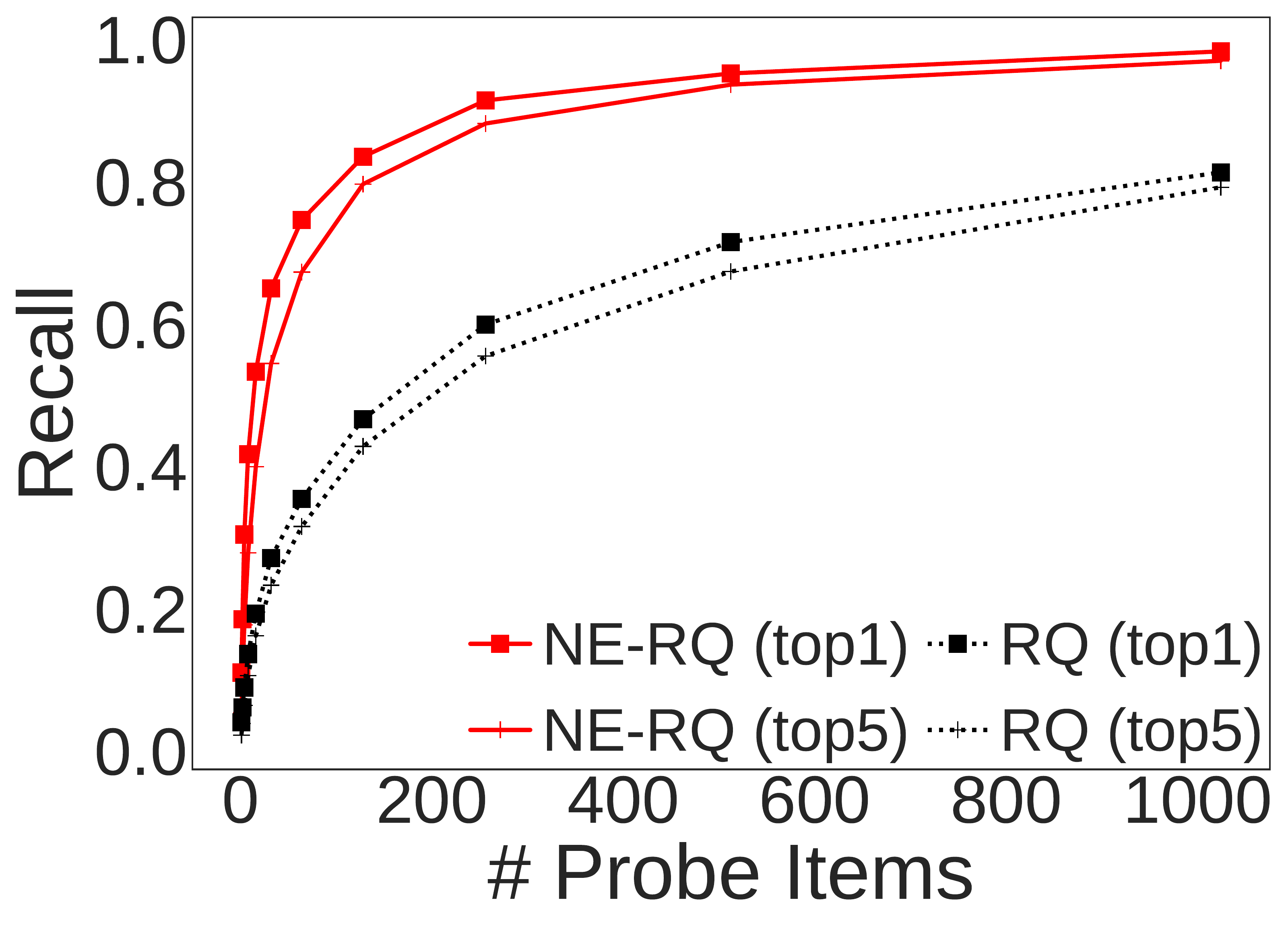}
		\includegraphics[width=0.49\textwidth]{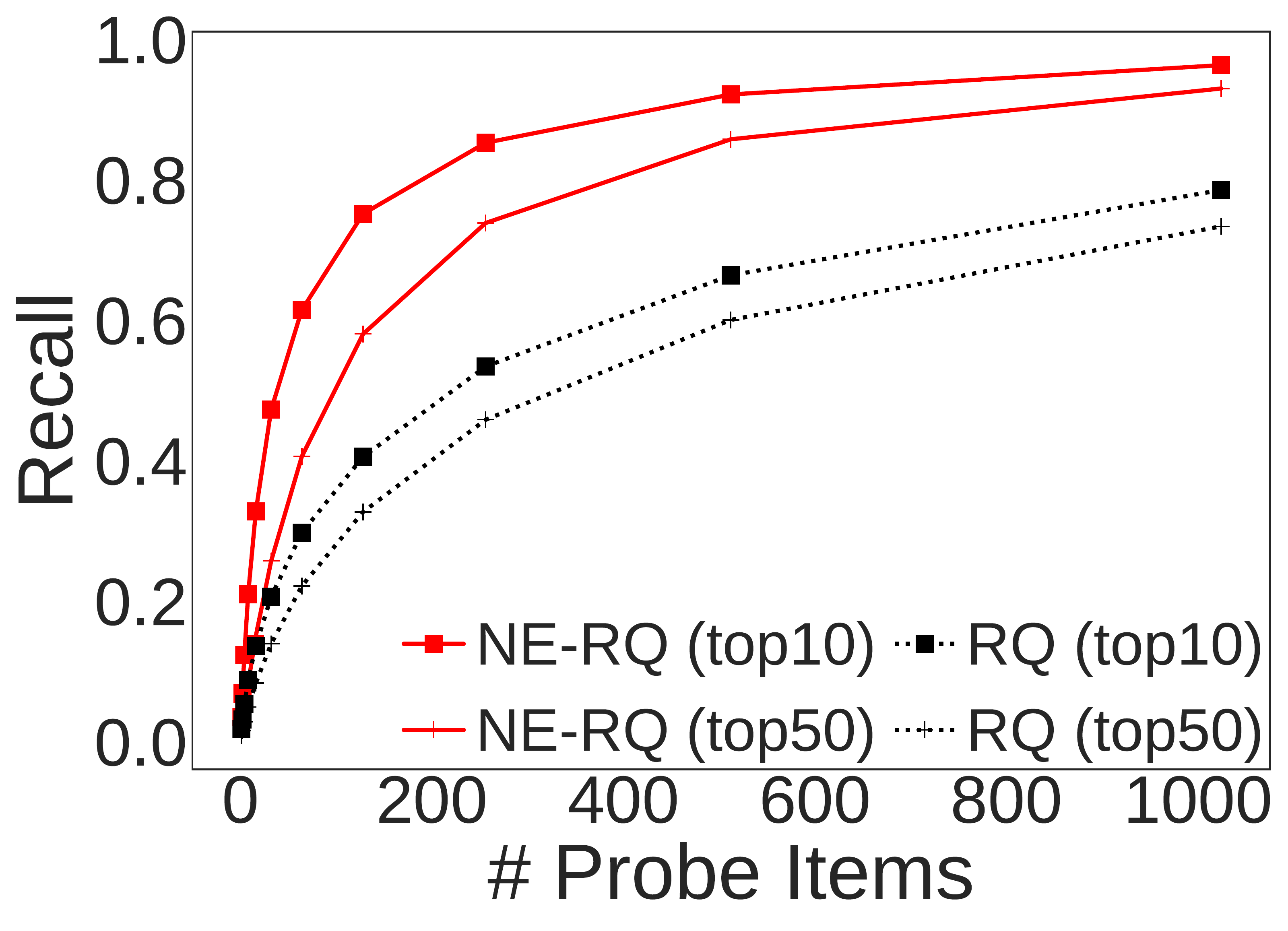}
		\caption{Different values of $k$ for RQ}\label{fig:value of k for RQ on imagenet}
	\end{minipage}
\end{figure}

\begin{figure}[!h]	
	\centering 
	\begin{minipage}[b]{0.49\textwidth}
		\includegraphics[width=0.49\textwidth]{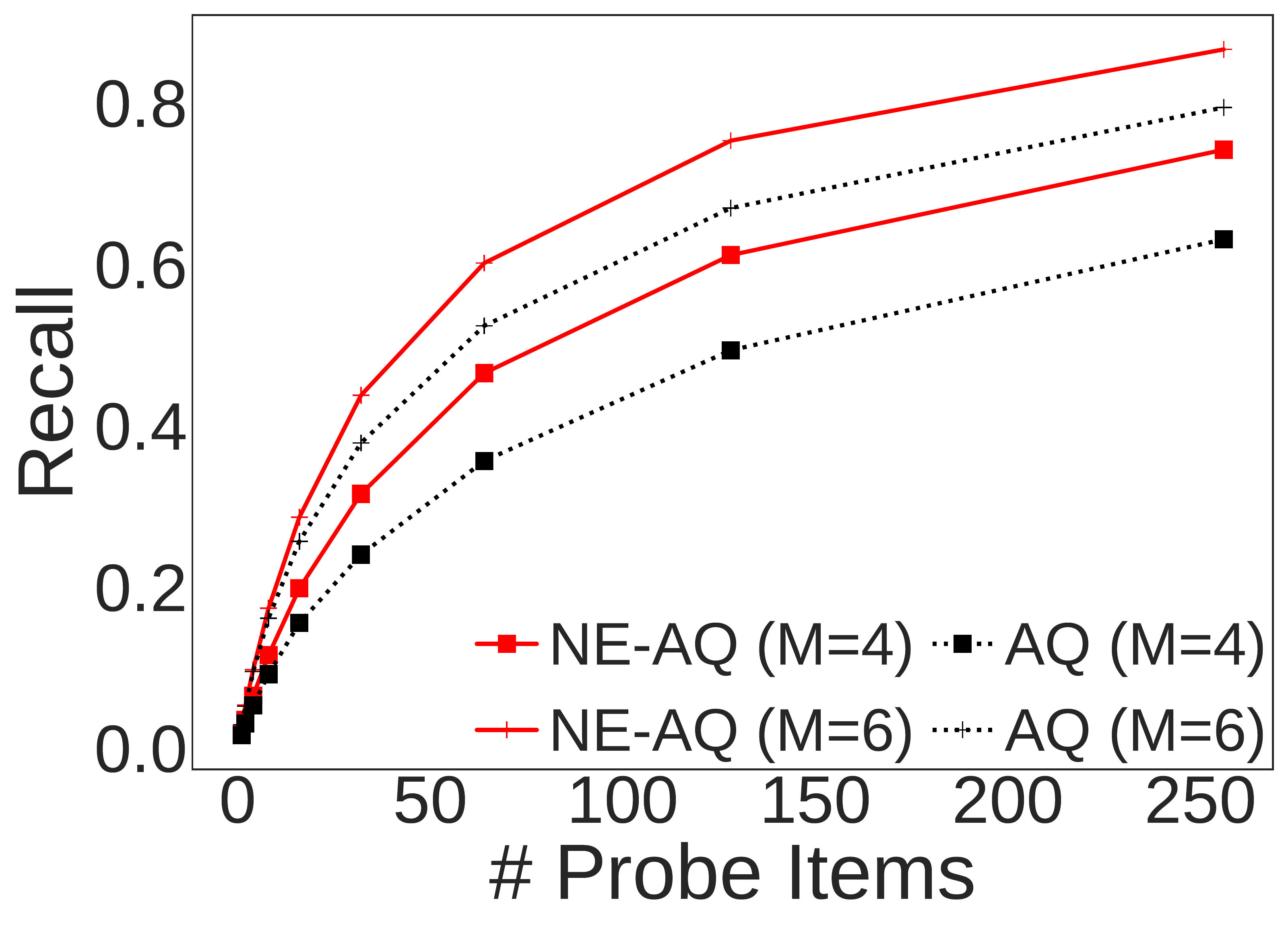}
		\includegraphics[width=0.49\textwidth]{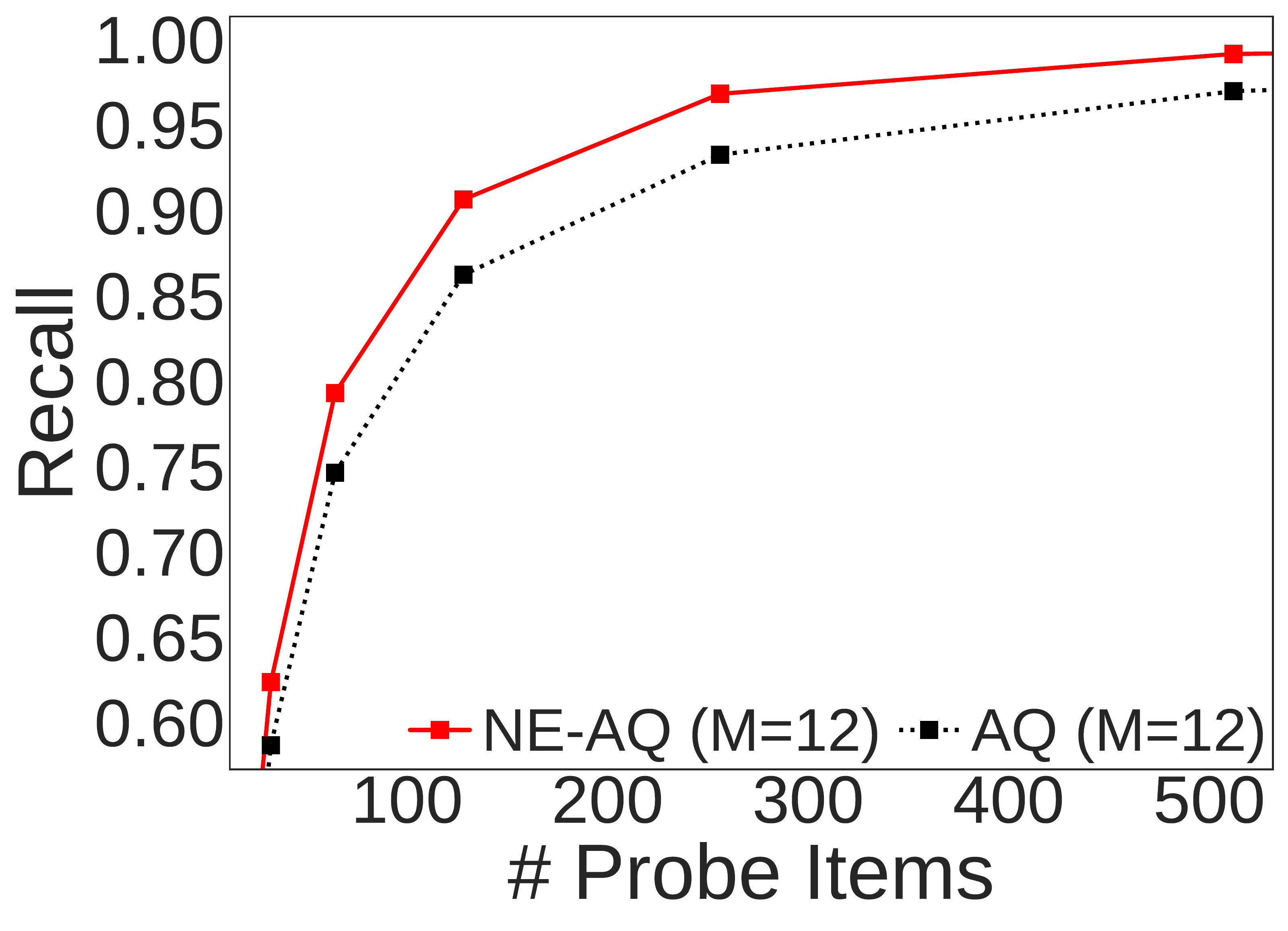}
		\caption{Different \# codebooks for AQ}\label{fig:number of codebooks for AQ on imagenet}
	\end{minipage}
	\begin{minipage}[b]{0.49\textwidth}
		\includegraphics[width=0.49\textwidth]{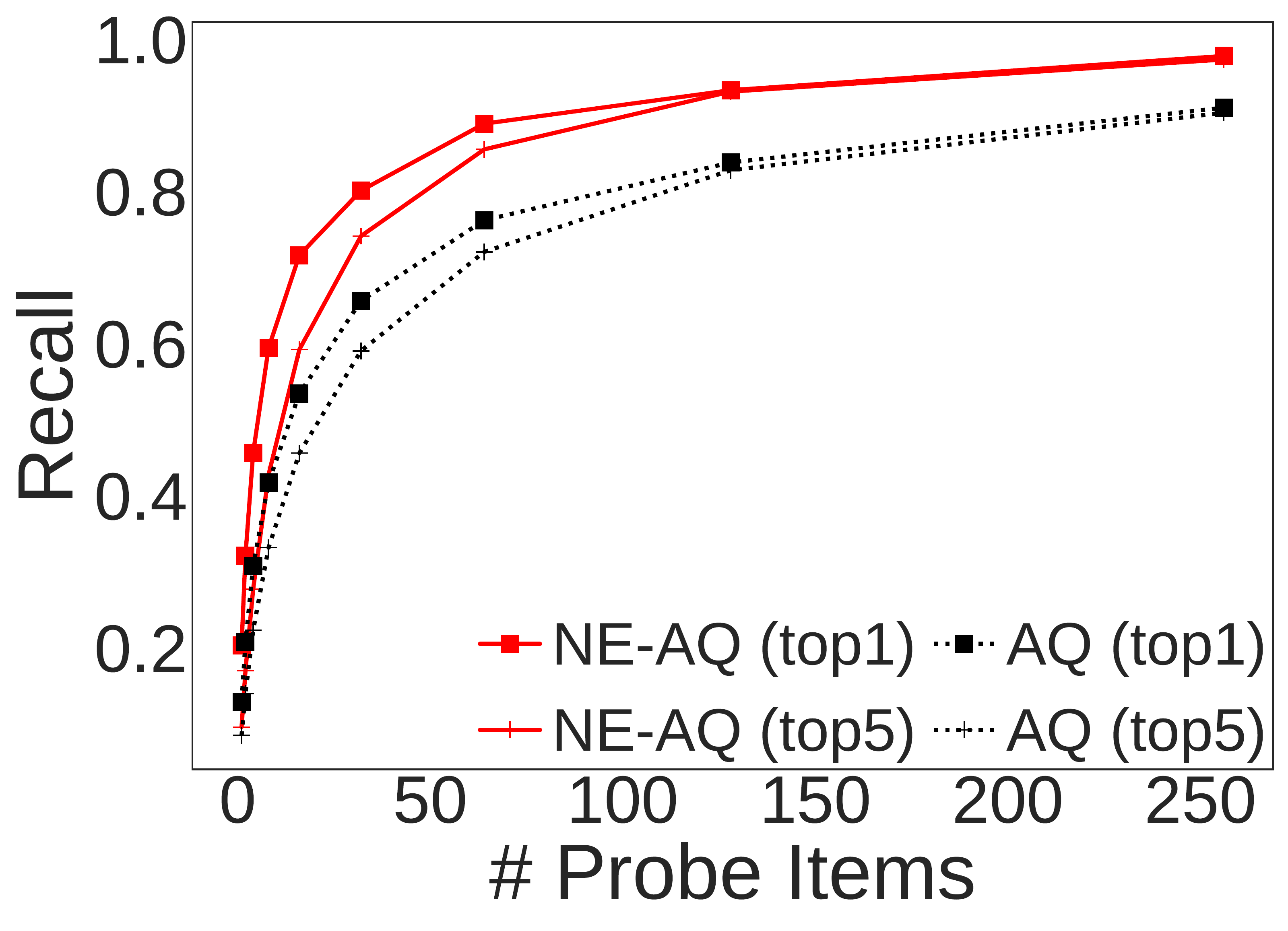}
		\includegraphics[width=0.49\textwidth]{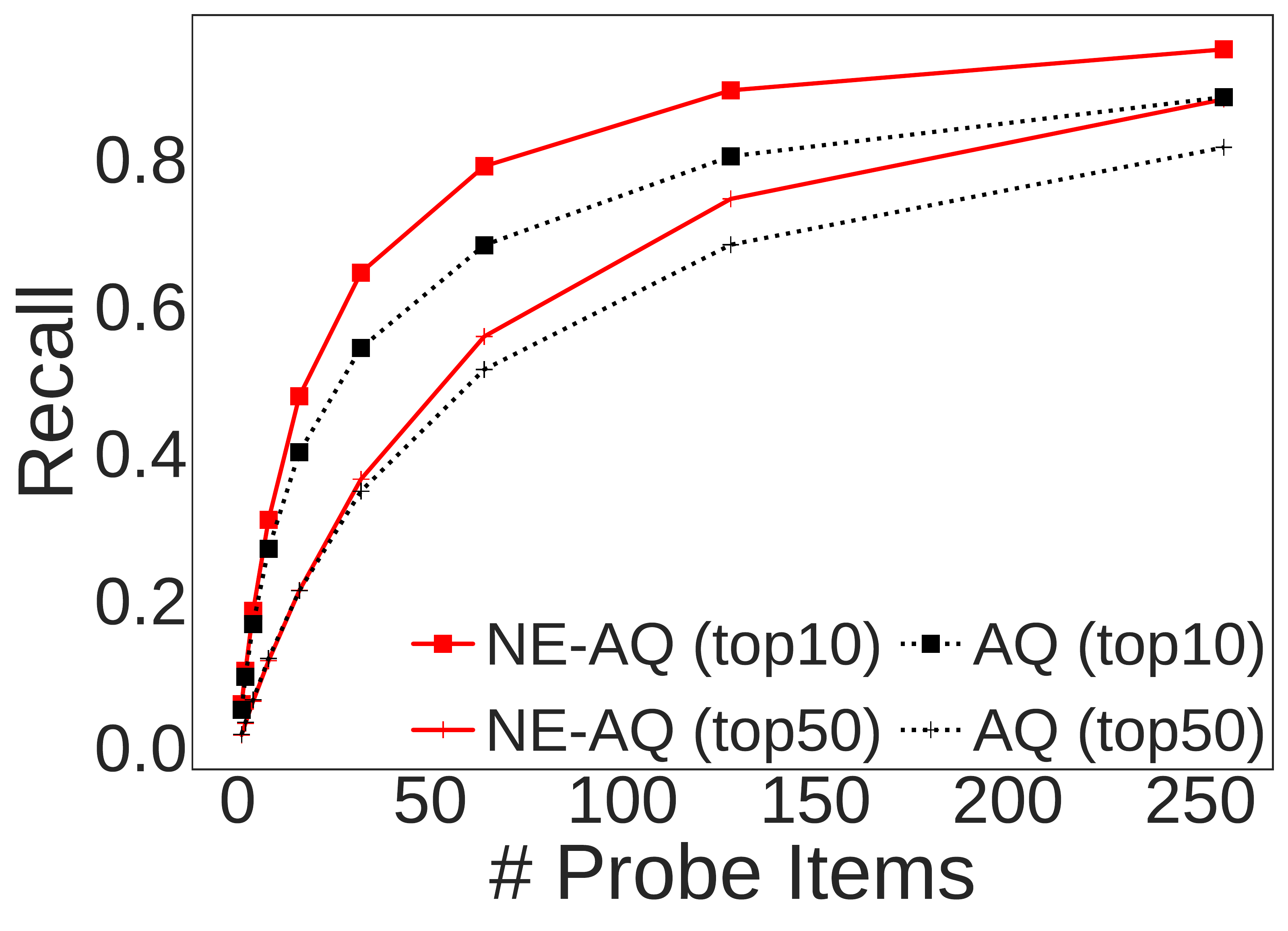}
		\caption{Different values of $k$ for AQ}\label{fig:value of k for AQ on imagenet}
	\end{minipage}
\end{figure}


\subsection{Recall vs. inner product computation when comparing with ip-NSW}

\begin{figure}[!h]	
	\centering 
	\includegraphics[width=0.3\textwidth]{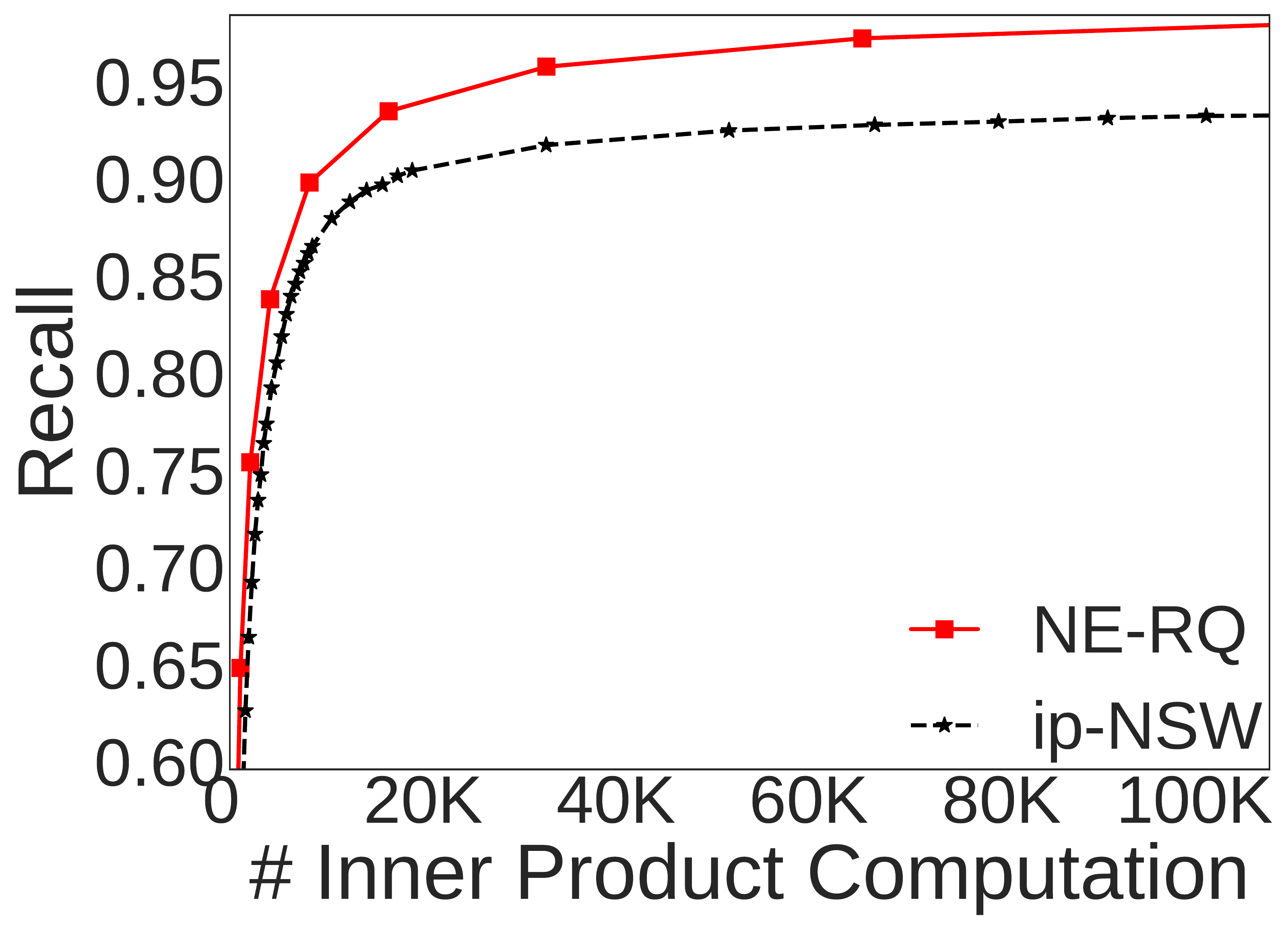}
	\caption{Recall vs. inner product computation comparison between NE-RQ and ip-NSW on ImageNet for top-20 MIPS}\label{fig:recall vs. item}
\end{figure}

In the main paper, we have shown that NE-RQ with two codebooks provides better recall-time performance than ip-NSW on the ImageNet dataset. We plot the recall vs. the number of inner product computation in Figure~\ref{fig:recall vs. item} to exclude the influence of the implementation on the running time.    

\subsection{Robustness across different runs}

\begin{figure*}[!t]	
	\centering 
	\includegraphics[width=0.3\textwidth]{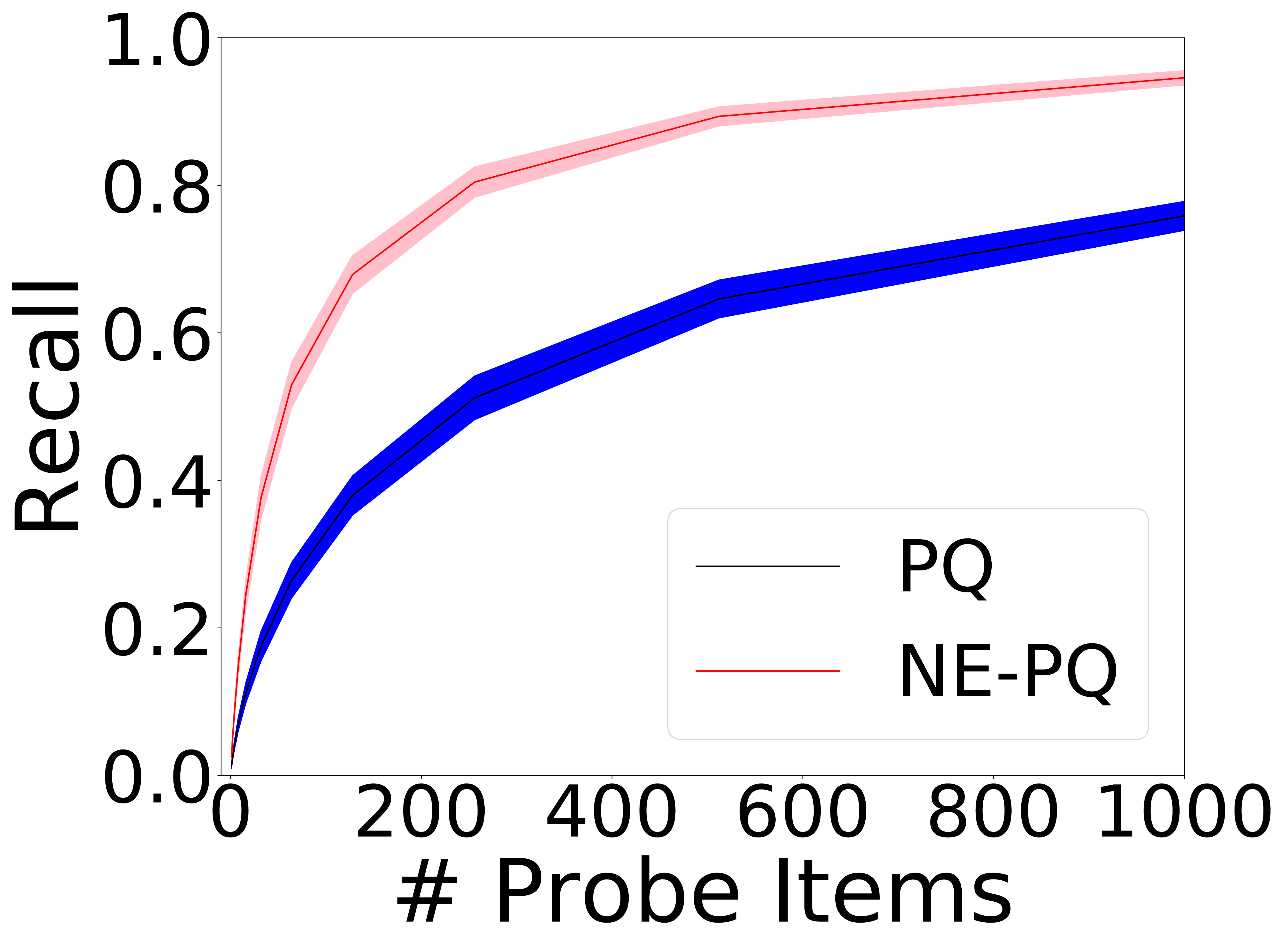}
	\includegraphics[width=0.3\textwidth]{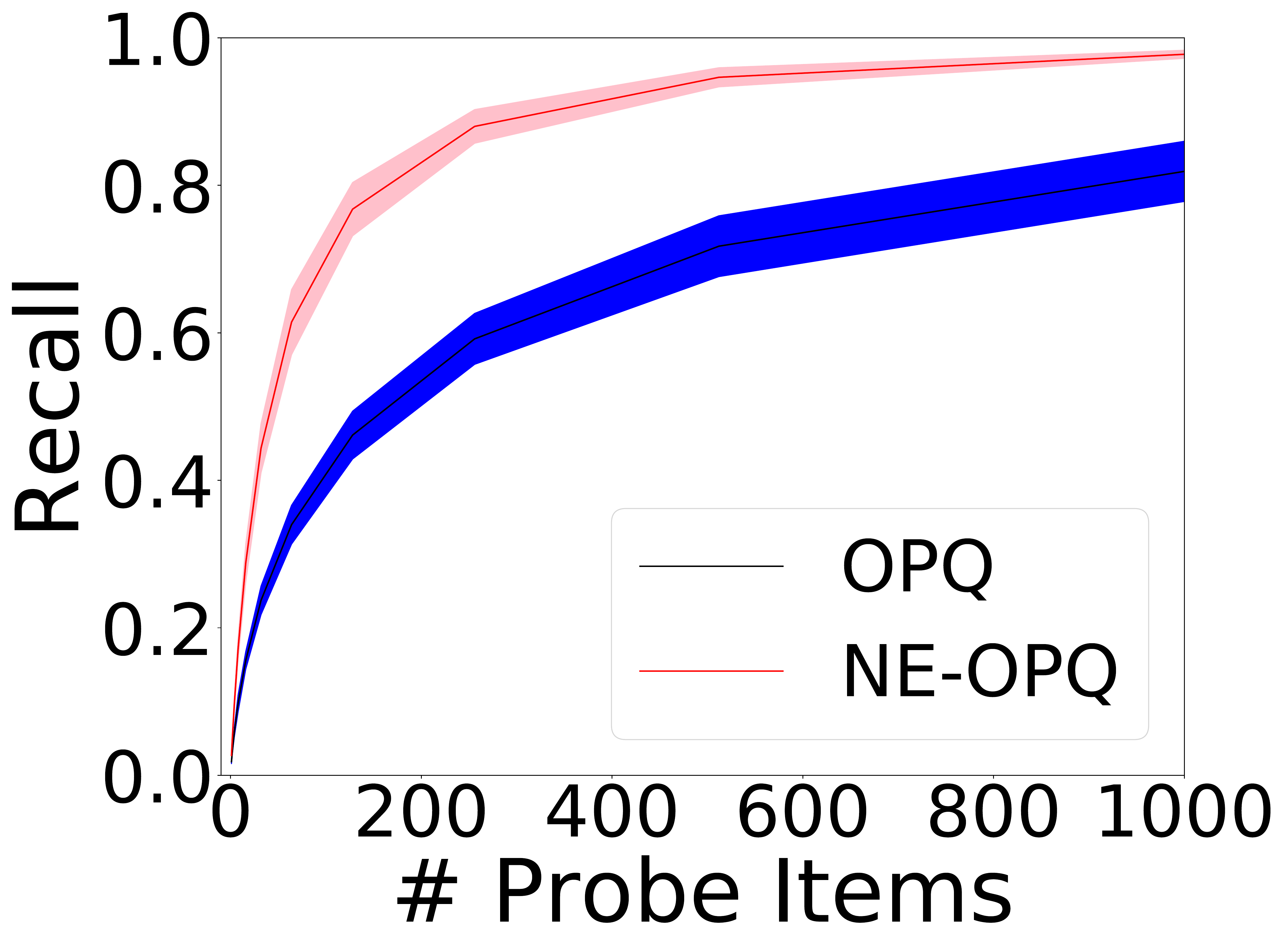}
	\includegraphics[width=0.3\textwidth]{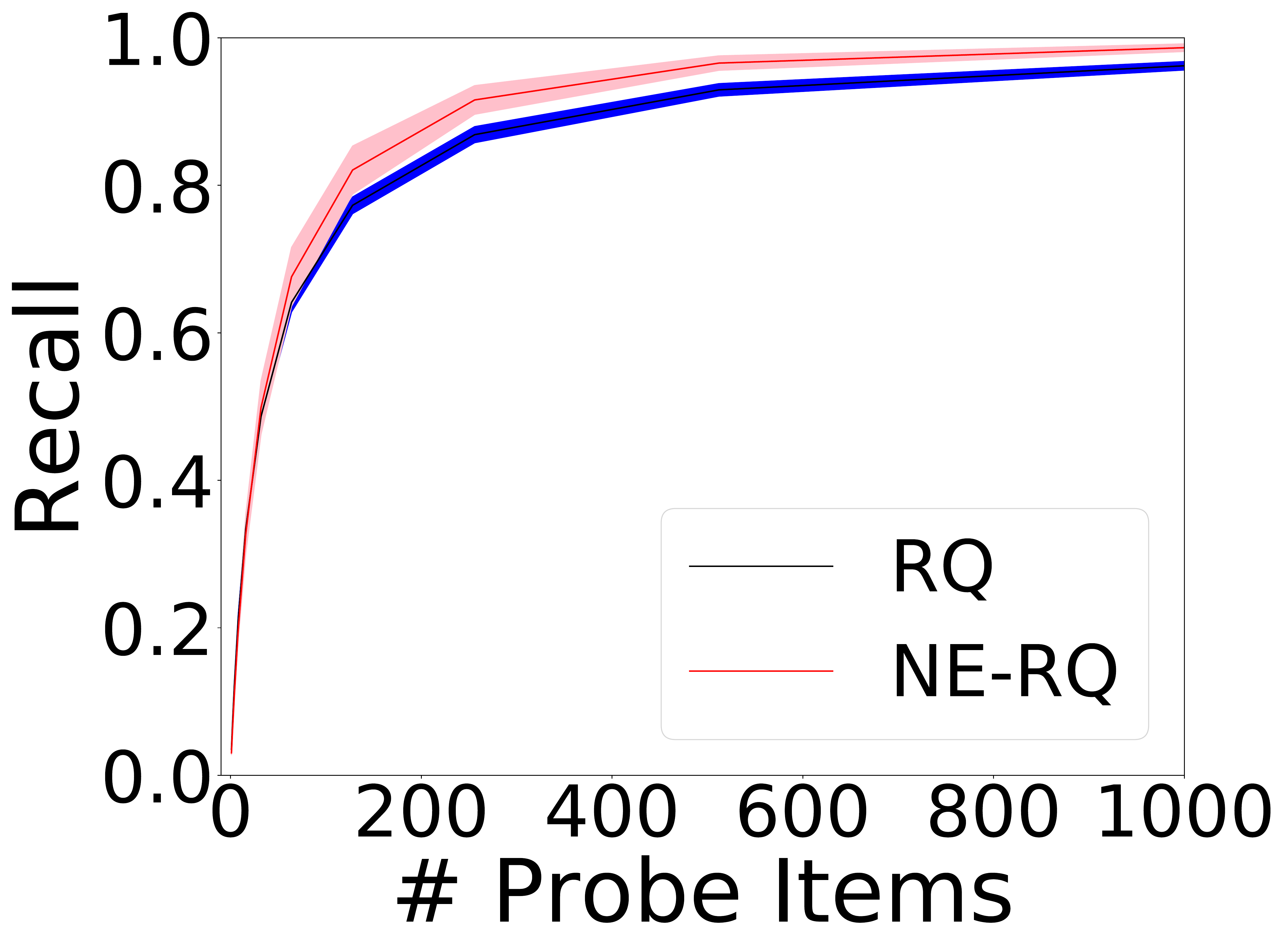}
	\caption{Robustness across different runs of the codebook learning algorithm for PQ (left), OPQ (middle) and RQ (right) on the ImageNet dataset for top-20 MIPS. The central line is the mean recall of 10 runs while the shadowed area indicates 3 times of the standard deviation}\label{fig:variance}
\end{figure*}

The vector quantization algorithms have randomness as they use k-means (which is random) to learn the codebooks. Therefore, in different runs of the codebook learning algorithm (producing different sets of codebooks), the same algorithm may produce different recalls when a fixed number of items are probed. To evaluate the robustness of NEQ, we conduct the codebook learning algorithm 15 times and obtain 15 recalls when a fixed number of items are probed. We plot the mean and 3 times of the standard deviation of the 15 runs in Figure~\ref{fig:variance}. The results show that NEQ-PQ and NE-OPQ are more robust than PQ and OPQ while NEQ-RQ and RQ are comparable.

\end{document}